\documentclass[aps,prb,twocolumn,eqsecnum]{revtex4}

\usepackage{graphicx}

\usepackage{bm}
\usepackage{amsmath}
\usepackage{amssymb}

\newcommand{\ocite}{\onlinecite}

\renewcommand{\i}{\infty}

\newcommand{\x}{\text}
\newcommand{\pd}{\partial}
\newcommand{\dg}{\dagger}
\renewcommand{\ln}{\langle}
\newcommand{\rn}{\rangle}
\newcommand{\lt}{\left}
\newcommand{\rt}{\right}

\newcommand{\f}{\frac}
\newcommand{\tf}{\tfrac}
\newcommand{\q}{\sqrt}
\newcommand{\lbl}{\label}

\newcommand{\nn}{\nonumber}
\newcommand{\cd}{\cdot}

\newcommand{\p}{\perp}
\newcommand{\n}{\nabla}

\newcommand{\um}{\hat{1}}

\newcommand{\tm}{\times}
\newcommand{\ot}{\otimes}
\newcommand{\os}{\varnothing}

\newcommand{\eq}[1]{Eq.~(\ref{eq:#1})}
\newcommand{\eqs}[2]{Eqs.~(\ref{eq:#1}) and (\ref{eq:#2})}
\newcommand{\eqss}[3]{Eqs.~(\ref{eq:#1}), (\ref{eq:#2}), and (\ref{eq:#3})}
\newcommand{\eqn}[1]{(\ref{eq:#1})}
\newcommand{\eqsn}[2]{(\ref{eq:#1}) and (\ref{eq:#2})}
\newcommand{\eqssn}[3]{(\ref{eq:#1}), (\ref{eq:#2}), and (\ref{eq:#3})}

\newcommand{\secr}[1]{Sec.~\ref{sec:#1}}
\newcommand{\secsr}[2]{Secs.~\ref{sec:#1} and \ref{sec:#2}}
\newcommand{\figr}[1]{Fig.~\ref{fig:#1}}
\newcommand{\figsr}[2]{Figs.~\ref{fig:#1} and \ref{fig:#2}}

\newcommand{\spc}{\mbox{ }}

\newcommand{\beq}{\begin{equation}}
\newcommand{\eeq}{\end{equation}}
\newcommand{\beqar}{\begin{eqnarray}}
\newcommand{\eeqar}{\end{eqnarray}}
\newcommand{\beqarn}{\begin{eqnarray*}}
\newcommand{\eeqarn}{\end{eqnarray*}}
\newcommand{\ba}{\begin{array}}
\newcommand{\ea}{\end{array}}
\newcommand{\bwt}{\begin{widetext}}
\newcommand{\ewt}{\end{widetext}}

\newcommand{\ra}{\rightarrow}

\newcommand{\rtarr}{\rightarrow}
\newcommand{\rarr}{\rightarrow}

\newcommand{\Lf}{\mathbb{L}}

\newcommand{\Kf}{\mathbb{K}}
\newcommand{\Ef}{\mathbb{E}}

\newcommand{\dx}{{\text d}}
\newcommand{\ex}{{\text e}}

\newcommand{\ix}{{\text i}}

\newcommand{\Ux}{{\text U}}

\newcommand{\ch}{\hat{c}}

\newcommand{\Dh}{\hat{D}}

\newcommand{\Hh}{\hat{H}}

\newcommand{\Uh}{\hat{U}}

\newcommand{\hr}{{\bar{h}}}

\newcommand{\xr}{{\bar{x}}}

\newcommand{\Der}{\bar{\Delta}}

\newcommand{\Ec}{\mathcal{E}}

\newcommand{\Kc}{\mathcal{K}}

\newcommand{\Nc}{\mathcal{N}}
\newcommand{\Oc}{\mathcal{O}}

\newcommand{\taub}{{\mbox{\boldmath{$\tau$}}}}

\newcommand{\Lt}{{\tilde{L}}}

\newcommand{\nb}{{\bf n}}

\newcommand{\rb}{{\bf r}}

\newcommand{\om}{\omega}

\newcommand{\al}{\alpha}

\newcommand{\de}{\delta}
\newcommand{\De}{\Delta}

\newcommand{\s}{\sigma}

\newcommand{\tht}{\theta}

\newcommand{\ka}{\varkappa}
\newcommand{\vphi}{\varphi}
\newcommand{\eps}{\varepsilon}
\newcommand{\e}{\epsilon}

\begin{document}
\title{Interplay of topology and interactions in quantum Hall topological insulators:\\
U(1) symmetry, tunable Luttinger liquid,
and interaction-induced phase transitions
}
\author{Maxim Kharitonov, Stefan Juergens, and Bj\"orn Trauzettel}
\address{Institute for Theoretical Physics and Astrophysics,
University of W\"urzburg, 97074 W\"urzburg, Germany}
\begin{abstract}

We consider a class of {\em quantum Hall topological insulators}:
topologically nontrivial states with zero Chern number
at finite magnetic field,
in which the counter-propagating edge states are protected by a symmetry (spatial or spin) other than time-reversal.
HgTe-type heterostructures and graphene are among the relevant systems.
We study the effect of electron interactions on the topological properties of the system.
We particularly focus on the vicinity of the topological phase transition,
marked by the crossing of two Landau levels,
where the system is a strongly interacting quantum Hall ferromagnet.
We analyse the edge properties using the formalism of the nonlinear $\sigma$-model.
We establish the symmetry requirement for the topological protection in this interacting system:
effective continuous U(1) symmetry with respect to uniaxial isospin rotations must be preserved.
If U(1) symmetry is preserved, the topologically nontrivial phase persists;
its edge is a helical Luttinger liquid with highly tunable effective interactions.
We obtain explicit analytical expressions for the parameters of the Luttinger liquid.
However, U(1) symmetry may be broken, either spontaneously or by U(1)-asymmetric interactions.
In either case, interaction-induced transitions occur to the respective topologically trivial phases
with gapped edge charge excitations.

\end{abstract}
\maketitle

\section{Introduction\lbl{sec:intro}}

\begin{figure}
\includegraphics[width=.40\textwidth]{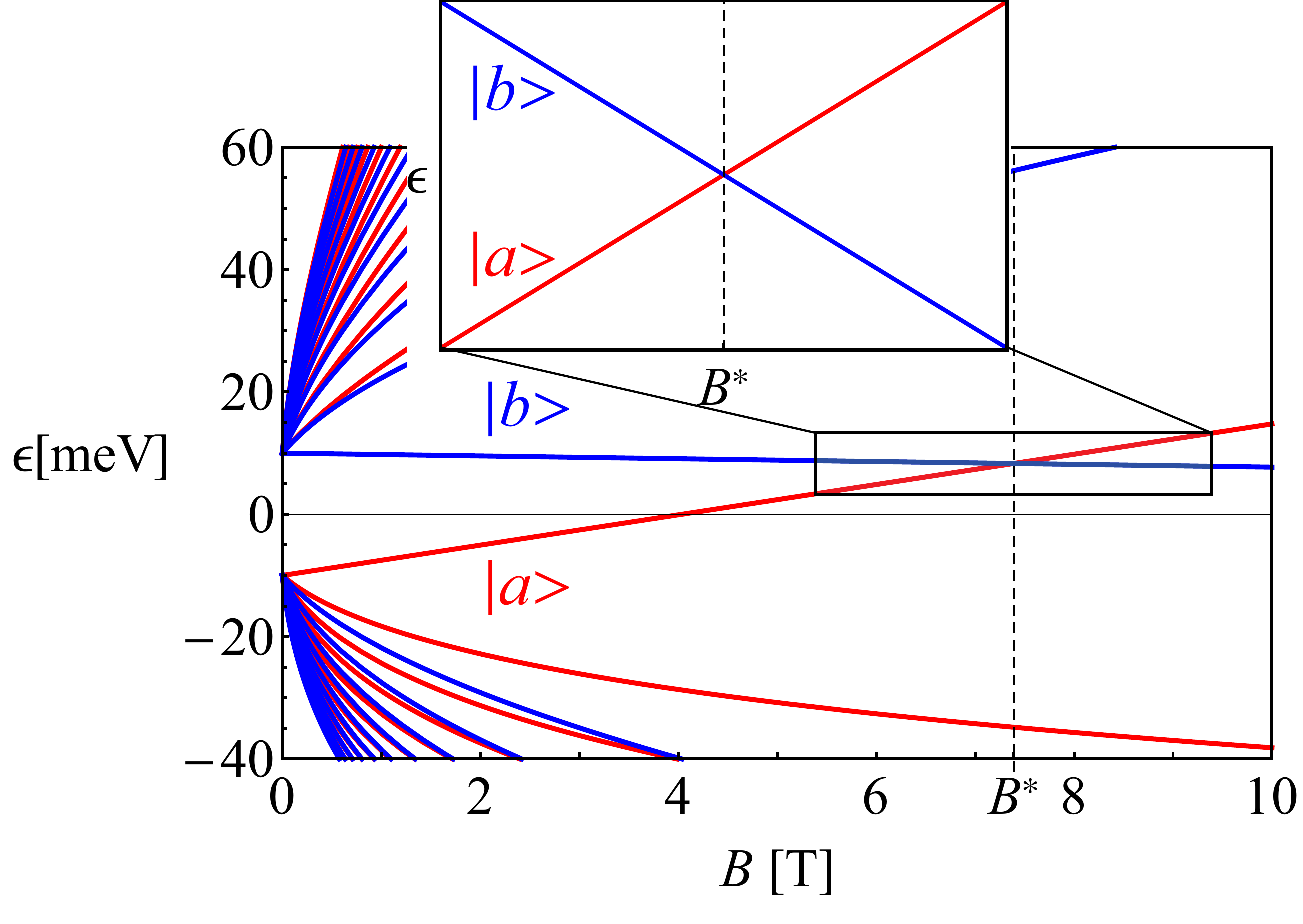}
\caption{(Color online)
One possible type of the Landau level (LL) structure of a quantum Hall topological insulator (QHTI).
At the single-particle topological phase transition point, LLs of different symmetries, labeled $a$ (red) and $b$ (blue), cross.
The spectrum shown was calculated for the Bernevig-Hughes-Zhang model~\cite{BHZ,Scharf} with a perpendicular orientation of the magnetic field;
in this case, LLs $a$ are $b$ are distinguished by the spatial inversion parity, even and odd.
}
\lbl{fig:LLs}
\end{figure}

\begin{figure}
\includegraphics[width=.35\textwidth]{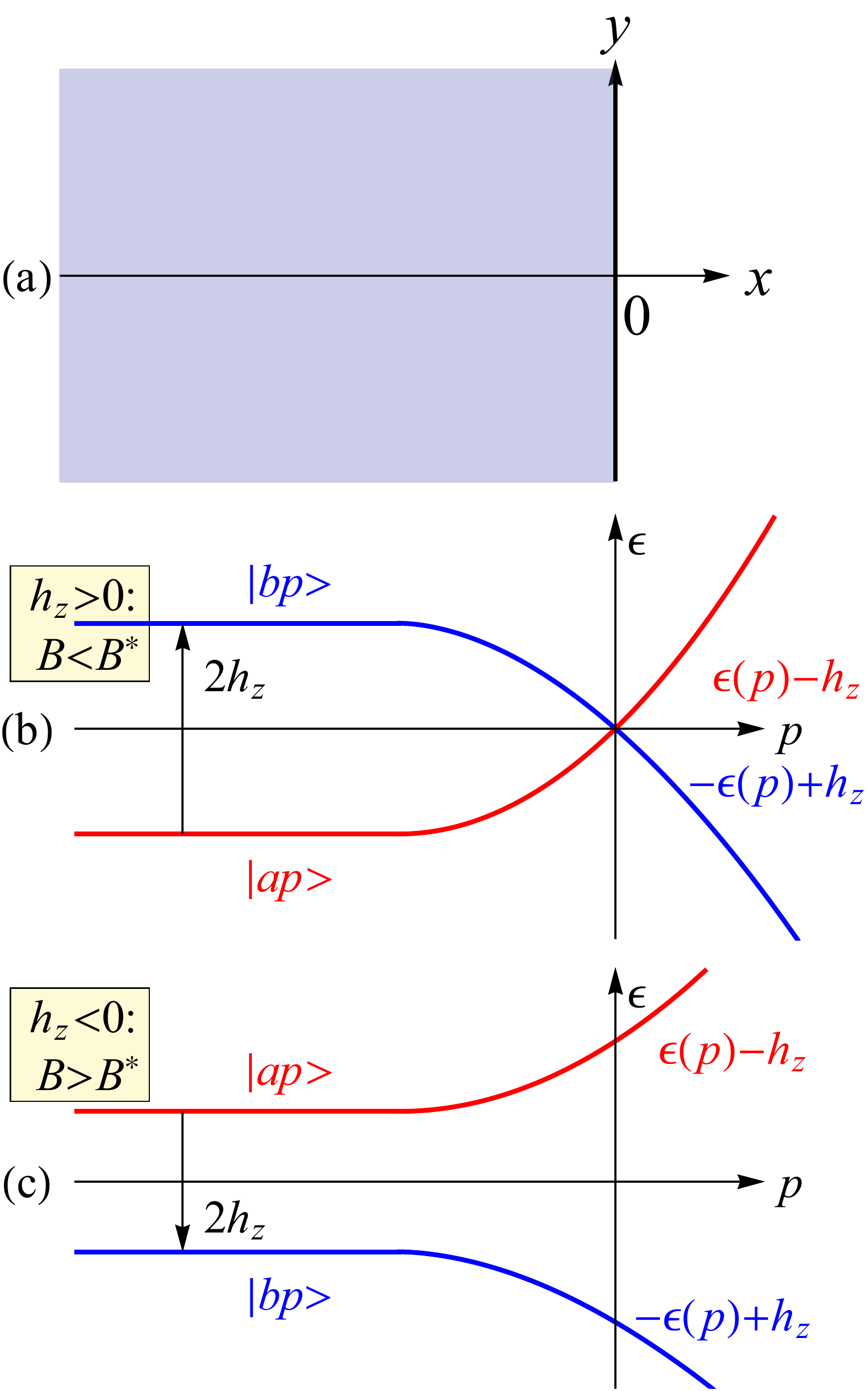}
\caption{(Color online)
(a) Half-infinite sample occupying the region $x\leq 0$.
(b) and (c) Schematics of the edge spectrum of a QHTI in the Landau gauge,
including only the two intersecting LLs $a$ and $b$ of interest, see \figr{LLs}.
The single-particle states are labeled by the conserved 1D momentum $p$ in the $y$ direction.
The states located in the bulk and at the edge correspond to the values $p\lesssim0$ and $p\gtrsim0$, respectively.
In the topologically non-trivial (TnT) phase at lower fields $B<B^*$ (b),
the edge states cross and are gapless.
In the topologically trivial (TT) phase at higher fields $B^*<B$ (c),
the edge states do not cross and are gapped.}
\lbl{fig:edgestates}
\end{figure}

Interacting topological~\cite{KaneMele,BHZ,Koenig,Hsieh,Xia,Knez,Li,HasanKane,QiZhang} systems
are currently an active area of research~\cite{HohenadlerAssaad,Pesin,Dzero1,Dzero2,Lemonik,
Cvetkovic,Amaricci,helicalLL}.
Of particular interest are the situations,
when interactions can change the single-particle picture in a qualitative way
and lead to effects not present in the non-interacting system.
Theoretical proposals of such a nontrivial behavior include a ``topological Mott insulator''~\cite{Pesin},
``Kondo topological insulator''~\cite{Dzero1,Dzero2},
interaction-induced topological phases in graphene~\cite{Lemonik,Cvetkovic},
and first-order topological phase transitions~\cite{Amaricci}.

Most of these predictions require sufficiently strong electron interactions
and were made for ``strongly correlated'' materials.
Meanwhile, most of the materials to date that have been experimentally firmly established
as topological insulators, such as  HgTe/CdTe~\cite{Koenig}, BiSb~\cite{Hsieh}, or BiSe~\cite{Xia} compounds,
are rather weakly interacting due to efficient screening of the Coulomb interactions.

It is thus desirable to expand the range of possibilities
to attain the regime of strong effective interactions in topological systems,
and it is even more desirable to be able to tune the strength of interactions by experimentally feasible means.

In this work, we have identified a class of topological systems,
in which such conditions can be realized even for weak bare interactions
by applying the orbital magnetic field. The interactions are tunable by the magnetic field
and their strength is controlled by the proximity to the topological phase transition.
The vicinity of the topological phase transition is automatically the regime of strong effective interactions,
in which Coulomb interactions are crucial for both bulk and edge properties
and lead to a nontrivial interplay of topological and interacting phenomena.

An important theoretical advantage of such a system is that it can be analyzed
in a well-controlled way. In particular, this allowed us to determine the symmetry requirements
for topological protection in this system,
which is one of the key questions raised in the studies of interacting topological systems.

\subsection{Quantum Hall topological insulators\lbl{sec:QHTIs}}

We consider a class of electron systems that we dub {\em quantum Hall topological insulators} (QHTIs):
(quasi) two-dimensional (2D) electron systems with zero Chern number $\nu=0$ at finite magnetic field $B$
that can still be topologically non-trivial (TnT) and exhibit counter-propagating edge states.
Since the time-reversal symmetry is broken by the magnetic field,
the TnT phase must be protected by some other symmetry.
In a system with appreciable spin-orbit interaction, such symmetry is some spatial symmetry (e.g., inversion, reflection, or rotation).
In a system with negligible spin-orbit interaction, axial spin rotation symmetry can also play the role of such symmetry.
We will refer to this symmetry responsible for the topological protection of a {\em noninteracting} QHTI
as the {\em physical symmetry}, in order to contrast it to the {\em effective}, or {\em emergent},
U(1) symmetry, which will be demonstrated to be central for an {\em interacting} system.

One possible type of the Landau level (LL) structure of a QHTI
would exhibit a crossing of two LLs at some value $B^*$ of the magnetic field:
one LL, to be labelled $a$, originates from the valence band and moves upward with increasing $B$,
and the other LL, to be labelled $b$, originates from the conduction band and moves downward, see \figr{LLs}.
This crossing is a point of the topological phase transition of a QHTI,
separating the TnT phase with counter-propagating states at lower $B<B^*$,
and the topologically trivial (TT) phase with gapped edge states at higher $B>B^*$, see \figr{edgestates}.
Other variants of the LL structure in QHTIs are also possible.

A number of previously studied theoretical models and real physical systems are relevant to the class of QHTIs.
The single-particle behavior of \figsr{LLs}{edgestates} has been identified~\cite{Scharf}
in the Bernevig-Hughes-Zhang model~\cite{BHZ}
for the direction of the magnetic field perpendicular to the 2D structure.
This behavior is likely to have topological origin and be protected a spatial symmetry.
This model is directly relevant to HgTe/CdTe~\cite{Koenig}
and InAs/GaSb~\cite{Knez,Li} heretostructures,
which are established 2D topological insulators at zero field, protected by the time-reversal symmetry.

Other likely QHTI systems are graphene single- and multi-layer structures.
Noninteracting graphene exhibits counter-propagating edge states~\cite{ALL}
at finite magnetic field due to spin splitting by the Zeeman field and the fact that graphene is a semimetal.
It can be seen as a QHTI protected by the continuous axial spin rotation symmetry~\cite{Young}.
Although directly relevant, graphene also has a few peculiarities and its LL structure differs from that in \figr{LLs};
the focus of the present work are the QHTIs with a spectrum of the type in \figr{LLs}.

To be clear, we emphasize that according to the above definition
QHTIs are not necessarily new topological systems symmetry-wise, in regard to the existing classifications~\cite{CFang,ShiozakiSato,LuLee,Chiu}.
The key requirement here is that the orbital magnetic field is explicitly present
and the system is in the quantum Hall (QH) regime.
This leads to physical phenomena, stemming mainly from the ``flat-band'' property of the LL spectrum in the bulk (\figr{edgestates}),
that are specific to the QH regime and otherwise may hardly be realized.

\subsection{Quantum Hall ferromagnet at the topological phase transition}

\begin{figure}
\includegraphics[width=.30\textwidth]{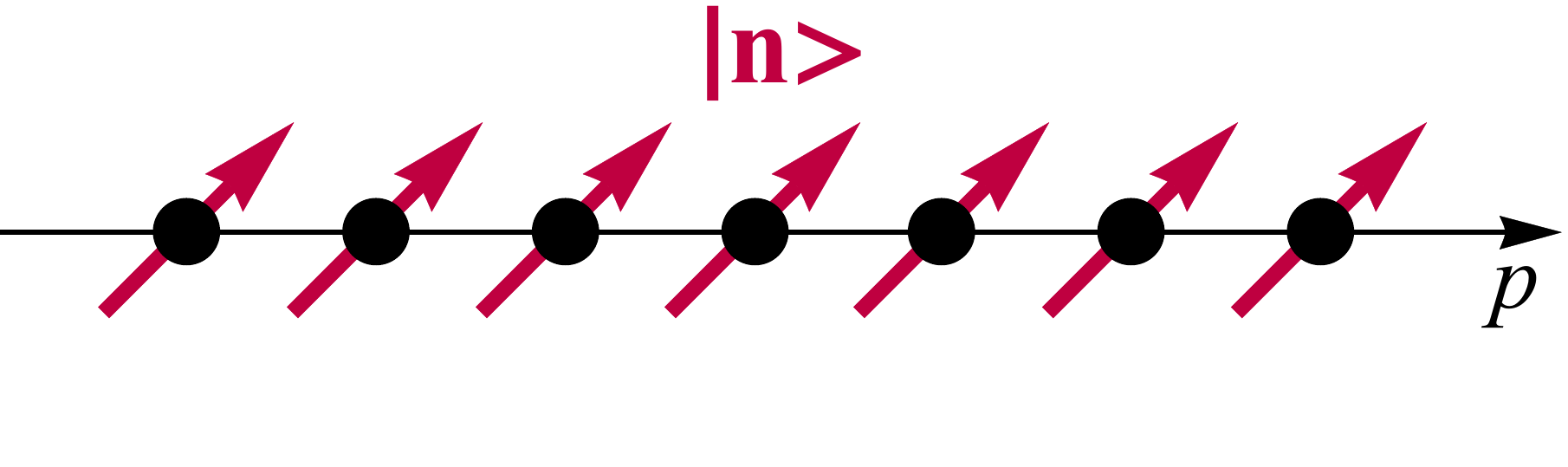}
\caption{(Color online)
Quantum Hall ferromagnet (QHFM) state realized at the crossing of LLs at half-filling.
For each momentum $p$, one electron occupies the state $|\nb\rn$ with given isospin $\nb$,
see \eqss{|n>}{chi}{n} and \figr{n}.}
\lbl{fig:QHFMstate}
\end{figure}

\begin{figure}
\includegraphics[width=.26\textwidth]{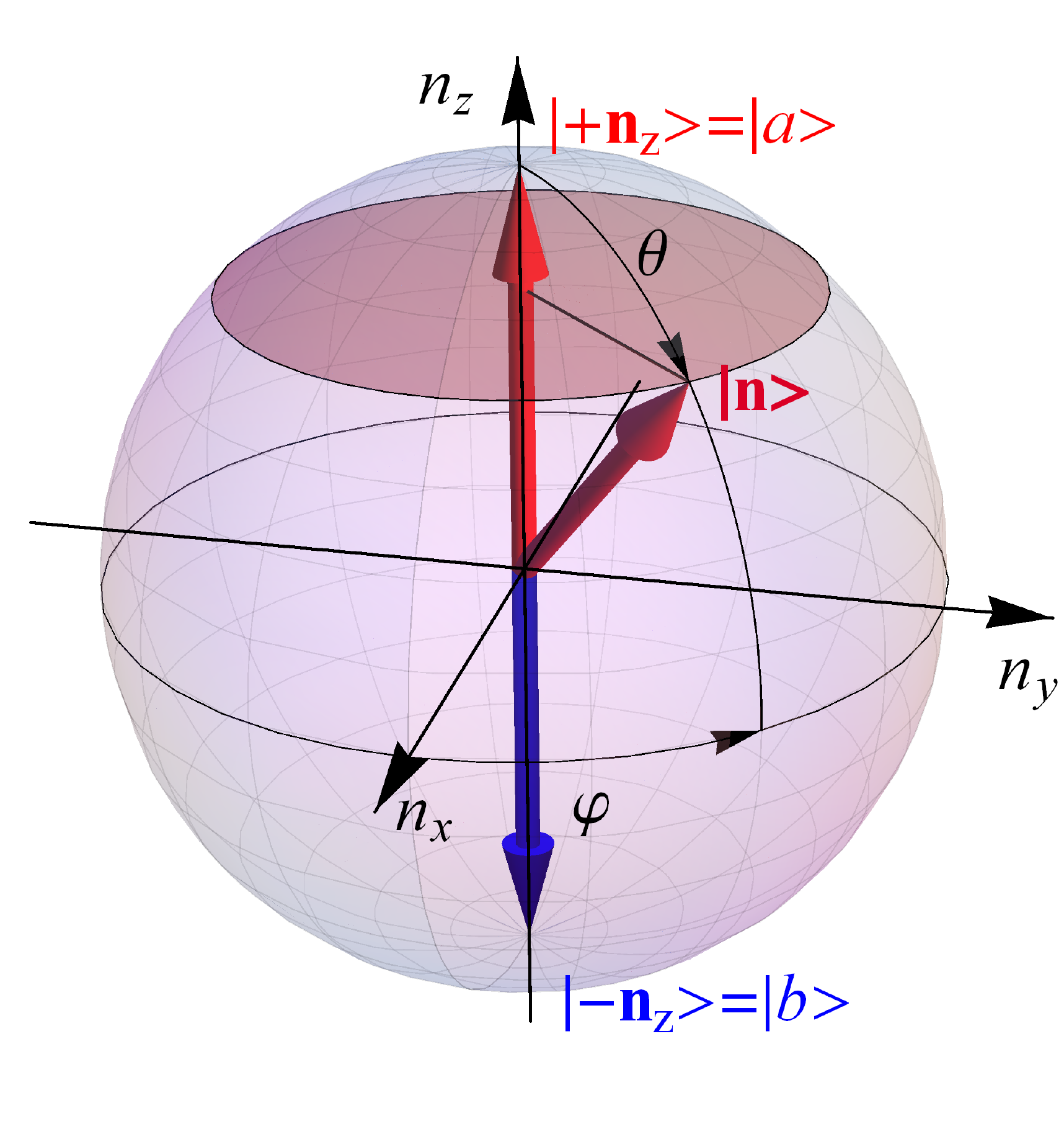}
\caption{(Color online)
Bloch sphere of the isospin $\nb$ in the 2D space of $a$ and $b$ LLs [\eqss{|n>}{chi}{n}].
The north $|\!+\!\nb_z\rn=|a\rn$ and south $|\!-\!\nb_z\rn=|b\rn$
poles corresponds to the occupation of the $a$ and $b$ LLs, respectively.
Any other state is a coherent mixture of the $|a\rn$ and $|b\rn$ states.
}
\lbl{fig:n}
\end{figure}

The QHTIs with the LL structure as in \figr{LLs}
are particularly appealing for the study of the interplay of interactions and topology:
due to the near degeneracy of the two LLs in vicinity of the single-particle topological phase transition point $B^*$,
electron interactions become the dominant effect that drives the physics there.
Thus, in QHTIs, effective interactions are tunable by the magnetic field,
and the regime of strong effective interactions is experimentally accessible even in a system with weak bare interactions.

In this work, we investigate the effect of electron interactions
on the topological properties of QHTIs with the LL structure shown in \figr{LLs},
particularly focusing on the regime of strong effective interactions
in the vicinity of the single-particle topological phase transition.

The zero Chern number $\nu=0$ corresponds to half-filling of the two crossing LLs $a$ and $b$,
with on average one electron per two states.
Analogous to the Hund exchange mechanism in atoms,
at such commensurate filling factor,
interactions make the electron system particularly prone to polarization in the 2D $ab$ space.
This results in the formation of the ``ferromagnetic'' ground state, see \figr{QHFMstate},
in which electrons at each orbital occupy exactly the same state
\beq
    |\nb\rn=\chi_a(\nb)|a\rn+\chi_b(\nb)|b\rn,
\lbl{eq:|n>}
\eeq
\beq
    \chi(\nb)
        =\lt(\ba{c}\chi_a(\nb)\\\chi_b(\nb)\ea\rt)
        =\lt(\ba{c}\ex^{-\f{\ix}2\vphi}\cos\f{\tht}2\\\ex^{\f{\ix}2\vphi}\sin\f{\tht}2\ea\rt),
\lbl{eq:chi}
\eeq
characterized by the unit-vector ``isospin''
\beq
    \nb=(n_x,n_y,n_z)=(\sin\tht\cos\vphi,\sin\tht\sin\vphi,\cos\tht)
    ,\spc
    \spc \nb^2=1,
\lbl{eq:n}
\eeq
\beq
    |\nb\rn\ln\nb|=\f12(\tau_0+\taub\cd\nb),
\lbl{eq:P}
\eeq
see \figr{n}.
Throughout, $\tau_0$ and $\taub=(\tau_x,\tau_y,\tau_z)$ will denote the unity
and Pauli matrices in the $ab$ space.

Exactly at the crossing point
and in the approximation of the SU(2)-symmetric interactions in the $ab$ space,
the isospin $\nb$ can be completely arbitrary and
the state describes spontaneous breaking of SU(2) symmetry.
This phenomenon is referred to as {\em quantum Hall ferromagnetism} (QHFMism)~\cite{QHFM}.

The bulk single-particle and interaction effects responsible for the deviation
from this fully degenerate SU(2)-symmetric situation,
as well the effect of the edge, can be taken into account within a low-energy field theory, the nonlinear $\s$-model,
for the isospin order parameter (OP) $\nb(\rb;t)$ generalized to configurations slowly varying in time and space.
We derive the closed form of such $\s$-model in the coordinate space.
Crucial for the description of the edge properties,
we incorporate the effect of the edge as a boundary condition for the order parameter.
This allows us to study most properties of interest analytically.
The analysis of the properties of the edge excitations follows
the approach developed in Refs.~\ocite{FB,F2,F3,F4,FI} for the $\nu=0$ QH state in graphene.
Their properties are governed by the spatially inhomogeneous textures of the order parameter at the edge.

We calculate the bulk phase diagram, ground state edge textures, and edge charge excitations.
For the interacting TnT phase, we derive and analyze the low-energy theory for the gapless edge excitations.

\subsection{Main findings}

As the central general result,
we find that topological properties of this
interacting many-body system
are directly tied to its effective symmetry:
the U(1) symmetry with respect to rotations about the isospin $z$ axis
is responsible for the topological protection.

We demonstrate that if U(1) symmetry is preserved,
the single-particle TnT phase with fully filled $|a\rn$ LL, corresponding to $\nb=\nb_z=(0,0,1)$ isospin
in the QHFM formalism, remains TnT in the presence of interactions in most of the range $0\leq B<B^*$.
The edge excitations remain gapless but take the form of collective excitations described by the helical Luttinger liquid.
We obtain explicit analytical expressions for the parameters of the Luttinger liquid in the QHFM regime.
We find that the effective interactions in this Luttinger liquid are highly tunable:
weak (with the interaction parameter~\cite{Gia} $\Kc\approx1$) at small magnetic fields $B\ll B^*$,
but strong ($\Kc\ll 1$) in the QHFM regime in the vicinity of the single-particle phase transition $B^*$.

More precisely, ``preserved U(1) symmetry'' means that {\em both} the bulk ground state
and the many-body Hamiltonian are U(1)-symmetric. According to the violation of one of these conditions,
we identify two mechanisms of the U(1) symmetry breaking,
which lead to the loss of topological protection and eventual transitions to the TT phases.

First, U(1) could be broken spontaneously:
upon increasing $B$, a second-order phase transition from the TnT phase $\nb=\nb_z$
to the phase with spontaneously broken U(1) symmetry may occur.
The gap in the edge excitation spectrum of this broken-U(1)-symmetry phase
grows monotonically upon further increasing $B$, starting from the zero value at the phase transition.
Also, upon approaching this phase transition from the TnT $\nb=\nb_z$ phase,
the edge Luttinger liquid becomes infinitely strongly interacting ($\Kc\rarr0$).

Second, the many-body Hamiltonian can be U(1)-asymmetric:
{\em interaction terms} can be present that fully respect the physical symmetry,
responsible for the topological protection of the non-interacting system, but break U(1) symmetry.
Such terms transform the Luttinger liquid model for the edge excitations of the TnT $\nb=\nb_z$ phase into the sine-Gordon model~\cite{Gia}.
As $B$ is increased, such terms will result in the phase transition to the state with broken U(1) symmetry at the edge and gapped edge excitations.

In either of the scenarios, the phase transitions from the TnT to the TT phases occur at the magnetic fields $B$
{\em lower} than the single-particle phase transition point $B^*$
and are thus {\em interaction-induced topological} quantum phase transitions.

The rest of the paper is organized as follows.
In \secr{H}, we present the projected Hamiltonian.
In \secr{smodel}, we derive the low-energy theory for the QHFM.
In \secr{bulk}, we obtain the bulk phase diagram.
In \secr{edgeground}, we obtain the ground state configurations for a system with an edge.
In \secr{edgeexc}, we study the edge charge excitations.
In \secr{lliq}, we derive the Luttinger liquid model for the edge excitations in the TnT phase.
In \secr{U(1)}, we establish U(1) symmetry as the requirement for topological protection.
In \secr{concl}, we present concluding remarks.

\section{Projected Hamiltonian\lbl{sec:H}}

\subsection{Restricted Hilbert space of two intersecting Landau levels\lbl{sec:hspace}}

We will work under the approximation where only the two intersecting LLs $a$ and $b$ in \figr{LLs}
are taken into account, while other LLs are neglected.
This is a standard approximation for quantum Hall systems,
justified for weak Coulomb interactions,
when the energy separation between the LLs of interest and other LLs is much larger than the
interaction energy scale set by the Coulomb energy
\beq
    \f{e_*^2}{l_{B_z}}=\f{e^2}{\ka\,l_{B_z}}.
\lbl{eq:eC}
\eeq
Here, $e_*$ is the electron charge   screened by the dielectric environment with the constant $\ka$,
and
\[
    l_{B_z}=\q{\f{c}{eB_z}}
\]
is the magnetic length, in which $e$ is the electron charge, $c$ is the speed of light,
and $B_z$ is the component of the magnetic field perpendicular to the sample plane, $B^2=B_z^2+B_\p^2$.
We assume arbitrary orientation of the magnetic field relative to the quasi 2D sample,
although this point will not be important.

We will work in the Landau gauge, in which the single-particle states are characterized by
the one-dimensional momentum $p$ along the edge $y$ direction.
The single-particle states of the LLs of interest are
\beq
    |ap\rn \mbox{ and } |bp\rn.
\lbl{eq:ab}
\eeq
We assume no discrete degeneracies (such as valleys) of these LLs.

We consider a half-infinite two-dimensional sample occupying the region $x<0$, see \figr{edgestates}.
The states with $p\lesssim 0$ are then the bulk states, for which the coordinate-momentum correspondence holds,
and the states with $p\gtrsim 0$ correspond to the edge states, localized over $l$ near the edge.
The electron annihilation operators will be denoted as $c_{ap}$ and $c_{bp}$;
in the formulas below, we join them into a two-component spinor
\beq
    \ch_p=\lt(\ba{c} c_{ap} \\ c_{bp} \ea\rt)
\lbl{eq:c}
\eeq
for compactness.

\subsection{$\Ux(1)$-symmetric projected Hamiltonian\lbl{sec:U1H}}

We first consider the following many-body ``projected'' Hamiltonian, operating within the states (\ref{eq:ab}) of the intersecting LLs:
\bwt
\beqar
    \Hh&=&\Hh_{1\circ}+\Hh_{1\circ}^\x{edge}+\Hh_{2\odot}+\Hh_{2\circ},
\lbl{eq:H}\\
    \Hh_{1\circ}&=&-h_z\sum_p \ch^\dg_p\tau_z\ch_p,
\lbl{eq:H1c}\\
    \Hh_{1\circ}^\x{edge}&=&\sum_p \e(p)\ch^\dg_p\tau_z\ch_p,
\lbl{eq:H1cedge}\\
    \Hh_{2\odot}&=&\f12\sum_{p_1+p_2=p_1'+p_2'} V(^0_0|^{p_1p_1'}_{p_2p_2'}) :[\ch^\dg_{p_1}\ch_{p_1'}][\ch^\dg_{p_2}\ch_{p_2'}]:,
\lbl{eq:H2od}\\
    \Hh_{2\circ}&=&\f12\sum_{p_1+p_2=p_1'+p_2'} \sum_{\al=x,y,z} V(^\al_\al|^{p_1p_1'}_{p_2p_2'})
    :[\ch^\dg_{p_1}\tau_\al\ch_{p_1'}][\ch^\dg_{p_2}\tau_\al\ch_{p_2'}]:
    ,\spc
    V(^x_x|^{p_1p_1'}_{p_2p_2'})=V(^y_y|^{p_1p_1'}_{p_2p_2'}).
\lbl{eq:H2c}
\eeqar
\ewt
The labels 1 and 2 designate single-particle and two-particle interaction terms, respectively;
the labels $\circ$ and $\odot$ are explained below.

The term (\ref{eq:H1c}) describes the energy spacing
between the two LLs of interest, equal to $2h_z$.
The energy $h_z(B)$ is a function of the magnetic field $B$ (\figsr{LLs}{edgestates}):
it decreases monotonically with increasing the magnetic field,
starting from positive values and changing to negative values at the crossing point $B=B^*$.
Close to the crossing point, one may expand it to the linear order as
\beq
    h_z(B)\approx-|\pd_B h_z(B^*)|(B-B^*).
\lbl{eq:hz(B)}
\eeq

Next, the term (\ref{eq:H1cedge}) describes the effect of the edge.
The dispersion function $\e(p)>0$ is shown schematically in \figr{edgestates};
it has a plateau $\e(p)\approx0$ in the bulk ($p\lesssim 0$)
and grows monotonically at the edge $p\gtrsim 0$.
Note that although the two branches at the edge do not have to be exactly particle-hole symmetric
and an additional energy term $\e_0(p)\tau_0$
could be added, it produces only a trivial $\nb$-independent term in the $\s$-model derived below;
so, we neglect it.

Crucially, due to the assumed topological protection by the physical symmetry,
the single-particle terms $\Hh_{1\circ}+\Hh_{1\circ}^\x{edge}$ do not couple the $|ap\rn$ and $|bp\rn$ states.
The single-particle spectrum $\pm[-h_z+\e(p)]$ of $\Hh_{1\circ}+\Hh_{1\circ}^\x{edge}$
describes two LL with counter-propagating edge states at $0<h_z$
(TnT phase) and a fully gapped spectrum, both in the bulk and at the edge, at $h_z<0$ (TT phase).

Due to this decoupling
of the $|ap\rn$ and $|bp\rn$ states, the single-particle Hamiltonian
$\Hh_{1\circ}+\Hh_{1\circ}^\x{edge}$ [\eqs{H1c}{H1cedge}]
possesses U(1) symmetry with respect to continuous rotations about the isospin $z$ axis, as described by the matrix
\beq
    \Dh(\phi)=\lt(\ba{cc} \ex^{-\ix\f{\phi}2}& 0\\0&\ex^{\ix\f{\phi}2}\ea\rt)
\lbl{eq:D}
\eeq
acting on the spinor (\ref{eq:chi}) in the $ab$ space as
\beq
    \Dh(\phi)\chi(\tht,\vphi)=\chi(\tht,\vphi+\phi).
\lbl{eq:Dchi}
\eeq
Here, $\tht$ and $\vphi$ are the angles of the spherical parametrization of the isospin [\eqs{chi}{n}]

As we shall find below, this effective continuous $\Ux(1)$ symmetry is central
to the properties of the edge charge excitations
of the interacting system and the associated topological properties.
For this reason, we consider
the form of two-particle interactions, presented in \eqs{H2od}{H2c}, that preserves this U(1) symmetry.
We also split these interactions into two parts, the SU(2)-symmetric part $\Hh_{2\odot}$ [\eq{H2od}]
and the SU(2)-asymmetric $\Ux(1)$-symmetric part $\Hh_{2\circ}$ [\eq{H2c}]
(The terms with the structure $\um\ot\tau_z$ are also $\Ux(1)$-symmetric,
but in the $\s$-model below they lead to an inconsequential shift of the position of single-particle transition point;
therefore, we discard them in order not to overburden the expressions.).
The exact form of the matrix elements $V(^\al_\al|^{p_1p_1'}_{p_2p_2'})$, $\al=0,x,y,z$,
will not matter for our considerations, only the presented structure of the terms
in the isospin space will. The only condition we assume is that
the $\Ux(1)$-asymmetric terms are much smaller than the SU(2)-symmetric ones,
in order to make the low-energy field theory a controlled expansion.
The SU(2)-symmetric interactions have the typical scale of the Coulomb energy,
\beq
    \sum_{p_2} V(^0_0|^{p_1p_2}_{p_2p_1})\sim \f{e_*^2}{l_{B_z}}.
\lbl{eq:V0est}
\eeq

The Hamiltonian $\Hh$ [\eq{H}] thus possesses $\Ux(1)$ symmetry.
We now consider possible U(1)-asymmetric terms.

\subsection{$\Ux(1)$-asymmetric terms \lbl{sec:U1nH}}

The mechanisms of (non-spontaneous) $\Ux(1)$ symmetry breaking could be grouped into two
categories according to an important symmetry distinction between them.

1) One category is when already the physical symmetry responsible
for the topological protection of a noninteracting system is violated.
Consequently, the $\Ux(1)$ symmetry is then broken already at the single-particle level.
The corresponding terms have the form of the isospin ``Zeeman'' field acting in the $xy$ plane:
\beq
    \Hh_{1\os}=-h_\p\sum_p \ch^\dg_p(\tau_x\cos\vphi_{1\os}+\tau_y\sin\vphi_{1\os})\ch_p.
\lbl{eq:H1p}
\eeq
Such terms result in the coupling between the $a$ and $b$ LLs.
The ``orientation'' of this field in the $xy$ plane, set by the angle $\vphi_{1\os}$,
depends on the choice of the relative phase factor between the $|ap\rn$ and $|bp\rn$ states and is largely arbitrary.

2) Another category is when the physical symmetry is preserved.
Then no single-particle terms breaking $\Ux(1)$ symmetry are allowed.
However, interactions that preserve the physical symmetry but break $\Ux(1)$ symmetry
could be present:
\beq
    \Hh_{2\os}
    =\f12\sum_{\substack{p_1+p_2\\=p_1'+p_2'}} {\sum_{\al_1\al_2}}^\os V(^{\al_1}_{\al_2}|^{p_1p_1'}_{p_2p_2'})
    :[\ch^\dg_{p_1}\tau_{\al_1}\ch_{p_1'}][\ch^\dg_{p_2}\tau_{\al_2}\ch_{p_2'}]:
\lbl{eq:H2U1n}
\eeq
where the sum $\sum_{\al_1\al_2}^\os$ contains only $\Ux(1)$-asymmetric terms.
The structure of such interactions depends on specific physical symmetry,
which does provide some constraints on the matrix elements $V(^{\al_1}_{\al_2}|^{p_1p_1'}_{p_2p_2'})$,
but for most physical symmetries such $\Ux(1)$-asymmetric interactions would be allowed.
For the $\s$-model approach we employ, however,
the detailed knowledge of their structure is not necessary,
only the corresponding anisotropy function is required.
The latter can be derived using symmetry considerations,
as we illustrate in \secr{U1nEc}.

\section{Low-energy nonlinear $\s$-model\lbl{sec:smodel}}

\subsection{Quantum Hall ``ferromagnet'' \lbl{sec:QHFM}}

Close to the crossing point $h_z=0$ of the LLs, the SU(2)-symmetric part $\Hh_{2\odot}$ [\eq{H2od}]
of the electron interactions is the dominant term in the Hamiltonian $\Hh$ [\eq{H}]:
its typical scale $e_*^2/l_{B_z}$ [\eq{V0est}] exceeds the energies of all other terms.
It is straightforward to show that, at half-filling of the two LLs $|ap\rn$ and $|bp\rn$,
the Slater determinant state
\beq
    \Psi(\nb)=\prod_{\x{bulk } p} c^\dg_{\nb p}|0\rn
\lbl{eq:Psi}
    \eeq
is an exact eigenstate of $\Hh_{2\odot}$.
Here, $|0\rn$ is the ``vacuum'' state with both LLs empty and
\beq
    c^\dg_{\nb p}=\chi_a(\nb)c^\dg_{ap}+\chi_b(\nb)c^\dg_{bp}
\lbl{eq:cn}
\eeq
is the operator creating an electron in the state $|\nb\rn$ [\eq{|n>}]
with the isospin $\nb$ [\eq{n}].
The isospin can be visualized by means of the Bloch sphere, see \figr{n}.
According to \eq{|n>}, the isospin at the ``poles'' of the Bloch sphere $\nb=\pm\nb_z$ ($\tht=0,\pi$),
corresponds to pure $|\nb_z\rn=|a\rn$ or $|-\nb_z\rn=|b\rn$ states.
Any other state with $-1<n_z<1$ ($0<\tht<\pi)$ is a coherent mixture of $|a\rn$ and $|b\rn$ states.

For a wide class of repulsive interactions, one can expect this eigenstate to be an exact ground state by the
Hund's rule argument. This is the main assumption of the QHFMism theory~\cite{QHFM}, also employed in this paper.

Importantly, the many-body wave-function $\Psi(\nb)$ is an eigenstate of $\Hh_{2\odot}$ for any choice of the isospin $\nb$.
It thus describes the state with spontaneously broken SU(2) symmetry;
the unit vector $\nb$ represents the OP of the family of degenerate ground states.

\subsection{$\Ux(1)$-symmetric nonlinear $\s$-model\lbl{sec:U(1)smodel}}

The effects of the other, SU(2)-asymmetric, terms in the Hamiltonian
on the ground state and excitations of the QHFM
can be taken into account within a low-energy field theory, the $\s$-model.
As long as the energy scales of these terms are much smaller than the
Coulomb scale (\ref{eq:V0est}) of the SU(2)-symmetric interactions,
the $\s$-model presents a controlled systematic low-energy expansion
about the exact ground state (\ref{eq:Psi}) of $\Hh_{2\odot}$.

For the Hamiltonian given by \eqss{H}{H1p}{H2U1n},
the derivation of the $\s$-model is rather standard and follows the general recipe~\cite{QHFM}.
The new aspect is incorporating the effect of the edge with counter-propagating states
into the real-space $\s$-model, which we perform below at the end of this section.

In the nonlinear $\s$-model, the homogeneous and static isospin OP $\nb$ of the state (\ref{eq:Psi})
is generalized to configurations $\nb(\rb;t)$ that vary slowly in time and space. The constraint
\[
    \nb^2(\rb;t)=1
\]
is satisfied locally.

The low-energy dynamics and energetics
is described by the Lagrangian functional;
for the bulk part (all terms except $\Hh_{1\circ}^\x{edge}$) of the U(1)-symmetric Hamiltonian \eqn{H}, it has the form
\beqar
    \Lf[\nb]&=&\Kf[\nb]-\Ef[\nb]=\int\f{\dx^2\rb}{s} L[\nb]
    ,\spc L[\nb]=K[\nb]-E[\nb],
\lbl{eq:L}\\
    \Kf[\nb]&=&\int\f{\dx^2\rb}s K[\nb]
    ,\spc
    K[\nb]=\f{\dot{\vphi}}2\cos\tht,
\lbl{eq:K}\\
    \Ef[\nb]&=&\int\f{\dx^2\rb}s E[\nb]
    ,\spc
    E[\nb]=\f{\rho}{2}(\n\nb)^2+\Ec(n_z),
\lbl{eq:E}
\eeqar
\beq
    \Ec(n_z)
    =\f{u}2 n_z^2-h_z n_z
    =\f{u}2 \cos^2\tht-h_z \cos\tht.
\lbl{eq:Ec}
\eeq
The Lagrangian $\Lf[\nb]$ is given by the difference of the kinetic $\Kf[\nb]$ and energy $\Ef[\nb]$ terms.
The spatial integration $\int\dx^2\rb\ldots$ is performed over the region $x<0$ occupied by the half-infinite sample.
We introduce the normalization factor
\[
    s=2\pi l^2_{B_z}
\]
equal to the area threaded by one magnetic flux quantum; $1/s$ is also the electron density per one LL.
This way, the respective densities $L[\nb]$, $K[\nb]$, $E[\nb]$ are defined per this area $s$ and have the dimension of energy.

The kinetic term $\Kf[\nb]$ [\eq{K}] contains the time derivative
and can be presented explicitly in terms of the spherical angles $\tht$ and $\vphi$
parameterizing the isospin [\eq{n}].
Note that the form of $K[\nb]$ is not unique, but is defined up to a full time derivative,
which results in an inconsequential constant contribution to the action $\int\dx t\,\Lf[\nb]$.

The energy functional $\Ef[\nb]$ [\eq{E}] consists of the gradient term $\f\rho2(\n\nb)^2$
and the energy function $\Ec(n_z)$ [\eq{Ec}].
The gradient term describes the energy cost of a spatially inhomogeneous configuration;
to the leading order, the stiffness
\[
    \rho=\f{l^4_{B_z}}{4}\sum_{p_2} V(^0_0|^{p_1p_2}_{p_2p_1})(p_1-p_2)^2
\]
is expressed in terms of the SU(2)-symmetric $\Hh_{2\odot}$ interactions.

The energy function $\Ec(n_z)$ describes the effect of the bulk
terms $\Hh_{1\circ}+\Hh_{2\circ}$ that have the symmetry lower than SU(2).
To derive it, it is sufficient to take the expectation value
\beq
    \Ec(n_z)=\f1\Nc\ln\Psi(\nb)|\Hh_{1\circ}+\Hh_{2\circ}|\Psi(\nb)\rn
\lbl{eq:Ecder}
\eeq
of the corresponding terms with respect to the state $\Psi(\nb)$ [\eq{Psi}]
($\Nc=\sum_p 1=\int\f{\dx^2\rb}{s}\,1$ is the number of orbital states,
equal to the number of flux quanta threading the sample.)
The term $\f{u}2 n_z^2$ quadratic in $\nb$ arises from the SU(2)-asymmetric two-particle interactions $\Hh_{2\circ}$
and can be referred to as the ``anisotropy'' term;
the anisotropy energy equals
\beqarn
    u&=&u_z-u_\p
,\spc
u_\p\equiv u_x=u_y,
\\
    u_\al&=&\sum_{p_2}[V(^\al_\al|^{p_1p_1}_{p_2p_2})-V(^\al_\al|^{p_1p_2}_{p_2p_1})]
,\spc
\al=x,y,z.
\eeqarn
We will consider the more interesting case of positive anisotropy energy
\[
    u>0,
\]
which is called ``easy-plane'' anisotropy, since the energy $\f{u}2 n_z^2$ alone
is minimized by the isospin in the plane $n_z=0$.

The only remaining term in the U(1)-symmetric Hamiltonian $\Hh$ [\eq{H}] that needs to be taken into account is
$\Hh_{1\circ}^\x{edge}$, which describes the edge.
Its effect can be presented as an effective boundary condition for the order parameter $\nb(\rb;t)$ as follows.
We first note that the edge states ($p\gtrsim 0$) are also ``half-filled'' (one electron per two states)
and thus their occupation can be described by the same isospin OP:
the filling factor remains the same for both bulk ($p\lesssim 0$) and edge ($p\gtrsim 0$) states.
At such $p$ that the energy $\e(p)$ becomes much greater than the energies $u$ and $h_z$ of the SU(2)-asymmetric terms,
electrons occupy the ``hole'' branch of the edge spectrum,
i.e., the states $|bp\rn$ with the negative energy $-\e(p)$, which corresponds to $\nb=-\nb_z$.
Since the edge states with $p\gtrsim 0$ are localized at spatial scales $\sim l_{B_z}$ near the edge $x=0$ of the sample
and $\nb(\rb;t)$, by assumption, varies at much larger scales,
the effect of the edge may be described in the real space by the boundary condition
\beq
    \nb(x=0,y;t)=-\nb_z, \spc \nb_z=(0,0,1).
\lbl{eq:bc}
\eeq
Thus, the effect of the edge amounts to ``pinning'' the OP in the state
that corresponds to the occupation of the ``hole'' branch of the edge spectrum.

Equations (\ref{eq:L})-(\ref{eq:Ec}) for the Lagrangian and \eq{bc}
for the boundary condition constitute the closed-form low-energy $\s$-model
in the coordinate space for the considered QHFM system with an edge, originating from the Hamiltonian $\Hh$ [\eq{H}].
Naturally, the model inherits the $\Ux(1)$ symmetry [\eqs{D}{Dchi}] of the Hamiltonian (\ref{eq:H})
and is invariant under the rotations of the isospin about the $z$ axis:
\[
    \vphi(\rb;t)\rarr\vphi(\rb;t)+\phi.
\]
The additional terms in the $\s$-model originating
from the $\Ux(1)$-asymmetric terms \eqn{H1p} and \eqn{H2U1n} in the full Hamiltonian, \secr{U1nH}, are considered below.

\subsection{$\Ux(1)$-asymmetric terms\lbl{sec:U1nEc}}

The single-particle term $\Hh_{1\os}$ [\eq{H1p}] that breaks $\Ux(1)$ symmetry
produces the following additional contribution to the energy function (\ref{eq:Ec}):
\beqar
    \Ec_{1\os}(\nb)
        &=&\f1\Nc\ln\Psi(\nb)|\Hh_{1\os}|\Psi(\nb)\rn\nn\\
        &=&-h_\p(n_x\cos\vphi_{1\os}+n_y\sin\vphi_{1\os})\nn\\
        &=&-h_\p\sin\tht\cos(\vphi-\vphi_{1\os}).
\lbl{eq:Ecp}
\eeqar

The structure of the anisotropy energy function arising from the U(1)-asymmetric two-particle interactions $\Hh_{2\os}$ [\eq{H2U1n}]
depends on the specific physical symmetry. It can be derived via group-theoretical considerations
without using any information about the interaction matrix elements in \eq{H2U1n}.
As an example, we will consider the inversion symmetry.

In this case, the LLs states $a$ and $b$ are characterized by opposite inversion parities $+$ and $-$.
Therefore, the isospin components transforming according to \eq{P}
as $n_{x,y}\sim |a\rn\ln b|$ and $n_z\sim |a\rn\ln a|-|b\rn\ln b|$ have $-$ and $+$ parities, respectively.

The anisotropy function arising from two-particle interactions is a quadratic function of $\nb$.
It must be invariant under inversion, i.e., have $+$ parity.
All quadratic functions with $+$ parity are
\[
    \{ n_z^2, n_x^2,2n_x n_y, n_y^2\}.
\]
The most general form of the anisotropy function
is an arbitrary linear combination of these terms.

It is convenient to choose the basis functions as
\[
    \{ n_x^2+n_y^2+n_z^2, n_z^2-n_x^2-n_y^2, n_x^2-n_y^2, 2n_xn_y\}.
\]
Then, the combination $n_x^2+n_y^2+n_z^2=1$
preserves SU(2) symmetry and, due to the constraint (\ref{eq:n}), is $\nb$-independent;
the combination $n_z^2-n_x^2-n_y^2=2n_z^2-1$ preserves $\Ux(1)$ symmetry and depends only on $n_z$.
These two functions produce, up to a constant,
the $\Ux(1)$-symmetric anisotropy function $\f12u n_z^2$ in $\Ec(n_z)$ [\eq{Ec}].
An arbitrary linear combination
\beqar
    \Ec_{2\os}(\nb)
        &=&\f12[u_+(n_x^2-n_y^2)+u_\tm 2 n_x n_y]\nn\\
        &=&\f12u_{2\os}\sin^2\tht\cos2(\vphi-\vphi_{2\os})
\lbl{eq:Ec2}
\eeqar
of the two remaining functions represents the $\Ux(1)$-asymmetric contribution to the anisotropy function.

\subsection{Outline of the approach}

The remaining part of the paper is devoted to the analysis of the obtained $\s$-model,
with the focus on the properties of the edge excitations.
The fact that the effect of the edge has been reduced to the boundary condition (\ref{eq:bc})
in the coordinate space is a technical advantage that will facilitate the analysis of the problem
and enable us to obtain explicit analytical expressions for many quantities of interest.

The approach we use to study the edge excitations follows that
developed in a series of papers~\cite{FB,F2,F3,F4,FI}
for the ferromagnetic (F) and canted antiferromagnetic (CAF)~\cite{Herbut,MKmlg,MKblg,MKsimp}
phases in the $\nu=0$ state in graphene with armchair-type boundary.
Although graphene has a few additional peculiarities
(most importantly, the presence of valley degrees of freedom, which makes the QHFM physics richer),
there are  mathematical and physical similarities to our model.
Another related system is a QH bilayer with an inverted band structure, studied theoretically in Ref.~\ocite{Pik}.
The model Hamiltonian considered in Ref.~\ocite{Pik}
essentially coincides with the U(1)-symmetric part of our model,
but the focus and methods of analysis of Ref.~\ocite{Pik} differ from ours in several respects.
We point out the analogies between our and these two systems as we move along.

As originally recognized in Ref.~\ocite{FB} for the F phase of the $\nu=0$ state in graphene with armchair-type boundary,
the physics of the edge in the QHFMs at $\nu=0$ is governed
by the fact that the order favored at the edge due to the propagating edge states
may be different from that favored in the bulk.
This leads to a spatially inhomogeneous OP texture at the edge, which connects the bulk and edge orders.
This ground state texture, which can be referred to as the {\em domain wall},
then determines the properties of the edge excitations.

Since, as already mentioned above, $\Ux(1)$ symmetry will turn out to be crucial for the existence
of the TnT phase in this interacting system, in the next Secs.~\ref{sec:bulk}-\ref{sec:lliq}
we perform the analysis of the $\Ux(1)$-symmetric model, Eqs. (\ref{eq:L})-(\ref{eq:Ec}), and (\ref{eq:bc}),
and consider the effect of the $\Ux(1)$-asymmetric terms (\ref{eq:Ecp}) and (\ref{eq:Ec2})
afterwards in \secr{U1nlliq}.

\section{Bulk phase diagram\lbl{sec:bulk}}

In this section, we obtain the bulk mean-field phase diagram
for the $\Ux(1)$-symmetric model [Eqs.~(\ref{eq:L})-(\ref{eq:Ec}), and (\ref{eq:bc})]
completely neglecting the edge [boundary condition (\ref{eq:bc})].
It is obtained by minimizing the energy function $\Ec(n_z)$ [\eq{Ec}].
The minimum isospin configuration will be denoted as $\nb^\i$
and referred to as the {\em bulk ground state}. The minimal energy will be denoted as
\beq
    \Ec^\i\equiv\Ec(n_z^\i=\cos\tht^\i)=\min_{n_z}\Ec(n_z).
\eeq

In the case $u>0$ of the easy-plane anisotropy we consider, minimization of $\Ec(n_z)$
within the interval $-1\leq n_z\leq 1$ gives the following phases
\bwt
\beqar
    \nb^\i&=&\nb_z=(0,0,1),\spc u<h_z, \lbl{eq:n u<h} \lbl{eq:n+<}\\
    \nb^\i&=&\nb^*(\vphi_0)=(\sin\tht^*\cos\vphi_0,\sin\tht^*\sin\vphi_0,\cos\tht^*),\spc -u<h_z<u, \lbl{eq:n-<<+}\\
    \nb^\i&=&-\nb_z=(0,0,-1),\spc h_z<-u. \lbl{eq:n<-}
\eeqar
\ewt
with the respective energy minima
\beqar
    \Ec^\i&=&\Ec(n_z=+1)=\f{u}2-h_z,\spc u<h_z, \lbl{eq:E+<}\\
    \Ec^\i&=&\Ec(n_z^*)=-\f{h_z^2}{2u}, \spc -u<h_z<u, \lbl{eq:E-<<+}\\
    \Ec^\i&=&\Ec(n_z=-1)=\f{u}2+h_z,\spc h_z<-u. \lbl{eq:E<-}
\eeqar

The phases are shown in \figr{nz}.
The phases $\nb^\i=\pm\nb_z$ at $u<h_z$ and $h_z<-u$, respectively,
are fully polarized along the direction $z$ of the field $h_z$.
According to the meaning of the isospin, see \eqss{|n>}{chi}{n},
the $\nb^\i=\pm\nb_z$ phases correspond to the occupation of either
$|\!+\!\nb_z\rn=|a\rn$ or $\spc |\!-\!\nb_z\rn=|b\rn$ LLs, respectively.
The Slater-determinant ground state $\Psi(\nb)$ [\eq{Psi}] in the $\nb^\i=\pm\nb_z$ phases
is thus the same as in the noninteracting system.

In the ``intermediate'' phase $\nb^\i=\nb^*(\vphi_0)$ at $-u<h_z<u$, the isospin has the optimal projection
\beq
    n_z^*=\cos\tht^*=\hr_z
\lbl{eq:nz*}
\eeq
on the $z$ direction and arbitrary orientation in the $xy$ plane, parameterized by the angle $\vphi_0$.
It is convenient to introduce the dimensionless field
\beq
    \hr_z=\f{h_z}u
\lbl{eq:hbrz}
\eeq
normalized by the anisotropy energy $u$. Thus, in the intermediate phase, electrons are in a coherent mixture
\[
    |\nb^*(\vphi_0)\rn=\ex^{-\ix\f{\vphi_0}2}\cos\f{\tht^*}2|a\rn+\ex^{\ix\f{\vphi_0}2}\sin\f{\tht^*}2|b\rn
\]
of the two LL states. The appearance of this intermediate phase is the first important distinction from the noninteracting picture.

The $\Ux(1)$ symmetry is thus preserved in the $\nb^\i=\pm\nb_z$ phases,
but it is spontaneously broken in the intermediate $\nb^\i=\nb^*(\vphi_0)$ phase.

The phase transitions at $h_z=\pm u$ are of the second order.
According to the dependence \eqn{hz(B)} of $h_z(B)$ on the magnetic field, the transition points $h_z=\pm u$,
correspond to the values
\beq
    B^*\mp\de B_u
    ,\spc
    \de B_u\equiv\f{u}{|\pd_B h_z(B^*)|},
\lbl{eq:dBu}
\eeq
of the magnetic field, respectively.

\begin{figure}
\includegraphics[width=.48\textwidth]{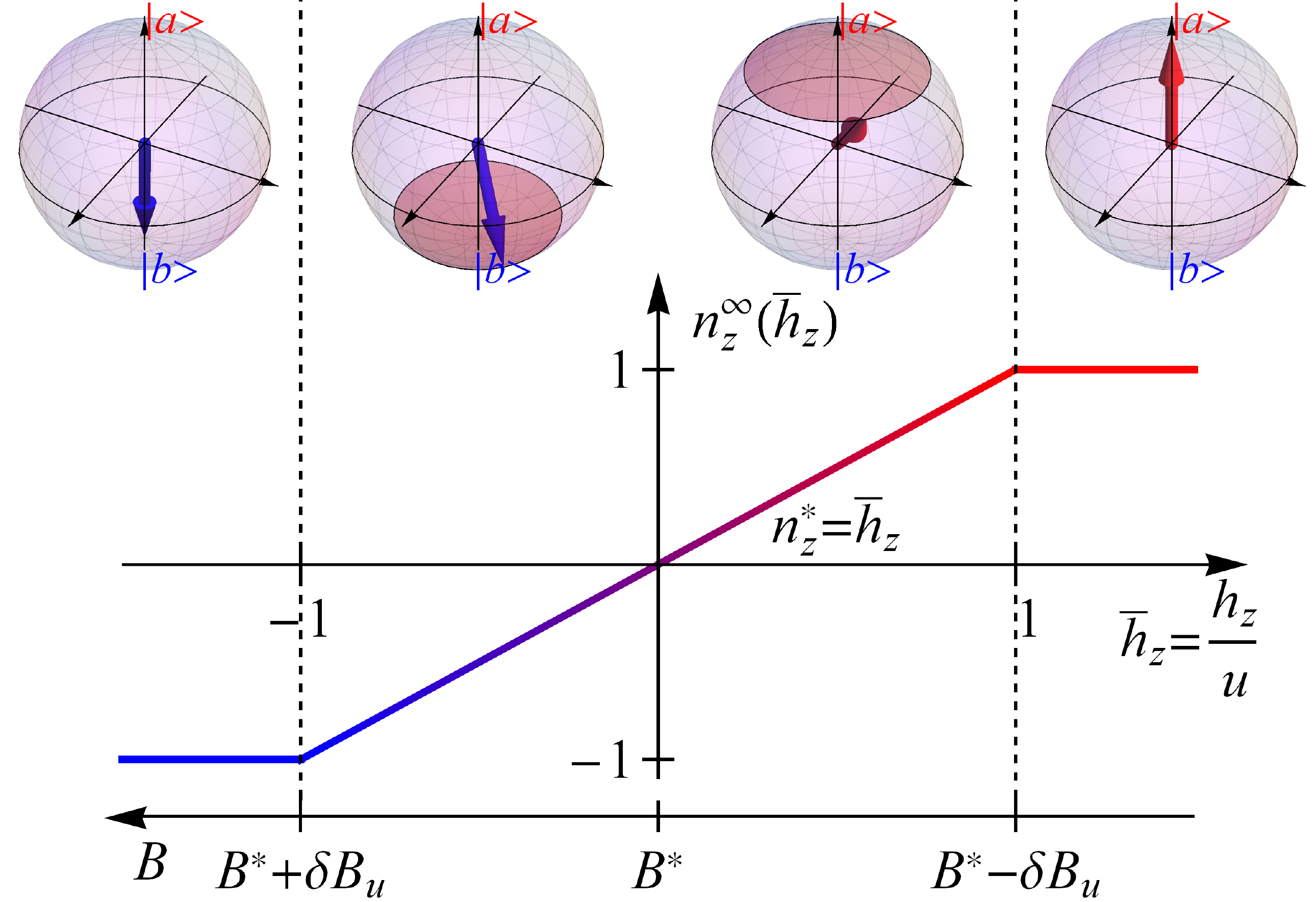}
\caption{(Color online)
Bulk phase diagram for the $\Ux(1)$-symmetric $\s$-model obtained by minimizing
the energy function $\Ec(n_z)$ [\eq{Ec}].
The plot shows the dependence $n_z^\i(\hr_z)$ of the optimal isospin projection
on the normalized field $\hr_z=h_z/u$;
the color code depicts the weight of the $a$ (red) and $b$ (blue) LL states.
The Bloch spheres depict
the corresponding isospin orders (\ref{eq:n+<}), (\ref{eq:n-<<+}), and (\ref{eq:n<-}).
}
\lbl{fig:nz}
\end{figure}

An analogous phase diagram was obtained for a double-layer system~\cite{Pik}.
Also, the region $0\leq h_z$
reproduces the part of the phase diagram for the $\nu=0$ state in graphene~\cite{Herbut,MKmlg,MKblg},
where the $\nb^\i=\nb_z$ and $\nb^\i=\nb^*(\vphi_0)$ phases correspond to the F and CAF phases, respectively.
For graphene, the isospin $\nb$ would correspond to the spin
polarization of one of the sublattices of the honeycomb crystal lattice.

\section{System with an edge, ground states\lbl{sec:edgeground}}

In this section, we obtain the ground state configurations of the OP $\nb(\rb)$ taking the effect of the edge into account.
Such configurations, to be denoted $\nb_0(\rb)$, minimize the energy functional (\ref{eq:E}),
\[
    \Ef[\nb_0]=\min_{\nb}\Ef[\nb],
\]
under the boundary condition (\ref{eq:bc}) constraint.
The ground state configuration $\nb_0(\rb)$ is a stationary point of the energy functional.
In terms of the spherical angles $\tht$ and $\vphi$ [\eq{n}],
the stationary-point equations read
\beq
    \f{\de\Ef[\tht,\vphi]}{\de\tht}=\rho[-\n^2\tht+\tf12\sin2\tht(\n\vphi)^2]+\pd_\tht\Ec(\tht)=0,
\lbl{eq:dEtht}
\eeq
\beq
    \f{\de\Ef[\tht,\vphi]}{\de\vphi}=-\rho\n(\sin^2\tht\n\vphi)=0.
\lbl{eq:dEphi}
\eeq
(Throughout, we will denote the energy dependence $\Ec(\tht)=\Ec(n_z=\cos\tht)$ on $\tht$
by the same function, since it should not lead to confusion.)

Since in the presence of the edge the translational symmetry along $y$ direction is still preserved,
the ground state configuration is $y$-independent, $\nb_0(x,y)\equiv\nb_0(x)$:
changes of the isospin with $y$ would only result in the rise of the gradient energy.

Away from the edge in the bulk, i.e., asymptotically at $x\rarr-\i$,
the ground state configurations must approach the constant value
\beq
    \nb_0(x\rarr-\i)=\nb^\i
\lbl{eq:n0iy}
\eeq
of the bulk ground state order $\nb^\i$,
which, depending on $h_z$, is one of the orders (\ref{eq:n+<}), (\ref{eq:n-<<+}), or (\ref{eq:n<-})
that minimize $\Ec(n_z)$, as obtained in the previous section.
Therefore, whenever $\nb^\i$ differs from the boundary order [\eq{bc}]
\beq
    \nb_0(x=0)=-\nb_z,
\lbl{eq:n0bc}
\eeq
$\nb_0(x)$ is a spatially inhomogeneous domain-wall configuration along $x$ that ``connects'' these orders.

Further, due to the $\Ux(1)$ symmetry of the energy function $\Ec(n_z)$ and boundary condition (\ref{eq:bc}),
it is clear that the angle $\vphi(x)\equiv\vphi_0$ in the spherical parametrization (\ref{eq:n}) of $\nb_0(x)$
is constant and arbitrary: similarly, changes in $\vphi$ would only result in the rise of the gradient energy.
Note that \eq{dEphi} is satisfied automatically by a constant $\vphi$.

Therefore, the ground state configuration has the following form
\beq
    \nb_0(x|\vphi_0)=(\sin\tht_0(x)\cos\vphi_0,\sin\tht_0(x)\sin\vphi_0,\cos\tht_0(x)).
\lbl{eq:n0}
\eeq
For the angle $\tht(x)$ dependent only on $x$ and the constant $\vphi_0$,
the energy functional (\ref{eq:E}) per unit length in the $y$ direction reduces to
\beq
    E^\x{1D}[\tht(x);u,h_z]
    =\int_{-\i}^0 \f{\dx x}{s} E_x[\tht],
    \spc
    E_x[\tht]=\f{\rho}2(\n_x\tht)^2+\Ec(\tht).
\lbl{eq:E1D}
\eeq
The ground state configuration $\tht_0(x)$ minimizes this functional and thus satisfies
its stationary point equation
\beq
    -\rho\n_x^2\tht+\pd_\tht\Ec(\tht)=0
\lbl{eq:thteq}
\eeq
(which is evidently equivalent to \eq{dEtht} under these assumptions).

Equation (\ref{eq:thteq}) needs to be supplemented by the boundary conditions
\beq
    \tht(x=-\i)=\tht^\i
    \mbox{ and }
    \tht(x=0)=\pi
\lbl{eq:thtbc}
\eeq
following from \eqs{n0iy}{n0bc}, where $\tht^\i$ is the angle of the bulk order $\nb^\i$.

The solution of this boundary problem can be facilitated by noticing the analogy of \eq{thteq}
with the Newton equation for a point particle in one dimension,
where $\tht$ and $x$ play the roles of coordinate and time, respectively. The equation has an integral of motion
\beq
    \f{\rho}{2}(\n_x\tht)^2-\Ec(\tht)=-\Ec^\i,
\lbl{eq:1Denergy}
\eeq
equivalent to the total energy of the effective particle.
It can be obtained by multiplying \eq{thteq} by $\n_x\tht$ and integrating once over $x$.
The gradient term $\f{\rho}{2}(\n_x\tht)^2$ in \eq{1Denergy} plays the role of the kinetic energy, while $-\Ec(\tht)$
plays the role of the potential energy. The value of this integral of motion
is set by its value $-\Ec^\i$ in the bulk [\eqss{E+<}{E-<<+}{E<-}], where $\n_x\tht\rtarr 0$.

Equation~(\ref{eq:1Denergy}) can be further integrated, which produces an implicit dependence of $\tht_0(x)$ on $x$ given by
\beq
    -x=\int_{\tht_0}^\pi \f{\dx\tht}{\q{\f2\rho\lt[\Ec(\tht)-\Ec^\i\rt]}}.
\lbl{eq:tht0int}
\eeq

The functional form (\ref{eq:Ec}) of $\Ec(n_z)$
allows for explicit integration of \eq{tht0int} in terms of elementary functions and subsequent inversion.
The explicit forms of the solutions are
\bwt
\beqar
    \tht_0(x)&=&\arccos\lt[1-\f{2(\hr_z-1)}{\hr_z\cosh^2(\q{\hr_z-1}\xr)-1}\rt],\spc u<h_z,
\lbl{eq:tht0+<}\\
    \tht_0(x)&=&2\arctan\lt[\q{\f{1-\hr_z}{1+\hr_z}}\f1{\tanh\lt(-\f{\q{1-\hr_z^2}}2\xr)\rt)}\rt], \spc -u<h_z<u,
\lbl{eq:tht0-<<+}\\
    \tht_0(x)&=&\pi,\spc h_z<-u.
\lbl{eq:tht0<-}
\eeqar
\ewt
Here, $\hr_z$ is the dimensionless field defined in \eq{hbrz} and
\beq
    \xr=\f{x}{l_u}
\lbl{eq:xbr}
\eeq
is the dimensionless coordinate normalized by the length scale
\beq
    l_u=\q{\f{\rho}u}
\lbl{eq:lu}
\eeq
set by the anisotropy energy $u$. The solutions, therefore, have the scaling form $\tht_0(x)=\tht_0(x;u,h_z)=\tht_0\lt(\xr;\hr_z\rt)$.

These solutions $\tht_0(x)$
minimize the functional $E^\x{1D}[\tht]$ [\eq{E1D}]
and the respective isospin configurations $\nb_0(x|\vphi_0)$ [\eq{n0}]
minimize the functional $\Ef[\nb]$ [\eq{E}]
in the presence of the edge, described by the boundary condition (\ref{eq:n0bc}).

\begin{figure}
\includegraphics[width=.40\textwidth]{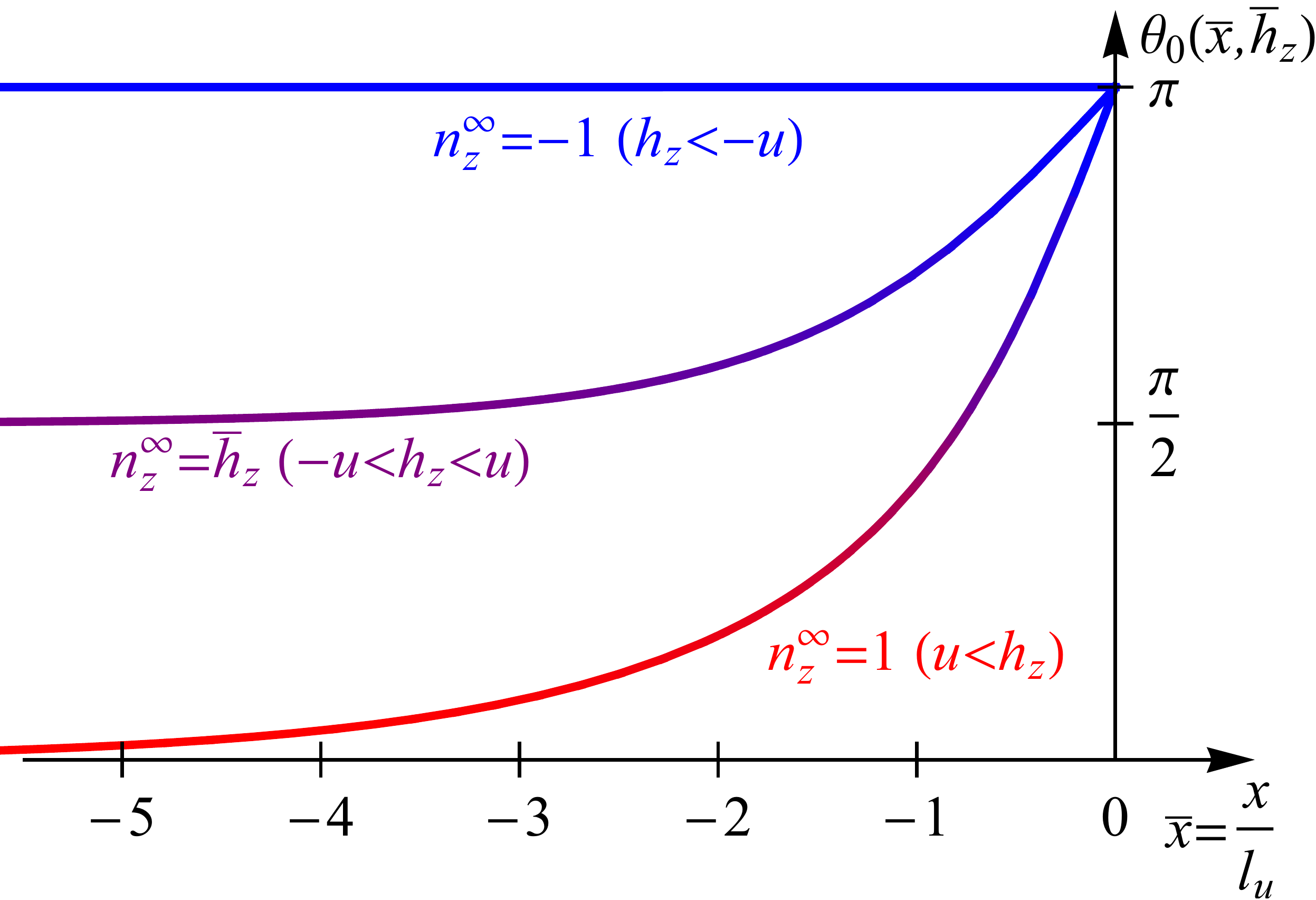}
\caption{(Color online)
The angle functions $\tht_0(x)=\tht_0(\xr,\hr_z)$ [\eqss{tht0+<}{tht0-<<+}{tht0<-}]
for the ground state solution (\ref{eq:n0}) for a system with an edge.
Three cases according to the three phases [\eqss{n+<}{n-<<+}{n<-}] are shown.
At $x\rarr-\i$, the solutions approach the asymptotic values $\tht^\i(\hr_z)$ for the bulk ground states.
At $x=0$, the solutions satisfy the boundary condition (\ref{eq:thtbc}) imposed by the edge.
The functions are color-coded according to the $n_z=\cos\tht_0(x)$ projection,
depicting the weight of the $a$ (red) and $b$ (blue) LL states.
}
\lbl{fig:tht0}
\end{figure}

The functions are plotted in \figr{tht0}.
In the $\nb^\i=\nb_z$ and $\nb^\i=\nb^*(\vphi_0)$ phases,
$\tht_0(x)$ grows monotonically from the bulk value $\tht^\i$ at $x=-\i$ to $\pi$ at the edge $x=0$.
Since $\tht^\i$ corresponds to the minimum of $\Ec(\tht)$
the effective particle starts at ``time'' $x=-\i$ at the maximum $-\Ec^\i$
of its potential energy $-\Ec(\tht)$,
i.e., in an unstable equilibrium position, and ``falls down'' the potential energy curve.
In the $\nb^\i=-\nb_z$ phase, the bulk and edge orders are the same and $\tht_0(x)\equiv\pi$ is a constant.

The solutions $\tht_0(x)$ approach the asymptotic bulk value $\tht^\i$ exponentially over the length scales
\[
   \f{l_u}{\q{\hr_z-1}}=\q{\f{\rho}{h_z-u}}, \spc u<h_z,
\]
\beq
   \f{l_u}{\q{1-\hr_z^2}}=\q{\f{\rho u}{u^2-h_z^2}}, \spc -u<h_z<u.
\lbl{eq:l-<<+}
\eeq
At both phase transitions $h_z=\pm u$, these length scales become infinite.

Exactly at the $h_z=u$ phase transition, the solution takes the form
\beq
    \tht_0(x)=2\arctan\lt(-\f1\xr\rt), \spc h_z=u,
\lbl{eq:tht0+}
\eeq
as follows from both \eqs{tht0+<}{tht0-<<+} in the limits $h_z\rarr u\pm0$.
It approaches the bulk value $\tht^\i=0$ as a power law $\tht_0(x)\approx -2/\xr$.
We emphasize that even exactly at the phase transition $h_z=u$,
the domain wall has the spatial scale $l_u$,
although the bulk asymptotic value is approached according to a power law, and not exponentially.
This point will be important for the considerations below in \secr{Deest}.

Close to the $h_z=-u$ transition, when $h_z+u\ll u$, the bulk value $\tht^\i$ is close to $\pi$,
and the solution simplifies to
\[
    \tht_0(x)=\pi-\q{2(1+\hr_z)}
    \tanh\lt(-\q{\f{1+\hr_z}2}\xr\rt).
\]

Most important for topological properties are, however,
the degeneracies of the isospin solutions $\nb_0(x|\vphi_0)$ [\eq{n0}].
According to the possible bulk phases, we have the following three cases.

1) In the $\nb^\i=\nb_z$ phase realized at $u<h_z$ [\eq{n+<}],
the ground state solution $\nb_0(x|\vphi_0)$
is degenerate according to the arbitrary angle $\vphi_0$.
Note though that the bulk order is nondegenerate
since at $\nb^\i=\nb_z$ the angle $\vphi_0$ is undefined
and the $\Ux(1)$ symmetry is not spontaneously broken in the bulk.
So, this degeneracy occurs (at the mean-field level) at the edge and not in the bulk.

2) In the $\nb^\i=\nb^*(\vphi_0)$ phase
realized at $-u<h_z<u$ [\eq{n-<<+}],
the ground state configuration $\nb_0(x|\vphi_0)$
is degenerate according to the arbitrary angle $\vphi_0$,
equal to the one of the asymptotic bulk configuration $\nb^*(\vphi_0)$.
So, this degeneracy describes the spontaneous breaking of $\Ux(1)$ symmetry
in the bulk and there is not extra degeneracy at the edge.

3) In the $\nb^\i=-\nb_z$ phase realized at $h_z<-u$ [\eq{n<-}],
the bulk and edge orders are exactly the same,
and the ground state solution for the system with an edge
is a constant $\nb_0(x)\equiv-\nb_z$
and thus nondegenerate ($\vphi_0$ is undefined).

The properties can also be illustrated with the help of the Bloch sphere, see \figr{geodesics}.
The ground state domain wall configurations $\nb_0(x|\vphi_0)$ can be visualized as geodesic paths
connecting the bulk $\nb^\i$ and edge $-\nb_z$ orders and parameterized by the coordinate $x$.
In the $\nb^\i=\nb_z$ phase, the bulk and edge orientations are exactly opposite,
and there is an infinite number of geodesics, parameterized by the angle $\vphi_0$.
In the $\nb^\i=\nb^*(\vphi_0)$ phase, for a given angle $\vphi_0$ in the bulk,
the geodesic connecting $\nb^*$ and $-\nb_z$ is unique: it is a path in the vertical plane of the constant $\vphi_0$.
As we show in the next \secr{edgeexc}, these degeneracy properties
of the ground state solutions are key to the properties of the charge edge excitations.

We calculate the ground state energy of the system with an edge.
It is sensible to subtract the asymptotic bulk contribution
and thus define the energy quantities
\bwt
\beq
    dE^\x{1D}[\tht(x);u,h_z]
    \equiv\int_{-\i}^0\f{\dx x}s\, dE_x[\tht(x);u,h_z]
    =E^\x{1D}[\tht(x);u,h_z]-E^{\x{1D}\i}(u,h_z),
\lbl{eq:dE1D}
\eeq
\beq
    dE_x[\tht(x);u,h_z]\equiv E_x[\tht(x);u,h_z]-\Ec^\i(u,h_z),
\lbl{eq:dEx}
\eeq
\[
    E^{\x{1D}\i}(u,h_z)\equiv\int_{-\i}^0\f{\dx x}s\, \Ec^\i(u,h_z)
\]
in terms of \eq{E1D}.
The corresponding ground-state energy
per unit length in the $y$ direction equals
\beq
    dE^\x{1D}_0(u,h_z)
    \equiv\min_{\tht(x)} dE^\x{1D}[\tht(x);u,h_z]
    =dE^\x{1D}[\tht_0(x;u,h_z);u,h_z]
\lbl{eq:dE1D0}
\eeq
and can be referred to as the {\em domain-wall energy}.
This quantity is not extensive in the $x$ direction and indeed describes
the energy associated only with the domain-wall isospin texture at the edge.

Exploiting the integral of motion \eqn{1Denergy},
the domain-wall energy can be presented as
\[
    dE^\x{1D}_0(u;h_z)=2\int_{-\i}^0\f{\dx x}{s}[\Ec(\tht_0(x))-\Ec^\i]
\]
and calculated explicitly using the expressions \eqsn{tht0+<}{tht0-<<+} for the ground state solutions.
For the $\nb^\i=\nb_z$ phase [the same can be done for the $\nb^\i=\nb_0^*(\vphi)$ phase],
we obtain
\beq
    dE^\x{1D}_0(u,h_z)=2 \f{l_u}{s} u F(\hr_z),
\lbl{eq:Edw}
\eeq
where
\beqar
    F(\hr_z)&=&-F_2(\hr_z)+\hr_z F_1(\hr_z)
        =\hr_z\arcsin\f1{\q{\hr_z}}+\q{\hr_z-1}, \spc u<h_z,
\lbl{eq:F}\\
    F_2(\hr_z)
        &=&\f12\int_{-\i}^0\dx\xr\,\sin^2\tht_0(\xr)
        =\hr_z\arcsin\f1{\q{\hr_z}}-\q{\hr_z-1},
\lbl{eq:F2}\\
    F_1(\hr_z)&=&\int_{-\i}^0\dx\xr\,[1-\cos\tht_0(\xr)]
        =2\arcsin\f1{\q{\hr_z}}
\lbl{eq:F1}
\eeqar
\ewt
are dimensionless functions of the normalized field $\hr_z$,
the latter two arising from the anisotropy $\f{u}2(n_z^2-1)$ and ''Zeeman'' $-h_z(n_z-1)$ contributions to $\Ec(\tht_0(x))-\Ec^\i$, respectively.
The function $F(\hr_z)$ is plotted in \figr{F}.

Importantly, as we will see below in \secsr{edgeexcgapless}{lliq},
the dependence \eqn{Edw} of the domain-wall energy on parameters $u$ and $h_z$
defines not only the ground state but also the properties of the low-energy edge excitations of the $\nb^\i=\nb_z$ phase.

\begin{figure}
\includegraphics[width=.35\textwidth]{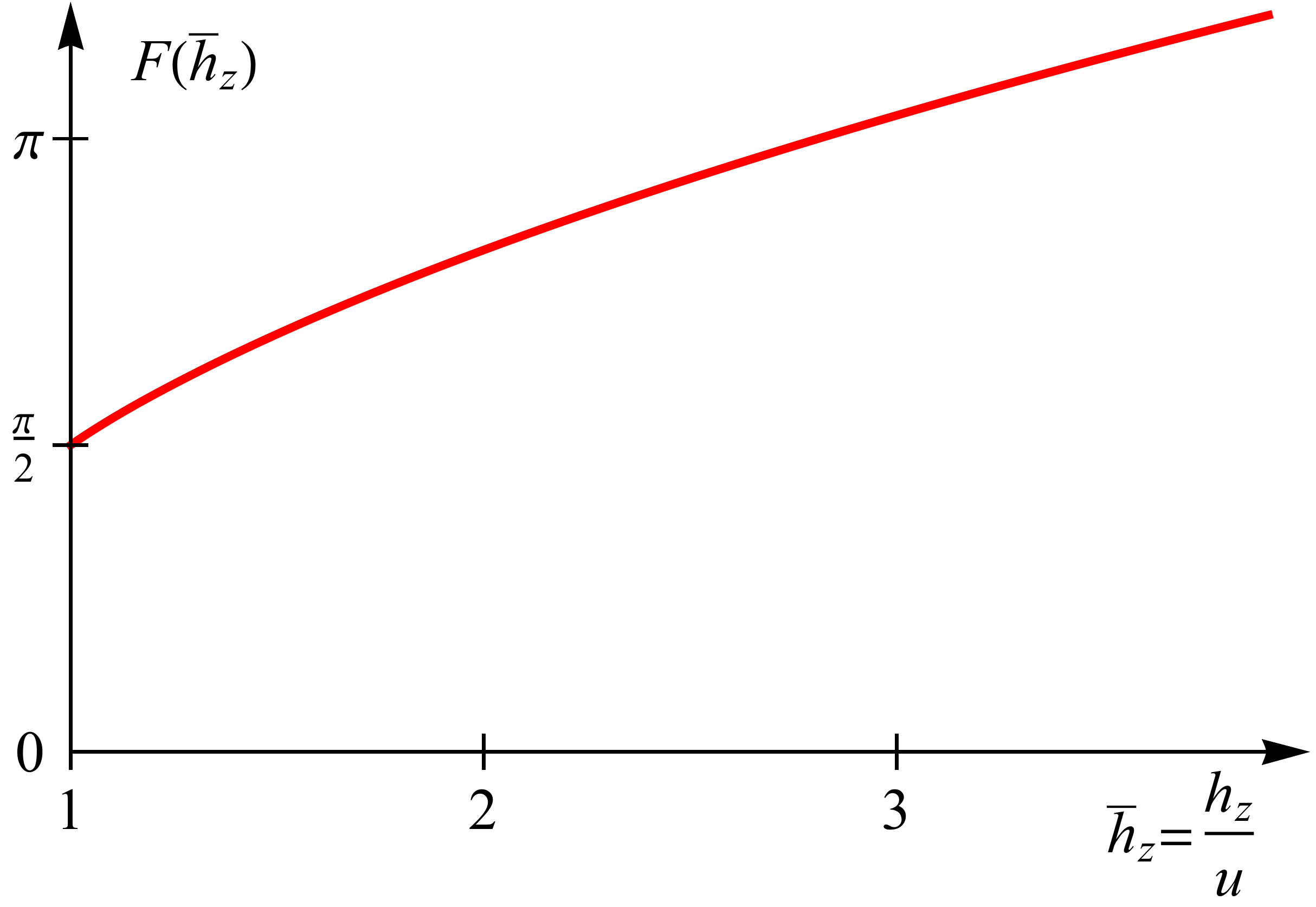}
\caption{(Color online)
The function $F(\hr_z)$ [\eq{F}], which determines the dependence of the ground-state domain-wall energy
$dE^\x{1D}_0(u,h_z)$ [\eq{Edw}] on the normalized field $\hr_z=h_z/u$ in the $\nb^\i=\nb_z$ phase at $u<h_z$.
}
\lbl{fig:F}
\end{figure}

\begin{figure}
\includegraphics[width=.23\textwidth]{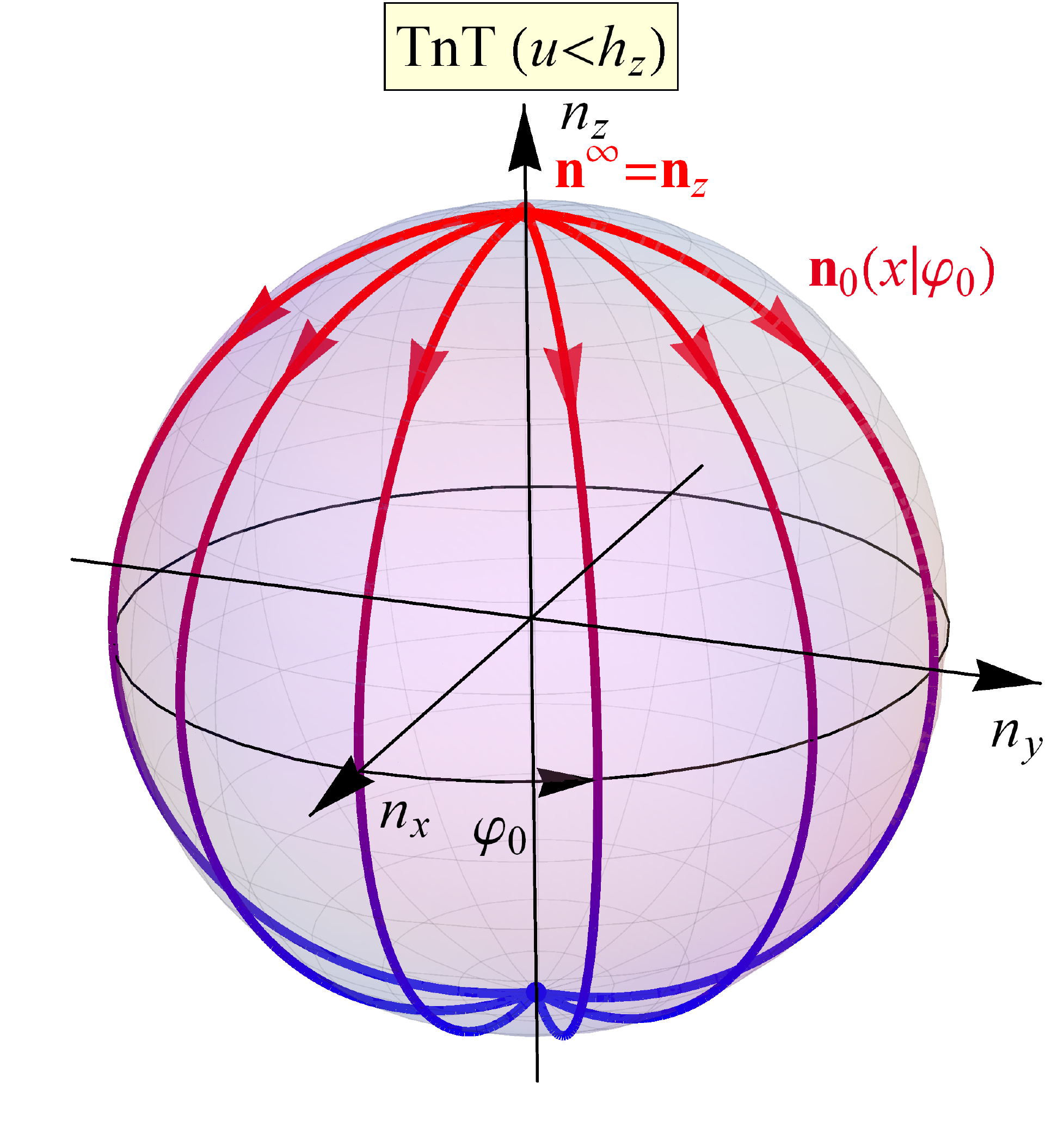}
\includegraphics[width=.23\textwidth]{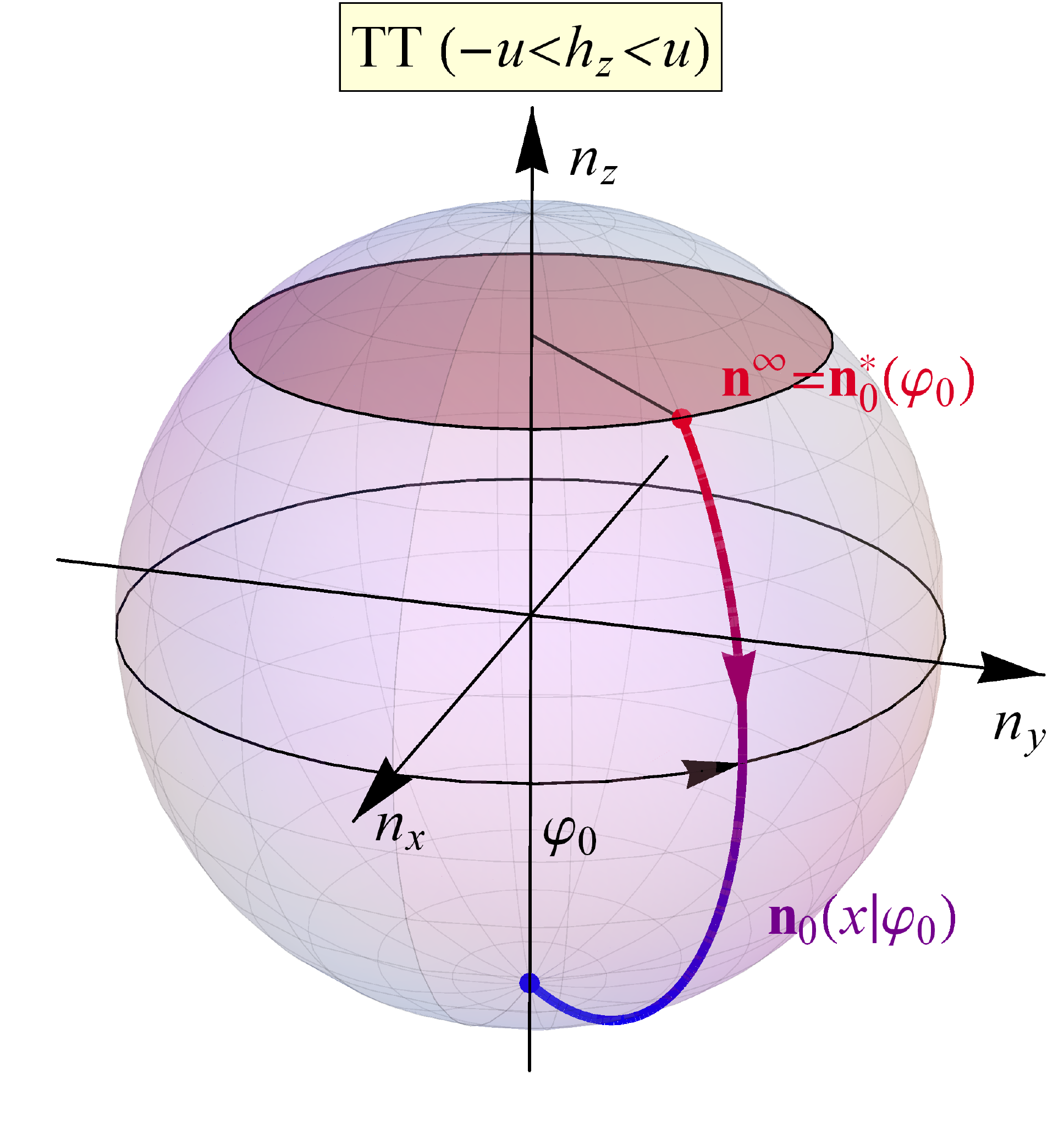}
\caption{(Color online)
The ground state geodesic paths $\nb_0(x|\vphi_0)$ [\eq{n0}] on the isospin Bloch sphere
for the $\nb^\i=\nb_z$ (left) and $\nb^\i=\nb_0^*(\vphi_0)$ (right)
phases, realized at $u<h_z$ and $-h_z<u<h_z$, respectively.
}
\lbl{fig:geodesics}
\end{figure}

\section{System with an edge, charge excitations \lbl{sec:edgeexc}}

\subsection{General considerations}

Having established the properties of the ground states of the system with an edge,
we now turn to the analysis of its charge excitations.

As is well-known~\cite{QHFM}, QHFM systems support charge excitations, which are described by the configurations of the OP
with nonzero topological charge, equal to the electric charge.
For the system in question, with the OP being the isospin-$\f12$,
the charge density of a configuration $\nb(\rb)$ is given by
\beqar
    \kappa[\nb](\rb)
        &=&\f1{4\pi}(\nb\cd[\n_x\nb\tm\n_y\nb]) \nn\\
        &=&\f1{4\pi}\sin\tht(\n_x\tht\n_y\vphi-\n_y\tht\n_x\vphi).
\lbl{eq:kappa}
\eeqar
The total net charge of the configuration is then given by the integral
\beq
    q[\nb]=\int\dx^2\rb\,\kappa[\nb](\rb).
\lbl{eq:q}
\eeq
This topological charge is the invariant of the mapping realized by $\nb(\rb)$ from the coordinate space of $\rb$ to the 2D Bloch sphere;
it can be visualized as the number of times the sphere is wound as the space of $\rb$ is explored.
In this article, we will be interested in the excitations with integer charge $q$,
whose boundary values are the same  as in the ground state.
In principle, excitations with non-integer charge $q$ are also possible if the ground state has broken continuous symmetry,
which is indeed the case for the intermediate phase $\nb^\i=\nb^*(\vphi_0)$,
but they are outside of the scope of this paper.

The difference
\beq
    \de\Ef[\nb]\equiv\Ef[\nb]-\Ef[\nb_0]
\lbl{eq:dE}
\eeq
between the energy of a configuration $\nb(\rb)$ and of the ground state configuration $\nb_0(x)$
will be called the {\em excitation energy} of $\nb(\rb)$.
The configuration $\nb^q(\rb)$ of charge $q$  will be further designated with a superscript.
Denote $\nb_0^q(\rb)$ the charge-$q$ configuration that minimizes the energy (\ref{eq:E})
among all charge-$q$ configurations $\nb^q(\rb)$,
\beq
    \Ef[\nb_0^q]=\min_{\nb^q}\Ef[\nb^q].
\lbl{eq:En0q}
\eeq
Clearly, the ground state configuration  $\nb_0(x|\vphi_0)$ has zero charge
(the charge density $\kappa[\nb_0]\equiv0$)
and hence $\nb_0=\nb_0^{q=0}$.

The excitation energy (\ref{eq:dE})
\beq
    \De^q
    \equiv \min_{\nb^q}\de\Ef[\nb^q]=\de\Ef[\nb_0^q]=\Ef[\nb_0^q]-\Ef[\nb_0]
\lbl{eq:Deqdef}
\eeq
of the minimal-energy configuration $\nb_0^q(\rb)$
will be called the {\em gap} of charge-$q$ excitations.
The minimum
\[
    \De=\min_{q\neq0}\De^q
\]
among all $q\neq0$ defines the gap of excitations of any charge.
Typically, $\De^q$ is a growing function of $q$,
and therefore the unit-charge $q=\pm1$ excitations, for which the Bloch sphere is covered once,
determine the gap, $\De=\De^{q=\pm 1}$.

The minimal-energy charge-$q$ configurations $\nb_0^q(\rb)$ satisfy
the same stationary-point \eqs{dEtht}{dEphi} as the ground state $\nb_0(x|\vphi_0)$,
since they describe any {\em local} minimum in the configuration space.
In order not to contain ``extensive'' contributions to the excitation energy (\ref{eq:Deqdef}),
proportional to the size of the system,
the configurations $\nb_0^q(\rb)$
must asymptotically approach the ground state configuration $\nb_0(x|\vphi_0)$:
\[
    \nb_0^q(x,y\rarr\pm\i)\rarr \nb_0(x|\vphi_0),
\]
\[
    \nb_0^q(x\rarr-\i,y)\rarr \nb^\i.
\]
The finite-size region,
where most of the winding of the Bloch sphere occurs, $\nb_0^q(\rb)$ differs from $\nb_0(x|\vphi_0)$,
and the charge density $\kappa[\nb^q_0](\rb)$ is located,
can be referred to as the {\em core} of the charge excitation.

For the considered system with an edge, one should distinguish between bulk and edge charge excitations.
In the bulk excitations, known as skyrmions,
the core is located deep in the bulk, away from the domain wall at the edge, such that its effect can be neglected.
In the edge excitations, the core is located in the domain wall.
At the qualitative level, it is clear that the energy of the edge charge excitation
will generally be smaller than that of the bulk skyrmion:
since in the ground state $\nb_0(x|\vphi_0)$ some changes in the isospin orientation are already present,
less additional changes are required in $\nb_0^q(\rb)$ to wind the whole Bloch sphere; hence, the smaller the energy cost.
The general properties of bulk skyrmions are well-understood~\cite{QHFM};
below, we concentrate on the edge excitations.

\subsection{Gapless edge excitations of the $\nb^\i=\nb_z$ phase \lbl{sec:edgeexcgapless}}

We first look at the $\nb^\i=\nb_z$ phase with preserved $\Ux(1)$ symmetry in the bulk.

According to the previous section,
in the $\nb^\i=\nb_z$ phase, the bulk isospin orientation is exactly opposite
to the edge isospin orientation and there is an infinite number of geodesics $\nb_0(x|\vphi_0)$,
parameterized by the angle $\vphi_0$, connecting these two orientations, see \figr{geodesics}.
This degeneracy allows one to construct a charge excitation
by winding the angle $\vphi_0$ in the $y$ direction along the edge~\cite{FB,FI}.
In fact, the Ansatz $(\tht(x),\vphi(y))$ for $\nb(\rb)$ [\eq{n}]
decouples the stationary point equations (\ref{eq:dEtht}) and (\ref{eq:dEphi}).
Equation (\ref{eq:dEphi}) reduces to
\beq
    -\n^2_y\vphi(y)=0.
\lbl{eq:dEphiy}
\eeq
Introducing the sample boundaries along the $y$ direction at $y=\pm\f{L_y}2$ and imposing the periodic boundary condition
\[
    \nb(x,y=+\f{L_y}2)=\nb(x,y=-\f{L_y}2),
\]
for the solution to \eq{dEphiy}, we obtain
\beq
    \vphi^q_0(y)=\vphi_0+2\pi q \f{y}{L_y}
    ,\spc
    \n_y\vphi^q_0=\f{2\pi q}{L_y}
    ,\spc
    u<h_z,
\lbl{eq:phi0q+<}
\eeq
where $q$ is integer. The integer $q$ is indeed the topological charge of the configuration,
as can be confirmed from either the geometric considerations or explicit calculation according to \eqs{kappa}{q}.

The energy functional then becomes
\bwt
\beq
    \Ef[\tht(x),\vphi_0^q(y)]
        =L_y\int_{-\i}^0\f{\dx x}s\,
        \lt\{\f{\rho}2[(\n_x\tht)^2+\sin^2\tht (\n_y\vphi^q_0)^2]+\Ec(\tht(x))\rt\}
    =L_y\lt\{ dE^\x{1D}[\tht(x);u-\rho(\n_y\vphi^q_0)^2,h_z]+E^{\x{1D}\i}\rt\}.
\lbl{eq:dE+<}
\eeq
\ewt
We notice that the gradient term $\f{\rho}2\sin^2\tht (\n_y\vphi^q_0)^2$
has the form of the anisotropy energy $\f{u}2(n_z^2-1)$
and the functional \eqn{dE+<} thus has the same form as the one \eqn{E1D} for the ground state
with the redefined anisotropy energy $u-\rho(\n_y\vphi_0^q)^2$ and can be expressed in terms
of the quantities in \eq{dE1D}.
Therefore, the functional \eqn{dE+<}
is minimized by the modified ground state configuration [\eq{tht0+<}]
\beq
    \tht_0^q(x)=\tht_0(x; u-\rho(\n_y\vphi_0^q)^2,h_z), \spc u<h_z,
\lbl{eq:tht0q+<}
\eeq
and the gap \eqn{Deqdef}
\bwt
\beq
    \De^q=L_y\lt[dE^\x{1D}_0(u-\rho\lt(2\pi q/L_y\rt)^2,h_z)-dE^\x{1D}_0(u,h_z)\rt], \spc u<h_z,
\lbl{eq:Deq+<}
\eeq
of charge-$q$ edge excitations of the $\nb^\i=\nb_z$ phase is expressed exactly
in terms of the domain-wall energy \eqn{Edw}.

The isospin configuration of the charge-$q$ excitation in the $\nb^\i=\nb_z$ phase
has the form
\beq
    \nb_0^q(x,y)=(\sin\tht_0^q(x)\cos\vphi_0^q(y),\sin\tht_0^q(x)\sin\vphi_0^q(y),\cos\tht_0^q(x)), \spc u<h_z,
\lbl{eq:n0q+<}
\eeq
and is shown in \figr{n01+<}.
\ewt

\begin{figure}
\includegraphics[width=.33\textwidth]{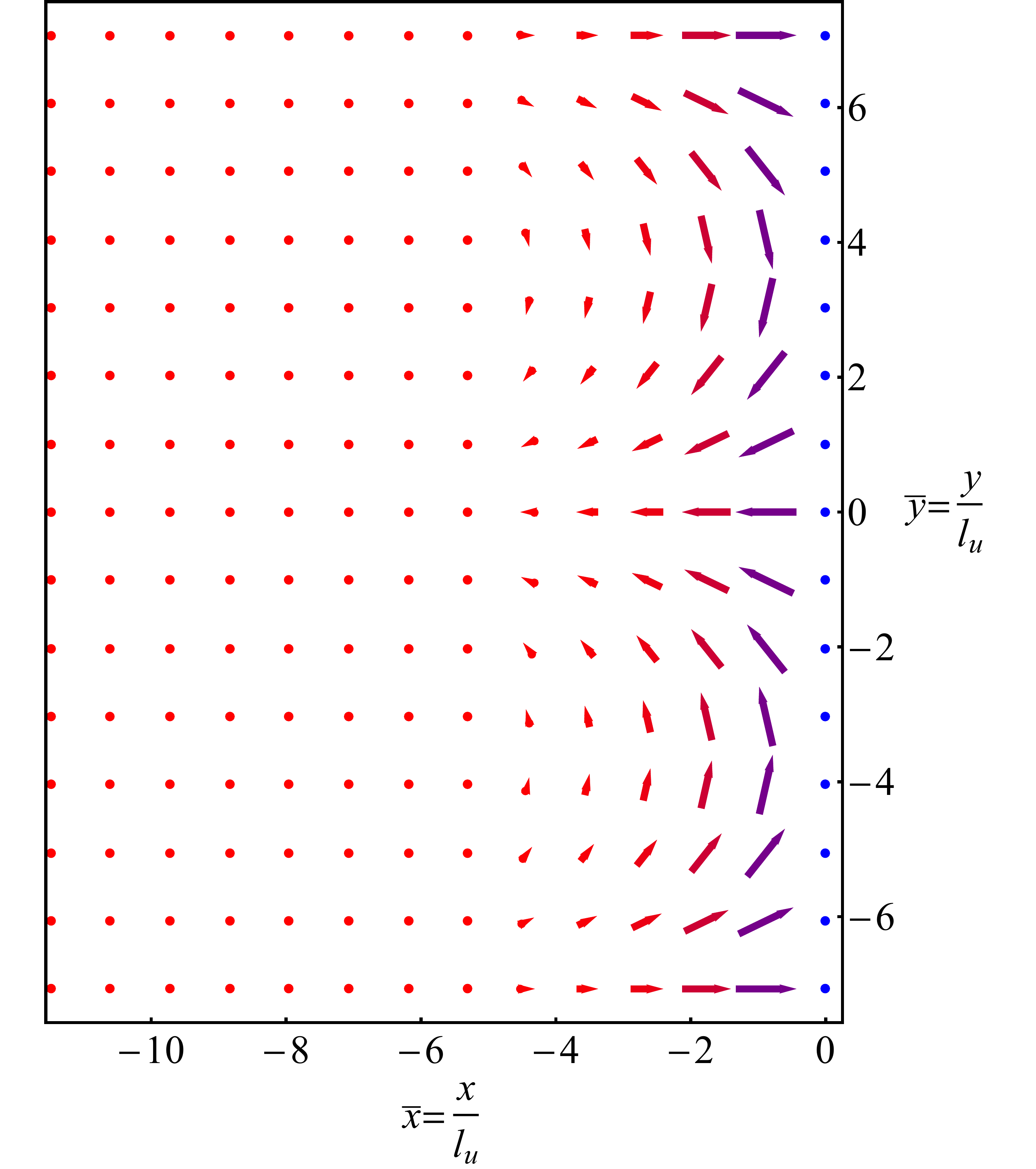}
\caption{(Color online)
The unit-charge isospin configuration $\nb_0^1(\rb)$ [\eqss{phi0q+<}{tht0q+<}{n0q+<}]
in the TnT $\nb^\i=\nb_z$ phase with preserved U(1) symmetry at $u<h_z$.
The arrows depict the 2D components $(n_x(\rb),n_y(\rb))$,
while the red-blue color code represents the value of the $n_z(\rb)$ component,
in accord with \figsr{nz}{tht0}. The paths on the Bloch sphere for varying $-\i<x\leq 0$
and several constant values of $y$ are shown in \figr{geodesics}(left).
}
\lbl{fig:n01+<}
\end{figure}

The gap \eqn{Deq+<} is finite only due to the finite size $L_y$ of the sample in the $y$ direction
and vanishes in the limit $L_y\rarr\i$.
Therefore, the phase $\nb^\i=\nb_z$ supports gapless edge charge excitations, similar to the findings of Refs.~\ocite{FB,FI}.
The leading term in the large-size limit $L_y\rarr\i$ can be obtained as an expansion
\beqar
    \De^q
        &\approx& -L_y \pd_u dE^\x{1D}_0(u,h_z) \rho\lt(\f{2\pi q}{L_y}\rt)^2 \nn\\
        &=&(2\pi q)^2 \f{l_u}{L_y}F_2(\hr_z)\eps_\odot
    \sim q^2 \f{l_u}{L_y}\eps_\odot, \spc u<h_z.
\lbl{eq:Deq+<exp1}
\eeqar
Here,
\beq
    \eps_\odot\equiv\f{\rho}{s}\sim\f{e_*^2}{l}
\lbl{eq:epsod}
\eeq
is the energy associated with the gradient term; it is due to SU(2)-symmetric interactions and is thus set by the Coulomb energy.
Equation~\eqn{Deq+<exp1} can also be obtained in a simpler way,
approximating $\tht_0^q(x)\approx \tht_0(x;u,h_z)$ [\eq{tht0q+<}]
by the ground state configuration \eqn{tht0+<}, i.e. taking $\nb^q_0(\rb)\approx\nb_0(x|\vphi_0^q(y))$,
in which case the gap contains only the gradient term
\beq
    \De^q\approx
    \int\f{\dx^2\rb}{s} \f{\rho}2 \sin^2\tht_0(x)(\n\vphi^q_0(y))^2,
\lbl{eq:Deq+<exp2}
\eeq
which does agree with \eq{Deq+<exp1}.

\subsection{Gapped edge excitations of the $\nb^\i=-\nb_z$ and $\nb^\i=\nb^*(\vphi_0)$ phases \lbl{sec:edgeexcgapped}}

The above construction of the gapless charge excitations
in the $\nb^\i=\nb_z$ phase is possible due to two conditions realized {\em in the ground state} $\nb_0(x|\vphi_0)$:
(i) preserved $\Ux(1)$ symmetry in the bulk
and (ii) continuous degeneracy of the ground state solution at the edge.
In the other two phases, one of these conditions is violated.
In the intermediate phase $\nb^\i=\nb^*(\vphi_0)$ [\eq{n-<<+}],
the $\Ux(1)$ symmetry is spontaneously broken in the bulk
and, {\em for a given bulk order}, characterized by the angle $\vphi_0$,
the ground state solution $\nb_0(x|\vphi_0)$ is unique.
In the $\nb^\i=-\nb_z$ phase [\eq{n<-}], the U(1) symmetry is preserved in the bulk,
but the bulk and edge orientations are exactly the same
and the ground state solution is just a constant $\nb_0(x)\equiv-\nb_z$.
These crucial differences in the ground state edge configurations of the phases
are visualized in \figr{geodesics}.
As a result, in both phases, the ground state solution is unique for a given bulk order
and analogous construction of the gapless charge excitations is not possible.
The edge charge excitations are therefore gapped.
In the intermediate $\nb^\i=\nb_0(\vphi_0)$ phase, the typical edge charge configuration $\nb_0^1(\rb)$
has the form shown in \figr{n01-<<+}, as we also demonstrate numerically, see \secr{num}.

\begin{figure}
\includegraphics[width=.33\textwidth]{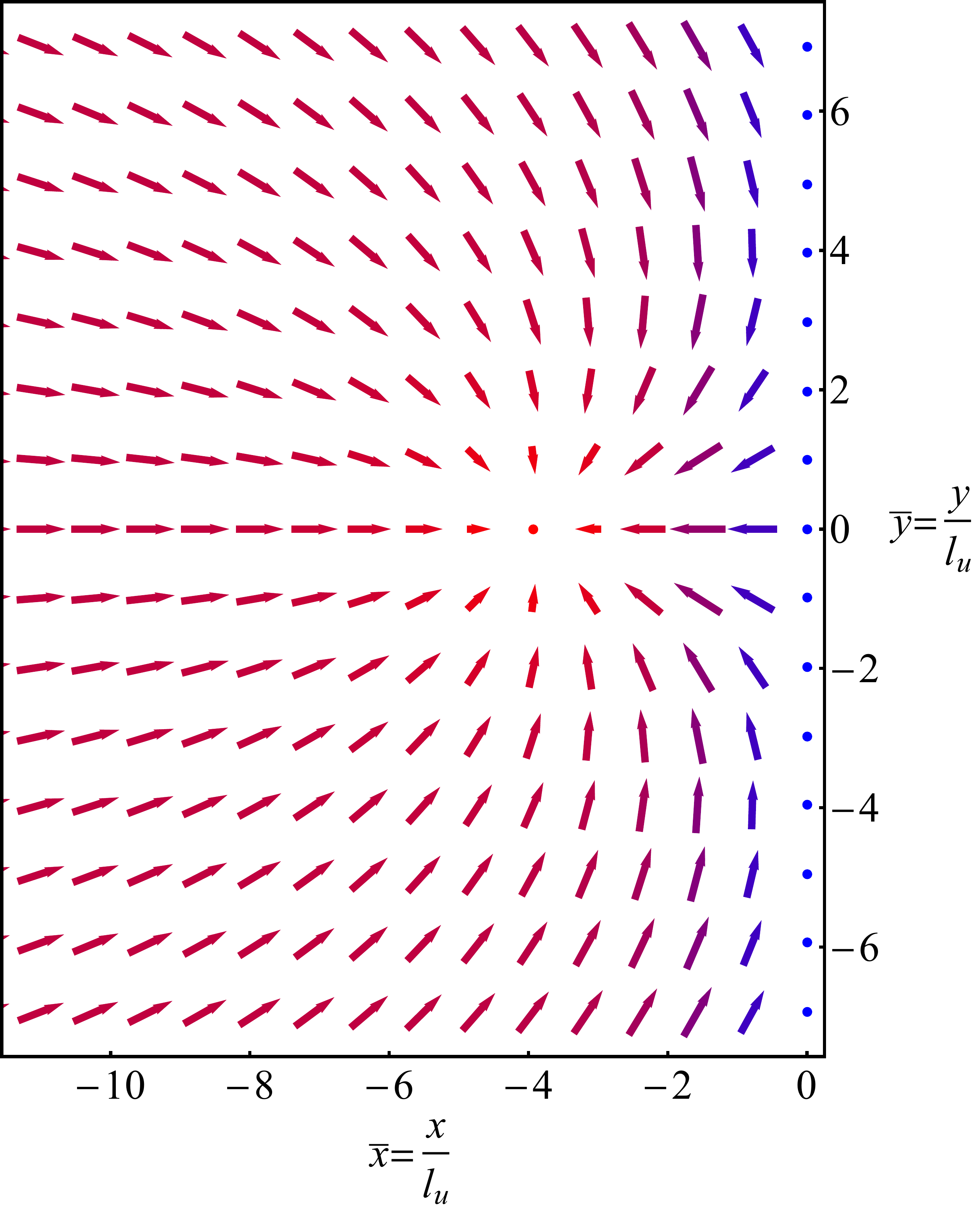}
\includegraphics[width=.28\textwidth]{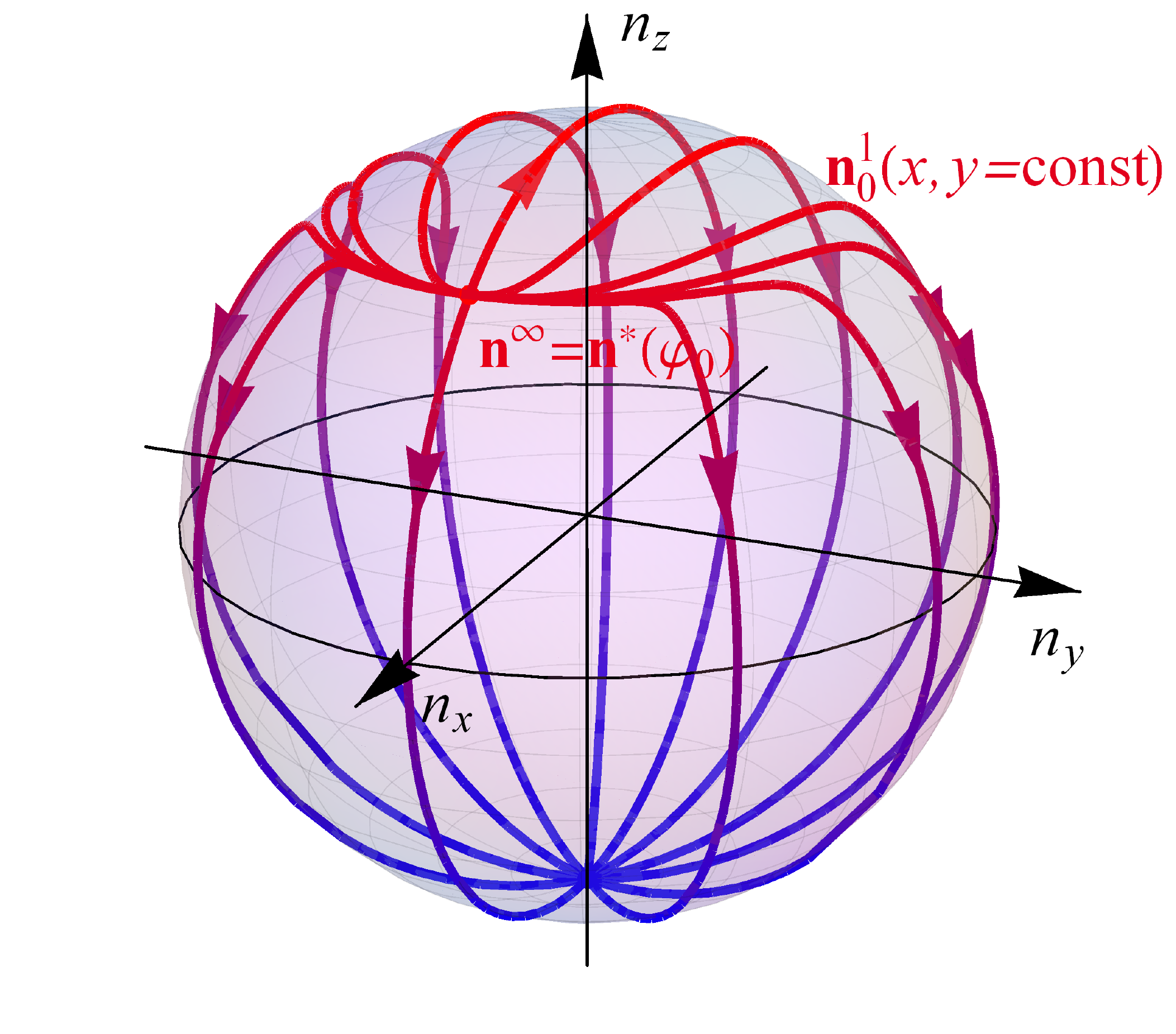}
\caption{(Color online)
The isospin unit-charge configuration $\nb_0^1(\rb)$ in the TT $\nb^\i=\nb^*(\vphi_0)$ phase
with spontaneously broken U(1) symmetry at $-u<h_z<u$,
obtained numerically for $\hr_z=0.5$ (top) and $\hr_z=0.75$ (bottom).
(top) The arrows depict the 2D components $(n_x(\rb),n_y(\rb))$,
while the red-blue color code represents the value of the $n_z(\rb)$ component.
(bottom) The paths on the Bloch sphere $\nb_0^1(x,y)$ for varying $-\i<x\leq 0$
and several constant values of $y$.
}
\lbl{fig:n01-<<+}
\end{figure}

In fact, in the phase $\nb^\i=-\nb_z$,
the lowest energy integer-charge excitations
are the bulk skyrmions, with the core infinitely far (relative to its size) away from the edge.
This can be understood  from the following argument.
The {\em bulk} skyrmions are the minimal-energy configurations
in an {\em infinite} sample among all charged configurations
with the constraint $\nb(r\rarr\i)=-\nb_z$.
For the half-infinite sample, when the boundary condition $\nb(x=0,y)=-\nb_z$ [\eq{bc}] is imposed,
one can continue the half-plane configurations to the whole plane with the constraint $\nb(x>0,y)=-\nb_z$
in the other half-plane. This would constrain the possible set of configurations,
which can only result in an increase of the excitation energy compared to that of the bulk skyrmions, for which such a constraint is absent.
In other words, placing the core closer to the edge in this case can only result in an energetically less favorable configuration.
Therefore, the $\nb^\i=-\nb_z$ phase has the largest energy of charge excitations among all three phases,
given by that of the bulk skyrmion,
\beq
    \De^q=\De_\x{sk}^q, \spc h_z<-u.
\lbl{eq:Deq<-}
\eeq
When the term (higher order in gradients)
describing the Coulomb self-interaction of the charge density $\kappa[\nb](\rb)$ [\eq{kappa}] is neglected,
as done in the $\s$-model we study, the skyrmion energy is given by
\beq
    \De^q_\x{sk}
    =4\pi\eps_\odot|q|
\lbl{eq:Deqsk}
\eeq
and its size is formally zero~\cite{Sondhi} due to the presence of the energy $\Ec(n_z)$ of the SU(2)-asymmetric terms.

The general qualitative behavior of the edge charge excitation gap $\De^q(h_z)$ (\ref{eq:Deqdef})
in the intermediate $\nb^\i=\nb^*(\vphi_0)$ phase can be understood from the continuity argument.
Since the transitions at $h_z=\pm u$ are continuous second-order transitions
and the intermediate phase continuously interpolates between the $\nb^\i=\nb_z$ and $\nb^\i=-\nb_z$ phases,
$\De^q(h_z)$ monotonically and continuously grows
upon decreasing $h_z$ in the range $-u<h_z<u$,
starting from zero value  at $h_z=u$ and reaching its
maximal value of $\De^q_\x{sk}$ at $h_z=-u$.
Since the bulk phase is controlled solely by the normalized dimensionless field $\hr_z$ [\eq{hbrz}],
the gap has the following scaling form
\[
    \De^q(h_z)=\eps_\odot\Der^q(\hr_z),
\]
where $\Der^q(\hr_z)$ is a dimensionless function of $\hr_z$.

An analogous continuous growth of the edge excitation gap
was earlier predicted for the CAF phase of the $\nu=0$ state in graphene,
originally employing a similar continuity argument~\cite{MKblg}
and within a simplified picture of single-particle edge excitations~\cite{MKsimp}.
The transport behavior well consistent with this scenario
was shortly after observed in both bilayer~\cite{Maher} and monolayer~\cite{Young} graphene.
More recently, an analytical estimate for the edge excitation gap of the CAF phase was obtained~\cite{F2}
within a low-energy theory approach analogous to the one employed here;
the estimate we make below is in accord with that result.

\subsection{Intermediate phase close to the phase transition $h_z=u$ \lbl{sec:Deest}}

In the intermediate phase $\nb^\i=\nb^*(\vphi_0)$ close to the phase transition $h_z=u$,
i.e., when the deviation
\beq
    \de\hr_z=\f{\de h_z}{u}=\hr_z-1
,\spc
    \de h_z=h_z-u
\lbl{eq:dhz}
\eeq
is negative and small, $|\de\hr_z|\ll1$,
the gap $\Der^q(\hr_z)\ll 1$ is also small and can be estimated analytically.

First, by analogy with the construction for the $\nb^\i=\nb_z$ phase,
consider the configuration $\nb_0(x|\vphi^q(y))$ obtained from the ground state configuration [\eqs{n0}{tht0-<<+}]
for the intermediate phase by slowly (i.e., at scales much larger than the domain-wall width $l_u$)
winding the angle $\vphi^q(y)$ $q$ times
as the $y$ direction is spanned. The shape of $\vphi^q(y)$ is to be optimized.
The excitation energy $\de\Ef[\nb_0(x|\vphi^q(y))]$ [\eq{dE}]
of such a configuration would also contain only the gradient contribution,
analogous to \eq{Deq+<}.
However, since, unlike the $\nb^\i=\nb_z$ phase,
the bulk asymptotic angle $\tht^\i=\tht^*\neq 0$ [\eq{nz*}]
is nonzero in the intermediate phase,
the integral $\int_{-\i}^0\dx x\,\ldots$ would not be constrained to the domain-wall region of size $l_u$,
but would also contain an extensive contribution proportional to the size of the sample in the $x$ direction.
Besides, due to the winding of $\vphi^q(y)$,
the asymptotic value $\nb_0(x=-\i|\vphi^q(y))=\nb^*(\vphi^q(y))$
would differ from that $\nb^q(x=-\i,y)=\nb^*(\vphi_0)$ of the bulk ground state.
Qualitatively, the charge-$q$ configuration $\nb^q(\rb)$ must have the form shown in \figr{n01-<<+} for $q=1$.

Nonetheless, for a given $\vphi^q(y)$, the energy is
still minimized well by the configuration $\nb_0(x|\vphi^q(y))$ in the domain-wall region.
It is in the bulk region, where the configuration needs to be modified.

The proximity to the phase transition allows one to efficiently separate the domain-wall and bulk contributions as follows.
The asymptotic bulk order $\nb^\i=\nb^*(\vphi_0)$ deviates only a little from $\nb_z$:
from \eq{nz*}, the optimal angle in the bulk ground state is given by
\[
    \tht^{*2} \approx 2|\de\hr_z| \ll 1.
\]
When the isospin $\nb(\rb)$ is close to $\nb_z$, such that its angle $\tht(\rb)\lesssim \tht^*$,
the energetics is governed by this smaller scale $|\de h_z|\ll u$
and the associated spatial scale
\[
    l_{\de h_z}\equiv \q{\f{\rho}{|\de h_z|}}=\f{l_u}{\q{|\de\hr_z|}}\gg l_u
\]
is much larger than the domain-wall width $l_u$ [see also \eq{l-<<+}].

We choose a length scale $x_0$ such that
\[
    l_{\de h_z}\gg x_0 \gg l_u.
\]
Since $x_0\gg l_u$, the ground state configuration $\nb_0(x=-x_0|\vphi_0)$ at $x=-x_0$
is already close to its bulk asymptotic value $\tht_0(x=-x_0)\approx \tht^*$.
We emphasize that even exactly at the phase transition $h_z=u$,
the domain wall width is $l_u$, only the bulk value is approached as a power law and not exponentially, see \eq{tht0+}.
So, for the sought charge-$q$ configuration $\nb^q(\rb)$,
we consider the above ground state configuration with the adiabatically changing angle $\vphi^q(y)$
only in the region up to this distance from the edge:
\[
    \nb^q(\rb)\approx \nb_0(x|\vphi^q(y)), \spc -x_0<x\leq0.
\]
Due to the other condition $l_{\de h_z}\gg x_0$,
the contribution to the excitation energy from the region $-x_0<x<0$
is not extensive and can be approximated as
\beqar
    \de\Ef[\nb^q]_{(-x_0,0)}
        &=&\int_{-x_0<x<0}\f{\dx^2\rb}s\,(E[\nb^q]-E[\nb_0]) \nn\\
        &\approx& \# \eps_\odot l_u \int_{-\i}^{+\i}\dx y\,[\n_y\vphi(y)]^2.
\lbl{eq:dE-<}
\eeqar
Here and below, $\#$ indicates undetermined numerical factors
that are beyond the accuracy of the considered approximation.

In the remaining ``bulk'' region $x<-x_0$, the configuration $\nb^q(\rb)$
must connect the boundary values $\nb^q(x=-\i,y)=\nb^*(\vphi_0)$ and $\nb^q(x=-x_0,y)=\nb_0(x|\vphi^q(y))\approx\nb^*(\vphi^q(y))$.
Since at both boundaries
the isospin is close to $\nb_z$,
the isospin $\nb^q(\rb)$ is close to $\nb_z$ within the whole region,
and therefore, according to the above, varies over the spatial scales on the order of $l_{\de h_z}$ or greater.

In fact, the bulk region $x<-x_0$ ``traps a vortex''
of charge $q$ (not to be confused with the skyrmion charge):
when going along its rectangular boundary, the phase $\vphi^q(x,y)$ winds the circle $q$ times (all at the $x=-x_0$ boundary),
while the angle $\tht\approx\tht^*$ remains almost constant.
The leading contribution to the energy of such a vortex configuration
comes from the region outside of its core -- the region
where the isospin covers the solid angle $\tht\lesssim\tht^*$.
This contribution is logarithmic; to obtain it, one may consider the radial form
$\tht(r,\phi)=\tht^*$, $\vphi(r,\phi)=q\phi$, where $\rb=r(\cos\phi,\sin\phi)$,
relative to the ``center'' of the vortex in the bulk region $x<-x_0$,
the point at which $\nb_0^q(\rb)=\nb_z$ in \figr{n01-<<+}.
This gives
\beqar
    \de\Ef[\nb^q]_{(-\i,-x_0)}
    &=&\int_{x<-x_0}\f{\dx^2\rb}s\,(E[\nb^q]-E[\nb_0])\nn\\
    &\approx& \int \f{2\pi r\dx r}{s} \f{\rho}{2} \sin^2\tht (\n\vphi)^2\nn\\
    &\approx&
    \int_{|q|l_{\de h_z}}^l \f{2\pi r\dx r}{s} \f{\rho}{2}\tht^{*2}\f{q^2}{r^2}\nn\\
    &=&\f{2\pi\rho}{s} |\de\hr_z| q^2 \x{ln}\f{l}{|q| l_{\de h_z}}.
\lbl{eq:dE<-}
\eeqar
The lower limit is determined by the size $|q| l_{\de h_z}$ of the vortex core.
In our case, the upper limit $l\gg|q|l_{\de h_z}$
is set by the distance from the vortex core to the edge.
This same scale $l$ has to match the extent of $\vphi^q(y)$ in the domain wall in the $y$ direction
(the size of the ``winding region'').
Estimating $\n_y\vphi^q(y)\sim q/l$
and adding the domain-wall (\ref{eq:dE-<}) and bulk (\ref{eq:dE<-})
contributions, for the excitation energy of the so-constructed configuration
one obtains
\beqar
    \de\Ef[\nb^q]&=&\de\Ef[\nb^q]_{(-\i,-x_0)}+\de\Ef[\nb^q]_{(-x_0,0)} \nn\\
        &\approx&
        \eps_\odot q^2 \lt( 2\pi|\de\hr_z| \x{ln} \f{l}{|q|l_{|\de h_z|}}+ \#\f{l_u}{l} \rt).
\lbl{eq:dEest}
\eeqar
The dimension $l$ is the only remaining variational parameter.
Minimization of this energy with respect to $L$
yields the leading terms of the asymptotics expansion
\beq
    \Der^q(\hr_z\ra1-0)
    =\pi q^2|\de\hr_z|
     \x{ln}\f{C}{q^2|\de\hr_z|}+o(|\de\hr_z|)
\lbl{eq:Deq@+}
\eeq
for the gap of charge edge excitations in the intermediate phase $\nb^\i=\nb^*(\vphi_0)$
close to the phase transition $h_z=u$.
The minimum (\ref{eq:Deq@+}) of \eq{dEest} is reached at the optimal length
\[
    l^*\equiv\f{l_u}{|\de\hr_z|}=\f{l_{\de h_z}}{\q{|\de\hr_z|}}.
\]
The numerical factor $C\sim 1$ cannot be determined within the accuracy of the considered logarithmic approximation.
For unit charge $q=\pm1$, the estimate (\ref{eq:Deq@+}) agrees with that of Ref.~\ocite{F2}.

\subsection{Numerical calculations\lbl{sec:num}}

\begin{figure}
\includegraphics[width=.40\textwidth]{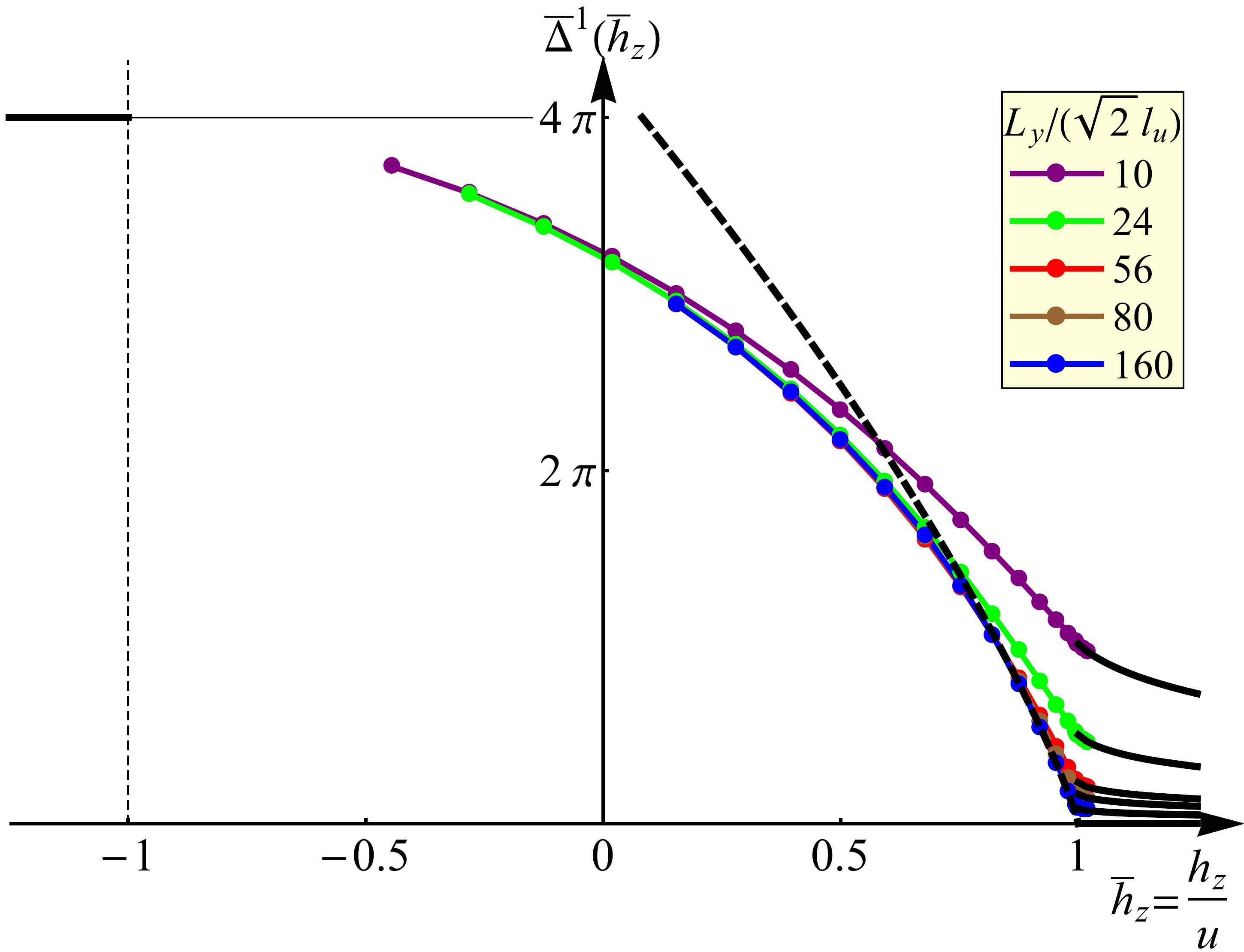}\\
\includegraphics[width=.40\textwidth]{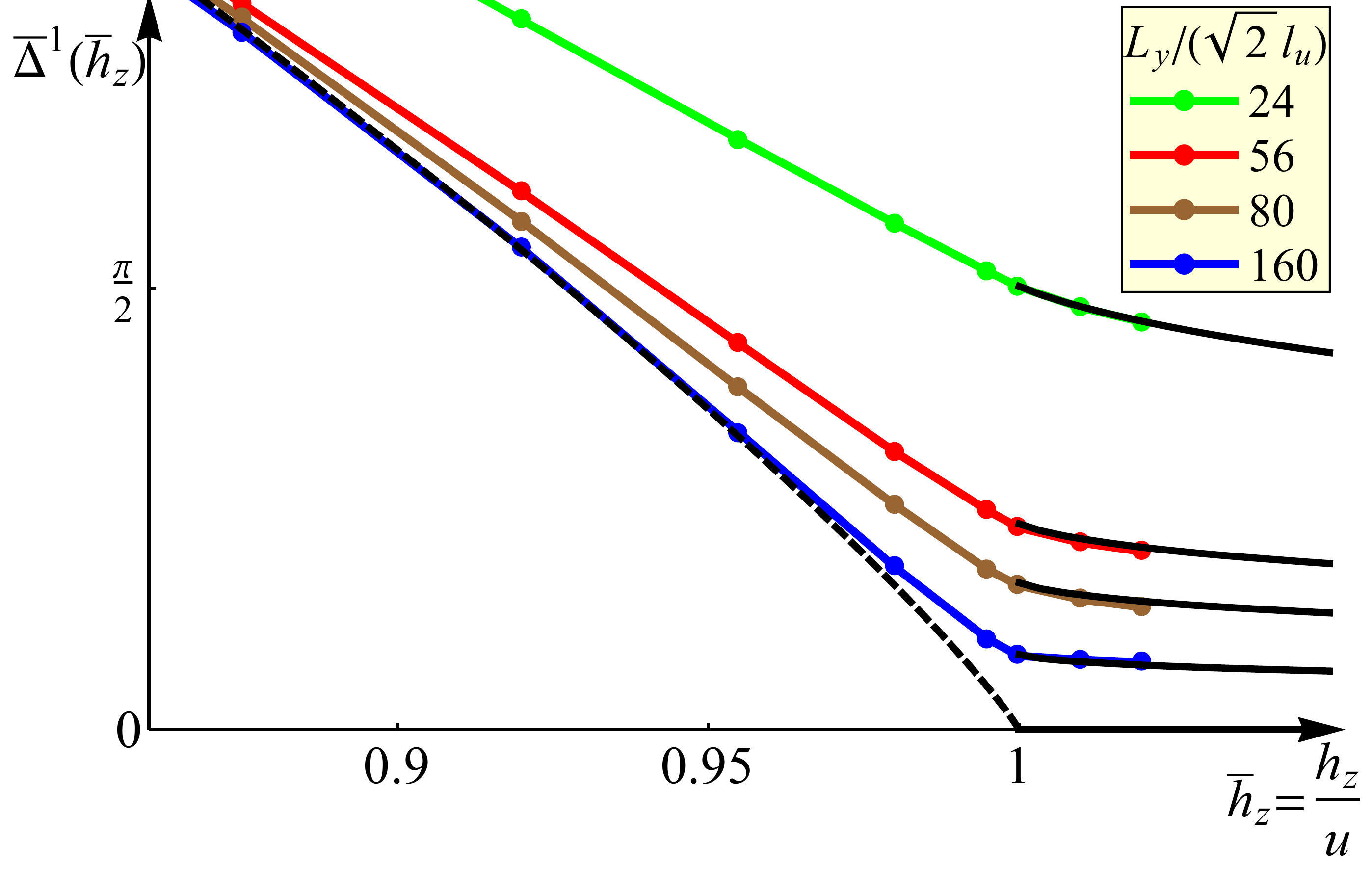}\\
\caption{(Color online)
(top) The dependence of the gap $\Der^1(\hr_z)$ of the edge charge excitations $\nb_0^1(\rb)$ on the normalized field $\hr_z$,
calculated numerically for various sample sizes.
The black solid lines in the $1<\hr_z$ region are the exact gap dependencies \eqn{Deq+<} for a finite-size system.
The black horizontal solid line in the $-1<\hr_z$ region is the analytic value \eqn{Deqsk} of the gap,
given by the energy of a free skyrmion. The black dashed line in the $-1<\hr_z<1$ region
is the asymptotic gap dependence \eqn{Deq@+} with the fitted parameter $\x{ln}\,C=4.27$.
(bottom) Zoomed-in region around the phase transition point $\hr_z=1$.
}
\lbl{fig:De}
\end{figure}

\begin{figure}
\includegraphics[width=.48\textwidth]{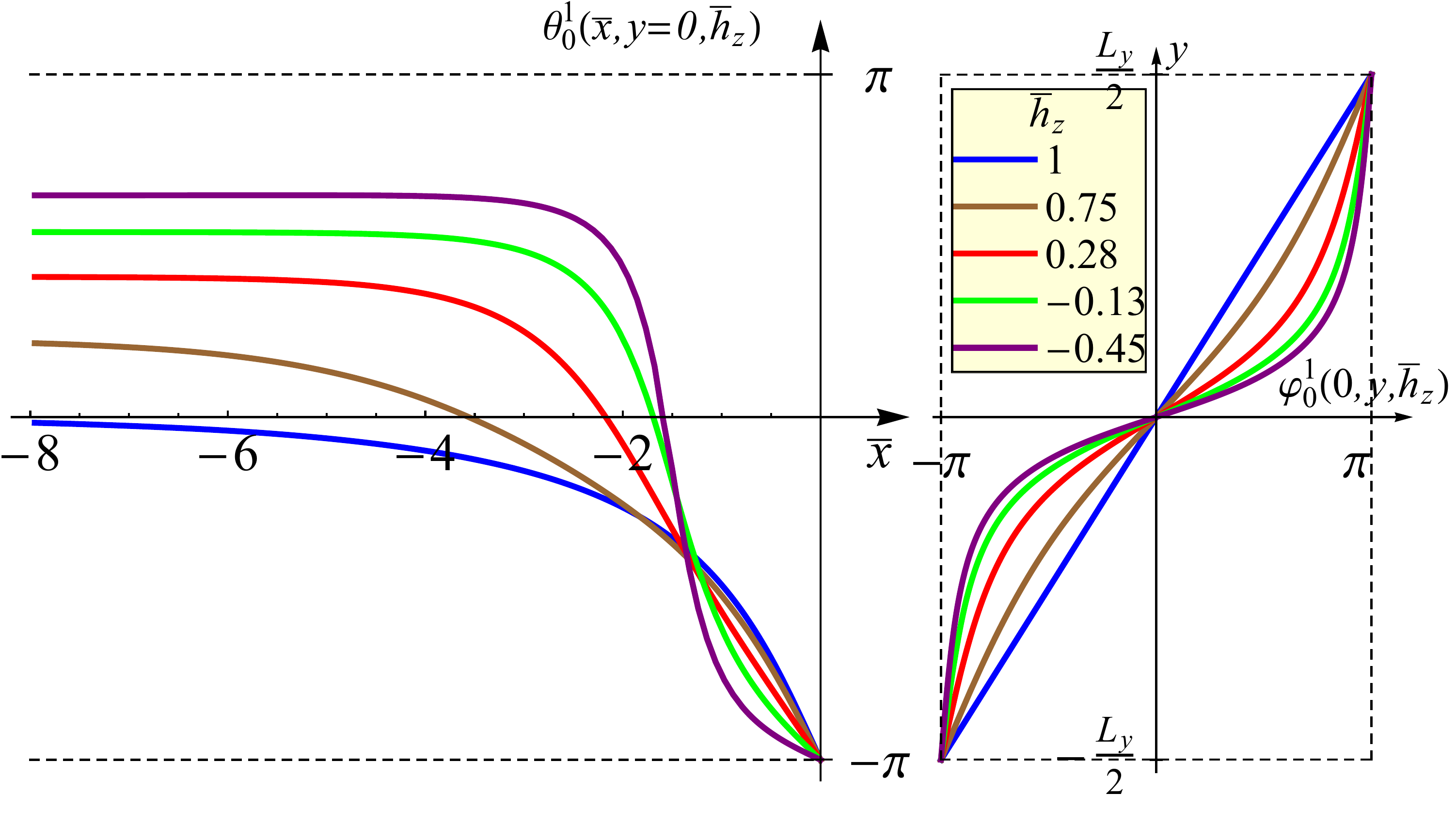}
\caption{(Color online)
The angle functions $\tht_0^1(x,y=0)$ and $\vphi_0^1(x=-0,y)$
of the unit-charge edge configurations $\nb_0^1(\rb)$ in the intermediate phase $\nb^\i=\nb^*(\vphi_0)$
at $-u<h_z<u$, obtained numerically for $L_x=L_y=10\q{2}l_u$ sample for various values of $h_z$.
The size of the configuration decreases with decreasing $h_z$.
}
\lbl{fig:tht01phi01-<<+}
\end{figure}

We also confirm the above-presented behavior by calculating
the unit-charge $q=1$ edge excitations numerically.
The configurations $\nb_0^1(\rb)$ that deliver the energy minimum (\ref{eq:En0q})
within the $q=1$ sector are found by solving the discretized version of the stationary-point \eqs{dEtht}{dEphi}.
We solve them using a variant of the multi-grid relaxation methods for boundary value problems~\cite{NR}.
The calculations are performed for a finite-size system $(x,y)\in(-L_x,0)\tm(-\f{L_y}2,+\f{L_y}2)$
with dimensions $L_x$ and $L_y$.
For all sizes $L_y$ indicated in \figr{De}, except the largest one $(L_x,L_y)=(120,160)\q2 l_u$,
we considered square samples with $L_x=L_y$.
The gap $\Der^1(\hr_z)$ is calculated using a discretized version of \eq{Deqdef}.
The plots for the gap $\Der^1(\hr_z)$ as a function of the normalized field $\hr_z$
for several different system dimensions $L_{x,y}/l_u$ are presented in \figr{De}.

In the region $u<h_z$ of the $\nb^\i=\nb_z$  phase,
the numerically calculated gap $\Der^1(\hr_z)$
accurately agrees with the exact analytic dependence (\ref{eq:Deq+<}) on $\hr_z$ and $L_y$,
thus confirming that the edge charge excitations are gapless in the infinite-size limit $L_y\rarr\i$.
The typical edge charge configuration $\nb_0^1(\rb)$ in the $\nb^\i=\nb_z$ phase
has the form shown \figr{n01+<}, in full agreement with the analytical expressions \eqssn{phi0q+<}{tht0q+<}{n0q+<}.

In the region $-u<h_z<u$ of the intermediate phase $\nb^\i=\nb^*(\vphi_0)$,
the gap becomes independent of the sample dimensions $L_{x,y}$,
as they become larger than the size of the configuration.
Close to the transition point $h_z=u$ in the intermediate phase,
for large enough $L_{x,y}/l_u$, the numerical data points fit to the analytical estimate (\ref{eq:Deq@+}):
for the largest-size sample $(L_x,L_y)=(120,160)\q2 l_u$ (with the smallest size effects),
fitting to the data points $(\hr_z,\Der^1(\hr_z))=(0.955,1.058),(0.920,1.715),(0.875,2.484)$,
we obtain $\x{ln}\,C=4.27$. 

The available numerical data for $\Der^1(\hr_z)$ in the intermediate phase range $-u<h_z<u$
visually extrapolate well to the value $\Der^1(\hr_z=-1)=4\pi$ [\eqs{Deq<-}{Deqsk}]
of the bulk skyrmion at the phase transition point $h_z=-u$.

The typical edge charge configuration $\nb_0^1(\rb)$ in the $\nb^\i=\nb^*(\vphi_0)$ phase is shown in \figr{n01-<<+}.
In \figr{tht01phi01-<<+}, the angle functions $\tht_0^1(x,0)$ and $\vphi_0^1(x=-0,y)$ of $\nb_0^1(\rb)$ are plotted for various $\hr_z$.
The regions where these functions vary determine the size of the charge excitation.
As $h_z$ decreases in the range $-u<h_z<u$, this size monotonically decreases, becoming smaller than $l_u$.

These behaviors of the gap and size of the excitation with decreasing $h_z$ in the $-u<h_z<u$ region
are in accord with the general arguments of \secr{edgeexcgapped}
that in the $\nb^\i=-\nb_z$ phase at $h_z<-u$,
the charge excitations are skyrmions with zero size
[for the considered model with neglected Coulomb self-interaction of the charge density \eqn{kappa}].

\subsection{Summary}

To summarize, in this section, we studied the edge charge excitations.
We found that the properties of the charge excitations of the phases $\nb^\i=\pm\nb_z$ with preserved $\Ux(1)$ symmetry
remain qualitatively the same in the presence of strong interactions:
the TnT phase $\nb^\i=\nb_z$ has gapless edge excitations
and the TT phase $\nb^\i=-\nb_z$ has gapped edge excitations.
However, collective charge excitations of the interacting system are microscopically quite different
from the single-electron excitations of the noninteracting system.

In the intermediate phase $\nb^\i=\nb^*(\vphi_0)$ with spontaneously broken $\Ux(1)$ symmetry
the edge charge excitations are gapped. This suggests
that the $\Ux(1)$ symmetry is responsible for the topological protection in this strongly interacting system.
This important point will be further substantiated
in the next \secr{lliq}, where we study the low-energy edge dynamics of the TnT phase $\nb^\i=\nb_z$,
including the effects of (non-spontaneous) $\Ux(1)$ symmetry breaking, \secsr{U1nH}{U1nEc}.

We point out here that, since
rigorous mathematical definitions of topological phases in interacting systems
are currently an actively researched subject,
in this article, we adopt an intuitive nomenclature, whereby
we refer to the phases with gapless and gapped edge excitations
as TnT and TT phases, respectively.

We also point out that while the properties of the edge excitations of the three phases are different,
their bulk charge excitations are qualitatively the same:
the bulk charge gap is finite in all three phases
and never closes during the transformation
from the TnT $\nb^\i=\nb_z$ to the TT $\nb^\i=-\nb_z$ phase with increasing $h_z$.
This is another qualitative distinction from the single-particle noninteracting picture,
where the topological phase transition is associated with the closing of the bulk gap.

\section{Helical Luttinger liquid\lbl{sec:lliq}}

\subsection{Derivation\lbl{sec:lliqd}}

In the previous \secr{edgeexc}, it was demonstrated that the $\nb^\i=\nb_z$ phase
at $u<h_z$ is characterized by gapless charge edge excitations.
In this section, we derive the effective low-energy theory
describing the dynamics of these edge excitations.
The criterion for the applicability of such a low-energy theory is quite clear at the physical level:
the nondegenerate bulk ground state $\nb^\i=\nb_z$ has a gapped excitation spectrum
and the theory is valid at energies below this gap $h_z-u$, i.e, as long as the bulk is not excited.
This criterion will be established more rigorously below.

The gapless edge excitations originate from the degeneracy of the (mean-field) ground state solution $\nb_0(x;\vphi_0)$
at the edge, characterized by an arbitrary angle $\vphi_0$.
Essentially, now we would like to include the slow variations of the angle $\vphi_0$ in space and time
and perform a gradient expansion.
Since some terms in the original Lagrangian $\Lf[\tht,\vphi]$ [Eqs.~(\ref{eq:L})-(\ref{eq:Ec})]
couple the $\vphi(\rb;t)$ and $\tht(\rb;t)$ variables, $\tht(\rb;t)$ cannot just be considered as static
and replaced by the ground state configuration $\tht_0(x)$.
However, the deviations from the ground state configurations due to slowly varying $\vphi(\rb;t)$ will be small,
and so the Lagrangian $\Lf[\tht,\vphi]$ may be expanded in deviations $\de\tht(\rb;t)$ about $\tht_0(x)$,
\beq
    \tht(\rb;t)=\tht_0(x)+\de\tht(\rb;t).
\lbl{eq:dtht}
\eeq
The deviation must satisfy the boundary condition
\beq
    \de\tht(x=0,y;t)=0.
\lbl{eq:dthtbc}
\eeq

For now, we assume a general, but slow, dependence of $\vphi(\rb,t)$ on $\rb$ and $t$; further approximations to follow.

To the leading order in gradients of $\vphi(\rb;t)$, it is sufficient to expand different terms to the lowest necessary order in $\de\tht(\rb;t)$.
This way, for the kinetic term (\ref{eq:K}), we have
\bwt
\beq
    K[\de\tht,\vphi]=\f{\dot{\vphi}}2[\cos\tht_0+\de\tht\pd_\tht\cos\tht_0+\Oc(\de\tht^2)]
        \rarr\f{\dot{\vphi}}2\de\tht\pd_\tht\cos\tht_0.
\lbl{eq:Kapprox}
\eeq
The term $\f{\dot{\vphi}}2\cos\tht_0$ is an inconsequential full time derivative and may be dropped.
For the reason to be provided below, we keep the derivative $\pd_\tht\cos\tht_0=-\sin\tht_0$ as is,
without explicitly differentiating it.

In the gradient term
\beq
    \f{\rho}2\sin^2\tht(\n\vphi)^2=\f{\rho}2[\sin^2\tht_0(x)+\Oc(\de\tht)](\n\vphi)^2
         \rarr\f{\rho}2 \sin^2\tht_0(x)(\n\vphi)^2
\lbl{eq:gradvphiapprox}
\eeq
in \eq{E}, keeping only the zero-order term is sufficient.

The remaining terms
\beq
    \int\f{\dx^2\rb}{s}
        \lt\{\f{\rho}2(\n\tht)^2+\Ec(\tht)\rt\}
    =\int\f{\dx^2\rb}{s}\lt\{E[\nb_0]+\f12\de\tht\Uh[\de\tht]+\Oc(\de\tht^3)\rt\}
    \rarr\int\f{\dx^2\rb}{s}\f12\de\tht\Uh[\de\tht]
\lbl{eq:Ethtapprox}
\eeq
\ewt
in \eq{E} do not depend on $\vphi(\rb;t)$ and need to be expanded to quadratic order
(the zero-order ground state energy $\Ef[\nb_0]$ may be dropped and
the linear order in $\de\tht(\rb;t)$ vanishes since $\tht_0(x)$ minimizes exactly this functional).
Here,
\beq
    \Uh=-\rho\n_y^2+\Uh_x,
\lbl{eq:U}
\eeq
\beq
    \Uh_x=-\rho\n_x^2+\pd_\tht^2\Ec(\tht_0(x)),
\lbl{eq:Ux}
\eeq
\beq
    \pd_\tht^2\Ec(\tht)
    =-u\cos2\tht+h_z\cos\tht.
\lbl{eq:D2Ec}
\eeq

Collecting these leading terms [\eqss{Kapprox}{gradvphiapprox}{Ethtapprox}],
we approximate the initial Lagrangian as
\[
    \Lf[\tht,\vphi]\rarr\Lf'[\de\tht,\vphi]
    =\int\f{\dx^2\rb}{s}\, L'[\de\tht,\vphi],
\]
\beq    -L'[\de\tht,\vphi]
        \equiv-\f{\dot{\vphi}}2 \de\tht \pd_\tht\!\cos\tht_0
        +\f12\de\tht\Uh[\de\tht]
        +\f{\rho}{2}\sin^2\!\tht_0(\n\vphi)^2.
\lbl{eq:L'}
\eeq

The structure of the Lagrangian (\ref{eq:L'}) allows for further approximations.

First, we observe that in both terms containing the $\vphi(\rb;t)$ variable, the function $\sin\tht_0(x)$ is present,
which constrains them to the domain-wall region of size $l_u$.
We split the field
\[
    \vphi(x,y;t)=\Phi(y;t)+\de_x\vphi(x,y;t)
\]
into an $x$-independent average
\beq
    \Phi(y;t)\equiv\ln \vphi(x,y;t)\rn_x
\lbl{eq:Phidef}
\eeq
along $x$ and fluctuations $\de_x\vphi(x,y;t)$, which are, by this construction,
zero on average,
\[
    \ln \de_x\vphi(x,y;t)\rn_x=0.
\]
There is certain freedom in the definition of this average.
Since the parametrization by the spherical angles becomes degenerate at $\tht=0,\pi$,
a meaningful average requires a weight function that takes that into account.
The most reasonable weight function seems to be $\sin\tht_0(x)$,
and so we define the average as
\[
    \ln f(x)\rn_x\equiv\int_{-\i}^0\dx x\, \sin\tht_0(x) f(x).
\]
Although other similar choices could also be used,
to the leading order, when $\de_x\vphi$ is just neglected, the exact definition of the average is not essential.

Since at low energies $\vphi(x,y;t)$ varies over spatial scales exceeding the domain wall size $l_u$,
the fluctuations of the field $\vphi(x,y;t)$ in the $x$ direction across the domain wall,
described by $\de_x\vphi(x,y;t)$,
will produce a parametrically smaller contribution than those in the $y$ direction along the domain wall, described by $\Phi(y;t)$.
Thus, to the leading order, $\de_x\vphi(x,y;t)$ may be neglected and the field
\beq
    \vphi(x,y;t)\rarr\Phi(y;t)
\lbl{eq:vphiapprox}
\eeq
may be approximated by a quasi-1D field $\Phi(y;t)$.
After this approximation, the gradient term becomes $(\n\vphi)^2\rarr(\n_y\Phi)^2$.

\begin{figure}
\includegraphics[width=.35\textwidth]{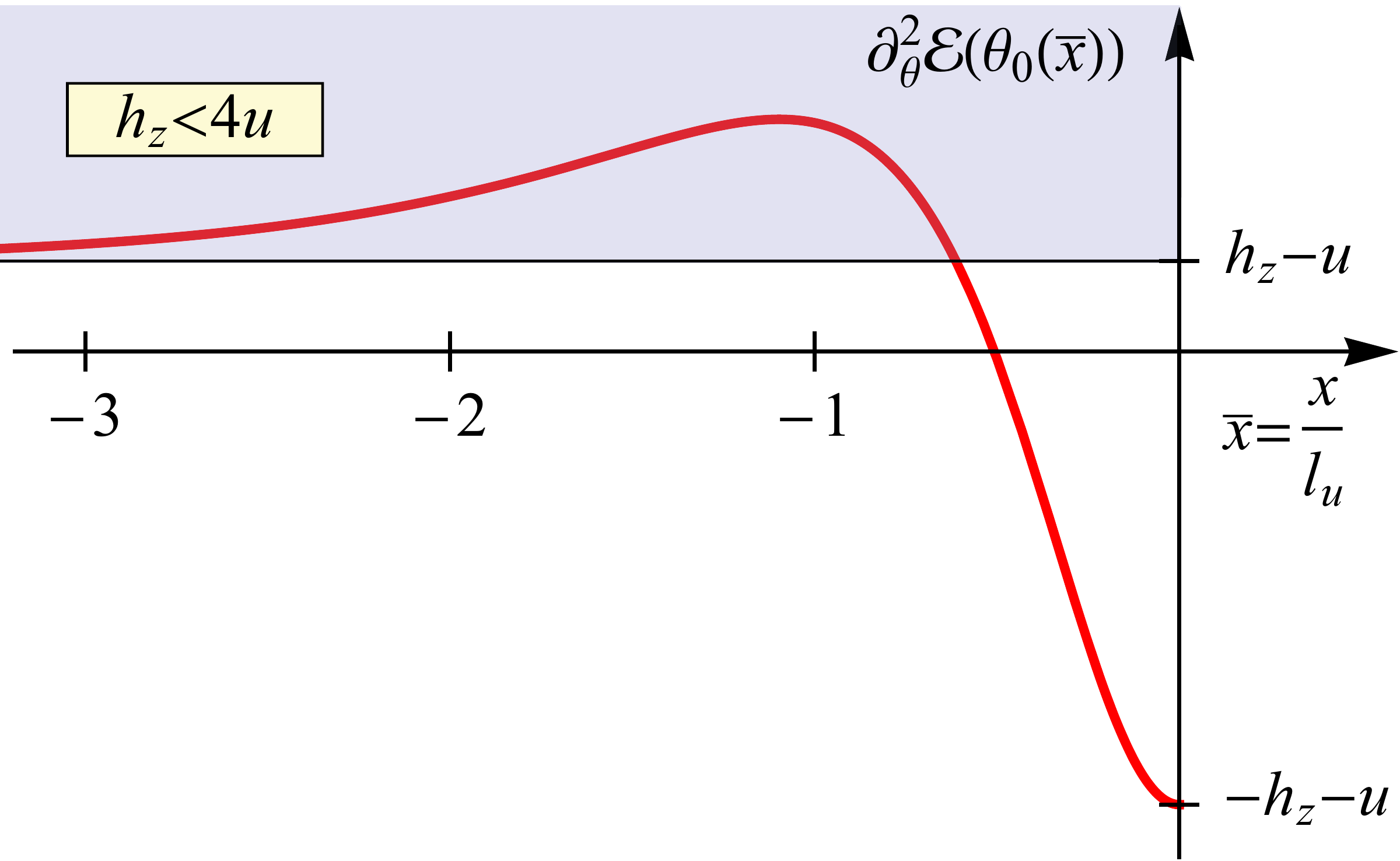}
\includegraphics[width=.35\textwidth]{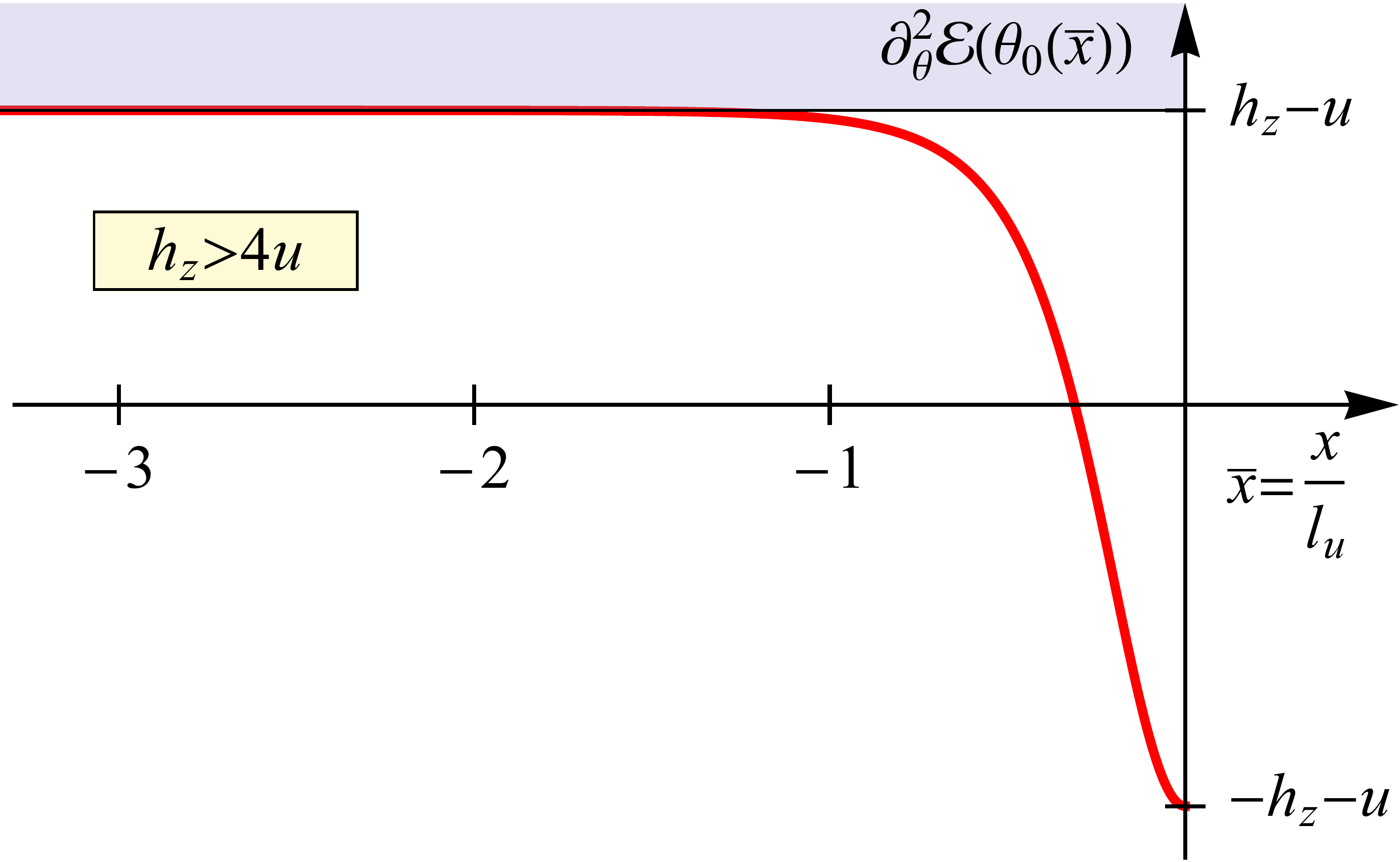}
\caption{The function $\pd_\tht^2\Ec(\tht_0(x))$ [\eq{D2Ec}],
which serves as an effective potential energy in the operator $\Uh_x$ [\eq{Ux}],
which describes the quadratic fluctuations $\de\tht$ [\eq{dtht}] about the ground state configuration $\tht_0(x)$ [\eq{tht0+<}]
in the $\nb^\i=\nb_z$ phase at $u<h_z$.
The operator $\Uh_x$ has only a continuous spectrum, marked by a shaded region,
that starts at the bulk asymptotic value $\pd_\tht^2\Ec(\tht_0(x=-\i))=h_z-u$, equal to the gap of the bulk isospin wave excitations,
and there are no discrete bound states.
}
\lbl{fig:Ux}
\end{figure}

Next, we analyze the properties of the operator $\Uh_x$ [\eq{Ux}].
The operator $\Uh_x$ has the form of the Hamiltonian for a Schr\"odinger particle in 1D
with the potential energy $\pd_\tht^2\Ec(\tht_0(x))$ [\eq{D2Ec}], plotted in Fig~\ref{fig:Ux},
and the ``hard wall'' boundary condition (\ref{eq:dthtbc}).
In the bulk,
\[
    \pd_\tht^2\Ec(\tht_0(-\i)=0)=h_z-u>0,
\]
which represents the mass of the ``isospin wave'' in an infinite system.
The minimum
\[
    \pd_\tht^2\Ec(\tht_0(0)=\pi)=-h_z-u<0
\]
is reached at the edge and is negative.
The ``potential energy'' $\pd_\tht^2\Ec(\tht_0(x))$ curve has a ``well'' in the region of the domain wall;
for $h_z<4u$, it also has a ``barrier''.
From this analogy with the quantum-mechanical problem,
we conclude that the operator $\Uh_x$ has a continuous eigenvalue spectrum in the energy range $(h_z-u,+\i)$,
but it {\em could} also have discrete levels in the range $(-h_z-u,h_z-u)$, which would correspond to
the ``states'' bound within the well. However, we have checked numerically that there are no such discrete eigenvalues.
In fact, only the positive range $(0,h_z-u)$ requires a check for bound states, since negative eigenvalues
are prohibited  because $\Uh_x$ is a quadratic form of the expansion about the minimum-energy configuration
and is therefore a positive-definite operator.

Thus, $\Uh_x$ has only a continuous ``massive'' eigenvalue spectrum that starts from $h_z-u>0$.
The eigenfunctions are extended and can basically be viewed as bulk isospin waves somewhat modified at the edge.

Due to these properties, the operator $-\rho\n_y^2$ in $\Uh$ [\eq{U}]
may be neglected compared to $\Uh_x$,
\beq
    \Uh=-\rho\n_y^2+\Uh_x\rarr\Uh_x.
\lbl{eq:Uapprox}
\eeq

The approximations (\ref{eq:vphiapprox}) and (\ref{eq:Uapprox})
allow us to further simplify the Lagrangian (\ref{eq:L'}) as
\[
    \Lf'[\de\tht,\vphi]
    \rarr
    \Lf''[\de\tht,\Phi]
    =\int\f{\dx^2\rb}{s}\, L''[\de\tht,\Phi],
\]
\beq
    -L''[\de\tht,\Phi]
    \equiv-\f{\dot{\Phi}}2\de\tht\pd_\tht\!\cos\tht_0+\f12\de\tht\Uh_x[\de\tht]
    +\f{\rho}2\sin^2\!\tht_0 (\n_y\Phi)^2.
\lbl{eq:L''}
\eeq

The Lagrangian (\ref{eq:L''}) is a second-order functional polynomial in $\de\tht(\rb;t)$.
Consider the configuration $\de\tht_0[\Phi](\rb;t)$ that delivers the minimum of $-\Lf''[\de\tht,\Phi]$
with respect to $\de\tht(\rb;t)$ for a given $\Phi(y;t)$.
It satisfies the stationary-point equation
\beq
    -\f{\de}{\de(\de\tht)}
    \int\dx t\,\Lf''[\de\tht,\Phi]
    =-\f{\dot{\Phi}}2\pd_\tht\!\cos\tht_0+\Uh_x[\de\tht]=0.
\lbl{eq:dtht0eq}
\eeq
This is  a differential-in-$x$ equation, and its solution can thus be formally written as
\beq
    \de\tht_0[\Phi](\rb;t)
    =\pd\tht_0(x) \f{\dot{\Phi}(y;t)}2,
\lbl{eq:dtht0}
\eeq
where
\beq
    \pd\tht_0(x)
    =\Uh^{-1}_x[\pd_\tht\!\cos\tht_0].
\lbl{eq:pdtht0}
\eeq
The reason for the notation $\pd\tht_0(x)$ will become clear shortly.

In terms of this minimum configuration and the deviation
\[
    \de\tilde{\tht}(\rb;t)=\de\tht(\rb;t)-\de\tht_0[\Phi](\rb;t)
\]
from it,
the Lagrangian (\ref{eq:L''}) can be rewritten identically as
\bwt
\beq
   -L''[\de\tht_0[\Phi]+\de\tilde{\tht},\Phi]
   =-\f12\lt(\f{\dot{\Phi}}2\rt)^2 \pd\tht_0(x) \pd_\tht\!\cos\tht_0(x)
   +\f{\rho}2\sin^2\tht_0(x)(\n_y\Phi)^2+\f12\de\tilde{\tht}\Uh_x[\de\tilde{\tht}].
\lbl{eq:L''2}
\eeq
\ewt
This procedure essentially amounts to completing the square
of the quadratic polynomial in the functional sense.
This way, decoupling of the field $\Phi(y;t)$ and the variables (``free modes'') $\de\tilde{\tht}(\rb;t)$
of the operator $\Uh_x$ has been achieved.
Since the latter have a gap $h_z-u$,
their contribution may be neglected below this energy scale.
In the end, this amounts to replacing $-\Lf''[\de\tht,\Phi]$ in \eq{L''} by its minimum with respect to $\de\tht(\rb;t)$.
This minimum is given by the first two terms in \eq{L''2} and represents the sought effective Lagrangian:
\[
    \Lf^\x{1D}[\Phi]\equiv\Lf''[\de\tht_0[\Phi],\Phi].
\]

The $y$-dependent 1D field $\Phi(y;t)$ is the only remaining variable,
while the rest is fixed functions of $x$.
This allows us to separate them and write the final Lagrangian for the edge excitations of the $\nb^\i=\nb_z$ phase in the form:
\beq
    \Lf^\x{1D}[\Phi]=\int \dx y\, L^\x{1D}[\Phi],\spc L^\x{1D}[\Phi]=\f1{2\pi\Kc}\lt[\f1v \dot{\Phi}^2-v(\n_y\Phi)^2\rt].
\lbl{eq:L1D}
\eeq
We recognize in \eq{L1D} the Lagrangian of a Luttinger liquid, with the phase field $\Phi(y;t)$
at the edge being the collective bosonic variable.
The parameters $v$ and $\Kc$ are given by
\bwt
\beq
    \f{1}{2\pi\Kc v}
    \equiv \f{l_u}{s u}F_t(\hr_z)
,\spc
    F_t(\hr_z)
    \equiv \f{u}8\int^0_{-\i}\dx\xr\, \pd_\tht\!\cos\tht_0(x) \pd\tht_0(x)
    =\f{u}8\int^0_{-\i} \dx\xr\,\pd_\tht\!\cos\tht_0\Uh^{-1}_x[\pd_\tht\!\cos\tht_0],
\lbl{eq:Ft}
\eeq
\beq
    \f{v}{2\pi\Kc}
    \equiv \f{\rho l_u}{s} F_y(\hr_z)
,\spc
    F_y(\hr_z)
    =\f12 \int^0_{-\i}\dx\xr\sin^2\!\tht_0(\xr;\hr_z).
\lbl{eq:Fy}
\eeq
\ewt
Due to the scaling form $\tht_0(x)=\tht_0(\xr;\hr_z)$ of the ground state solution (\ref{eq:tht0+<}),
the parameters can be expressed in terms of the dimensionless functions
$F_{t,y}(\hr_z)$ of the normalized field $\hr_z=h_z/u$.

We recognize that $F_y(\hr_z)=F_2(\hr_z)$ [\eq{F2}],
which is not surprising, since this quantity has already appeared in an essentially identical calculation of \eq{Deq+<exp1}
in \secr{edgeexcgapless}. On the other hand, without additional insights, calculating $F_t(\hr_z)$
would require first finding the solution $\pd\tht_0(x)$ [\eq{pdtht0}] to the differential equation (\ref{eq:dtht0eq})
and then calculating the integral in \eq{Ft}.
Below we provide a more elegant and streamlined way of deriving the low-energy model \eqn{L1D},
which not only allows us to obtain the explicit expression for $F_t(\hr_z)$,
but also uncovers the origin of the low-energy model in the degenerate ground state solution.
The above derivation is nonetheless useful for justifying the employed approximations.

We make two crucial observations about the general structure of the Lagrangian \eqn{L}-\eqn{Ec}.

First, we observe that the kinetic term $\f{\dot{\vphi}}2\cos\tht$ [\eq{K}]
has the same structure as the ''Zeeman'' term $-h_z\cos\tht$ in \eqs{E}{Ec},
in which $\f{\dot{\vphi}}2$ plays the role of an additional (in general, time- and coordinate-dependent) ''Zeeman'' field.

Second, as already noticed in \secr{edgeexcgapless},
we observe that the gradient term $\f{\rho}2\sin^2\tht(\n\vphi)^2$ has
the form of the anisotropy $\f{u}2(n_z^2-1)$,
in which $-\rho(\n\vphi)^2$ plays the role of an additional anisotropy energy.

These two observations allow us the rewrite the Lagrangian density \eqn{L} identically as [\eqs{dE1D}{dEx}]
\bwt
\[
     -L[\tht,\vphi;u,h_z]
     =E[\tht,\vphi;u,h_z+\f{\dot{\vphi}}2]
     =dE_x[\tht;u-\rho(\n\vphi)^2,h_z+\f{\dot{\vphi}}2]+\f{\rho}2(\n_y\tht)^2+\Ec^\i(u,h_z)
\]

After the two key approximations,
(i) considering only-$y$-dependent configurations $\vphi(\rb;t)\rarr\Phi(y;t)$ [\eq{vphiapprox}]
and (ii) neglecting the gradient terms $\n_y\tht$ [\eq{Uapprox}],
the Lagrangian density per unit length in the $y$ direction
\beq
     -\int\f{\dx x}s\, L[\tht,\vphi;u,h_z]
     \rarr
     dE^\x{1D}[\tht;u-\rho(\n_y\Phi)^2,h_z+\f{\dot{\Phi}}2]+E^{\x{1D}\i}(u,h_z)
\lbl{eq:L'''}
\eeq
becomes equivalent to the functional \eqn{E1D} for the ground state
with the modified parameters. The functional \eqn{L'''} is minimized by the modified ground state configuration
$\tht_0(x;u-\rho(\n_y\Phi)^2,h_z+\f{\dot{\Phi}}2)$ [\eq{tht0+<}]
and the minimum
\beq
    -\Lt^\x{1D}[\Phi]=dE^\x{1D}_0\lt(u-\rho(\n_y\Phi)^2,h_z+\f{\dot{\Phi}}2\rt)-dE^\x{1D}(u,h_z)
\lbl{eq:L1Dnlin}
\eeq
is expressed in terms of the domain wall energy \eqn{Edw}.
Its expansion up to the quadratic order in the derivatives yields the form
\beq
    -\Lt^\x{1D}[\Phi]
    \approx-\pd_u dE^\x{1D}_0(u,h_z)\rho(\n_y\Phi)^2
        +\pd_{h_z} dE^\x{1D}_0(u,h_z) \f{\dot{\Phi}}2
        +\f12 \pd_{h_z}^2 dE^\x{1D}_0(u,h_z) \lt(\f{\dot{\Phi}}2\rt)^2
    =-L^\x{1D}[\Phi]+\pd_{h_z} dE^\x{1D}_0(u,h_z) \f{\dot{\Phi}}2
\lbl{eq:L1Dnlinexp}
\eeq
\ewt
of the Luttinger liquid Lagrangian \eqn{L1D}.
As in \eq{Kapprox}, the term linear in $\dot{\Phi}$ is an inconsequential full time derivative
and may be dropped. This allows us to express the parameters
\[
    \f1{2\pi\Kc v}=-\f18\pd_{h_z}^2 dE^\x{1D}_0(u,h_z)
,\spc
    \f{v}{2\pi\Kc}=-\rho\pd_u dE^\x{1D}_0(u,h_z)
\]
of the Luttinger liquid in terms of the derivatives of the domain-wall energy and
thus calculate the coefficients $F_{t,y}$ [\eqs{Ft}{Fy}] explicitly as
\beqar
    F_t(\hr_z)
    &=&
        -\f14\pd_{\hr_z}^2 F(\hr_z)
        =\f18\f1{\hr_z\q{\hr_z-1}}
\lbl{eq:Ftf}
    ,\\
    F_y(\hr_z)
    &=&
        F_2(\hr_z)=
        \hr_z\arcsin\f1{\q{\hr_z}}-\q{\hr_z-1}.
\lbl{eq:Fyf}
\eeqar
The functions are plotted in \figr{F}.
Their asymptotic expressions at the transition point and at large $h_z$ are
\beqar
    F_t(\hr_z)
    &=& \f18\lt\{ \ba{ll} \f1{\q{\hr_z-1}},& h_z\rarr u+0,
    \\ \f1{\hr_z^{\f32}},& h_z\gg u,\ea\rt.
\lbl{eq:Ftfa}
    \\
    F_y(\hr_z)
    &=& \lt\{\ba{ll} \f{\pi}2,& h_z\rarr u+0,\\ \f23\f{1}{\q{\hr_z}},& h_z\gg u.\ea \rt.
\lbl{eq:Fyfa}
\eeqar

Also, according to this approach,
the function $\de\tht_0[\Phi](\rb;t)$ [\eqs{dtht0}{pdtht0}]
can be identified as the linear term of the expansion of $\tht_0(x;u,h_z+\f{\dot{\Phi}}2)$ in $\f{\dot{\Phi}}2$.
Therefore, the function (\ref{eq:pdtht0})
\beq
    \pd\tht_0(x)
    =\pd_{h_z}\tht_0(x;u,h_z)
\lbl{eq:pdhtht0}
\eeq
is the derivative of the ground state solution with respect to $h_z$.
Inserting this form into \eq{Ft} for $F_t(\hr_z)$,
we then observe that the integrand is a full derivative with respect to $h_z$
(which is the reason for not differentiating $\pd_\tht\cos\tht_0$ explicitly),
and the coefficient can alternatively be expressed in terms of $F_1(\hr_z)$ [\eq{F1}] as
\beq
    F_t(\hr_z)
    =\f18 \int_{-\i}^0\dx\xr\, \pd_{\hr_z}\!\cos\tht_0(\xr;\hr_z)
    =-\f14 \pd_{\hr_z} F_1(\hr_z),
\lbl{eq:Ft'}
\eeq
which, of course, agrees with \eqn{Ftf}.

\begin{figure}
\includegraphics[width=.35\textwidth]{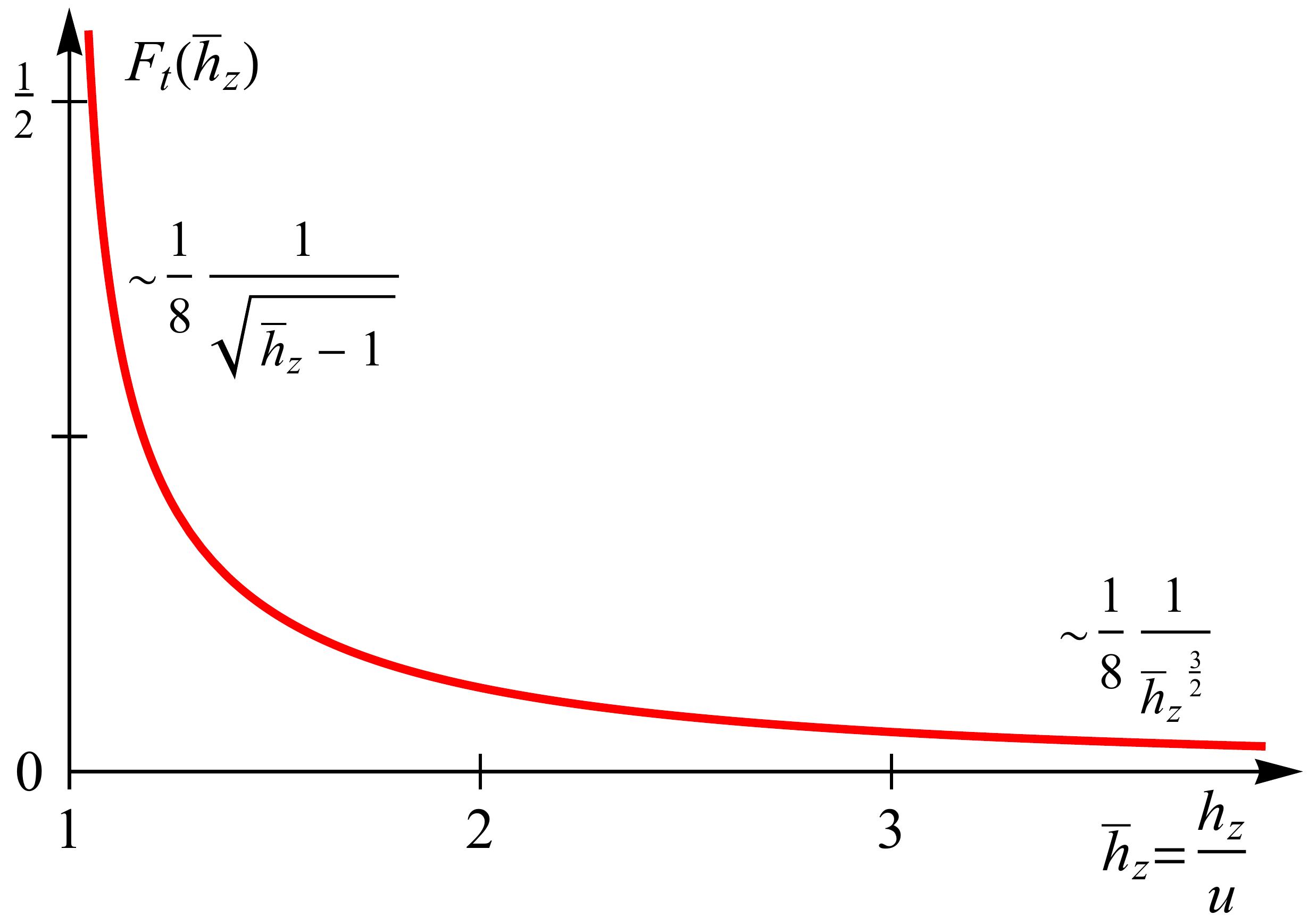}
\includegraphics[width=.35\textwidth]{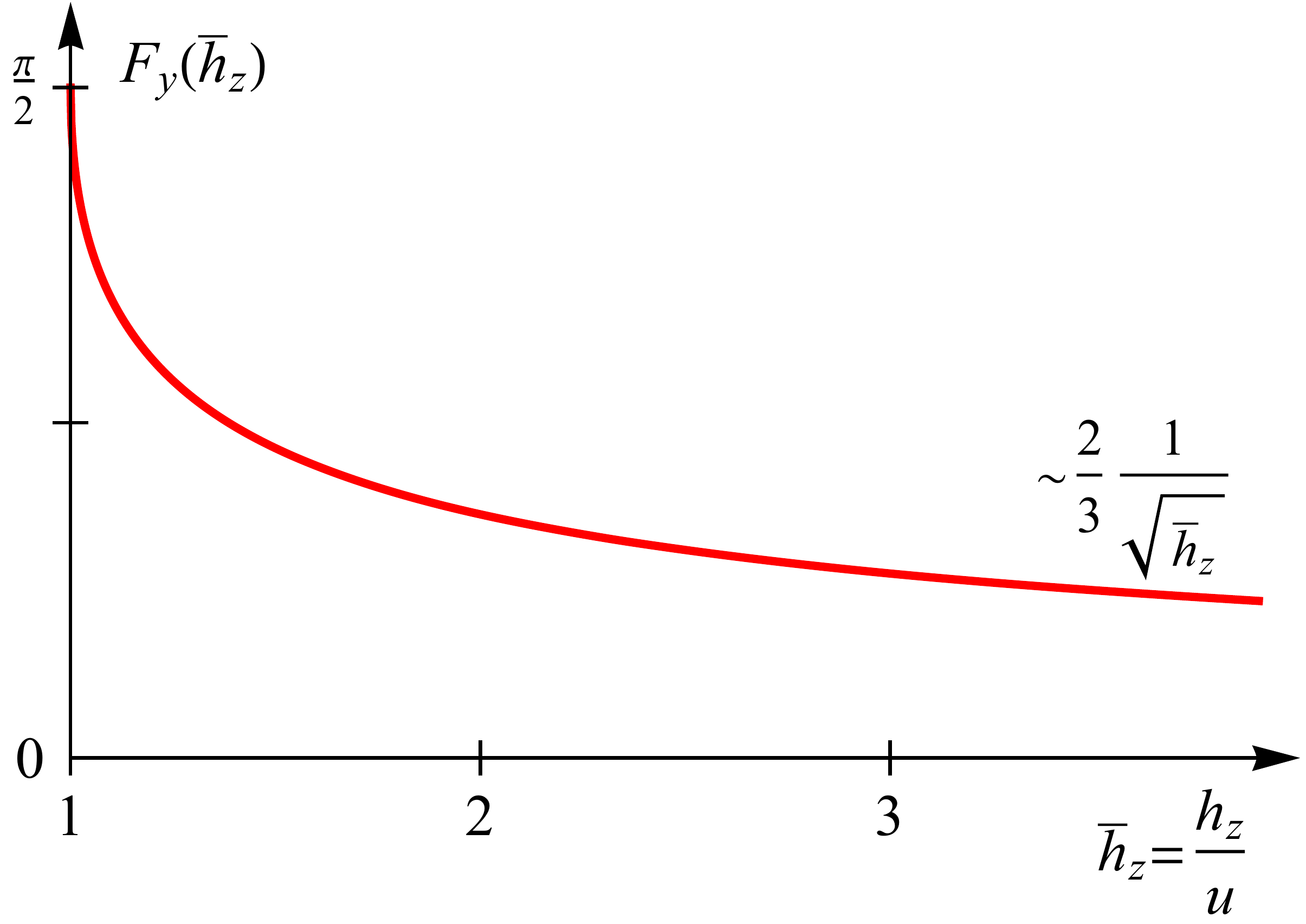}
\caption{(Color online)
The functions $F_t(\hr_z)$ [\eq{Ft}] and $F_y(\hr_z)$ [\eq{Fy}]
that determine the coefficients in front of the the time- and coordinate-derivative terms in the Luttinger liquid model \eqn{L1D}
for the edge excitations of the TnT $\nb^\i=\nb_z$ phase.
The asymptotic functions (\ref{eq:Ftfa}) and (\ref{eq:Fyfa})
at the transition point $\hr_z=1$ and at large $\hr_z\gg 1$ are indicated.
}
\lbl{fig:F}
\end{figure}

This method of the derivation of the Luttinger liquid model \eqn{L1D} as an expansion of the modified domain-wall energy
also allows us to determine the restrictions on the allowed magnitude of fluctuations.
The domain-wall energy $dE^\x{1D}_0(u,h_z)$ [\eq{Edw}] contains a nonanalytic square-root dependence $\q{\hr_z-1}$.
As a result, the power expansion \eqn{L1Dnlinexp} is valid
as long as the fluctuation energies
\beq
    \dot{\Phi}
,\spc
    \rho (\n_y\Phi)^2 \ll h_z-u
\lbl{eq:L1Dapplic}
\eeq
are much smaller than the deviation $h_z-u$ from the transition point.
This deviation, as one would expect, coincides with the gap of the neutral bulk excitations (isospin waves).

In this regard, we caution about using the unexpanded functional \eqn{L1Dnlin}
at fluctuation energies $\dot{\Phi}$ and $\rho (\n_y\Phi)^2$ comparable to $h_z-u$:
while this is an essentially exact expression {\em under the made approximations},
these approximations amount to neglecting other isospin configurations,
such as bulk excitations, which become relevant at energies $\sim h_z-u$.

In the next \secr{lliqa}, we analyze the main properties of the obtained Luttinger liquid model.

\subsection{Analysis\lbl{sec:lliqa}}

\begin{figure}
\includegraphics[width=.35\textwidth]{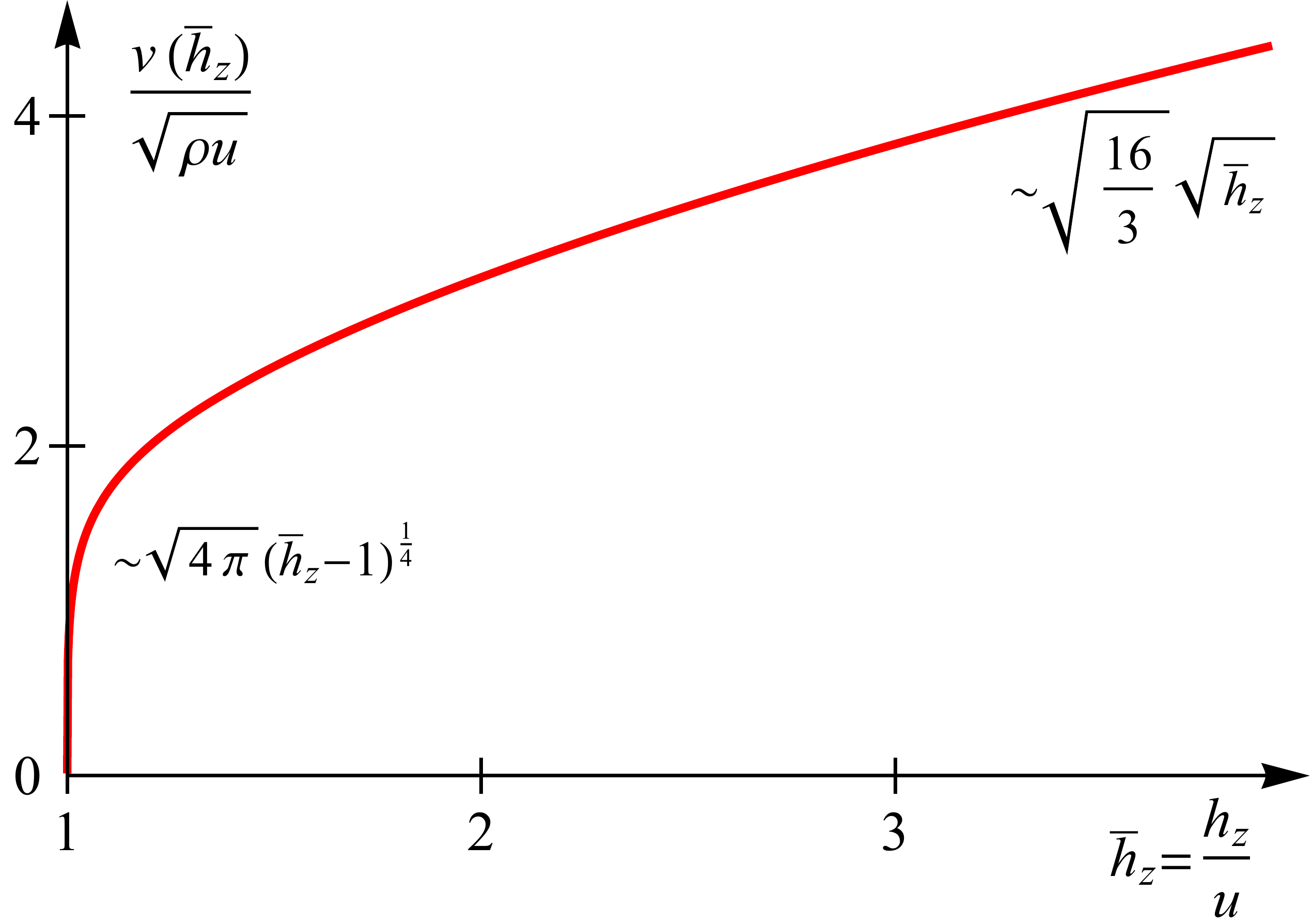}
\includegraphics[width=.35\textwidth]{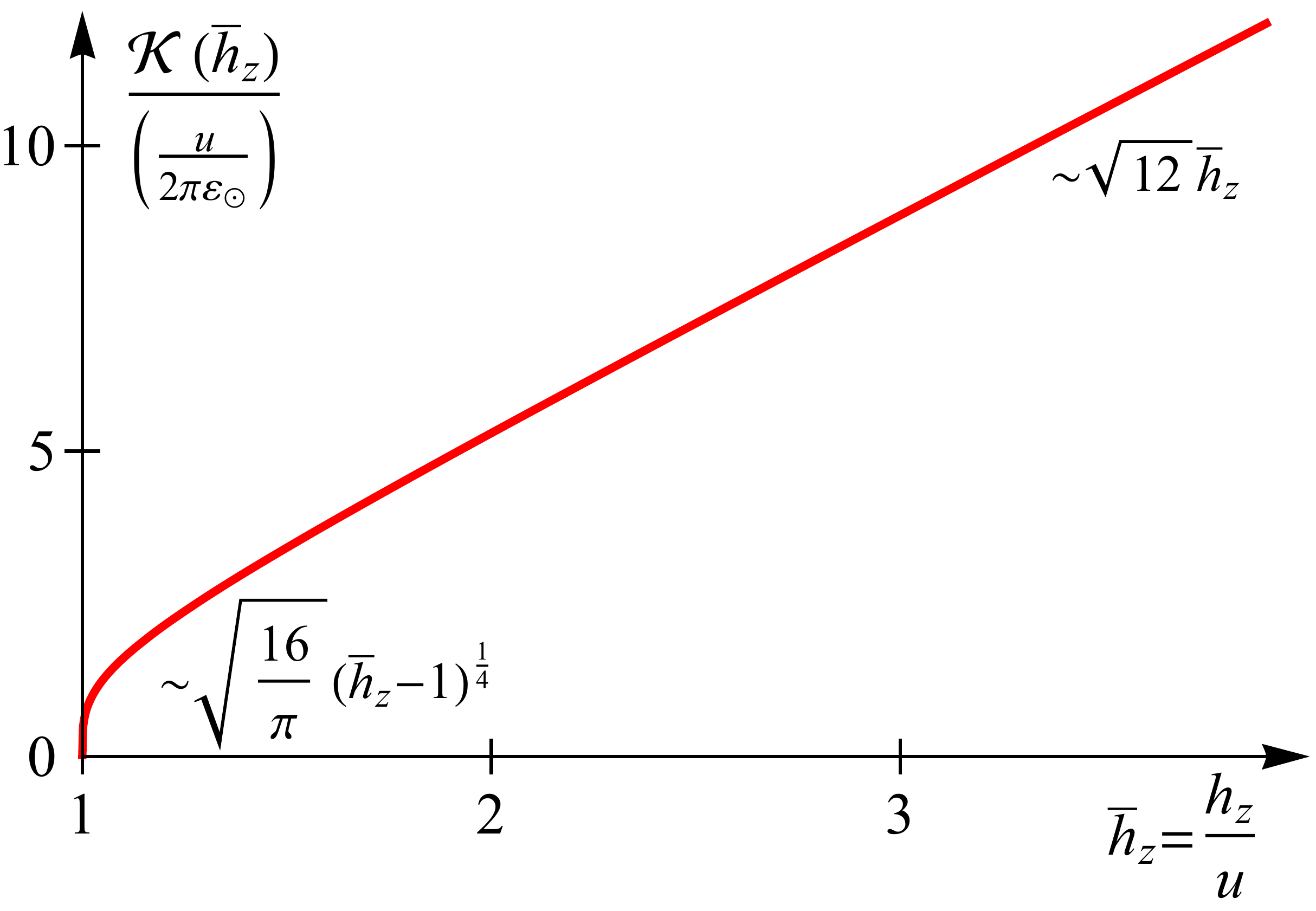}
\caption{(Color online)
The velocity $v(\hr_z)$ [\eq{v}] and interaction parameter $\Kc(\hr_z)$ [\eq{Kc}]
of the Luttinger liquid model (\ref{eq:L1D})
for the edge excitations of the TnT $\nb^\i=\nb_z$ phase.
The asymptotic functions (\ref{eq:va}) and (\ref{eq:Kca})
at the transition point $\hr_z=1$ and at large $\hr_z\gg 1$ are indicated.
}
\lbl{fig:vK}
\end{figure}

The Luttinger liquid model \eqn{L1D} describes the edge excitations of the $\nb^\i=\nb_z$ phase, realized at $u<h_z$;
its collective variable is the angle $\Phi(y;t)$ varying with time $t$ and coordinate $y$ along the edge.
To the leading approximation, the isospin texture associated with $\Phi(y;t)$ is given by
the ``deformed'' ground state configuration $\nb_0(x|\Phi(y;t))$ [\eqs{n0}{tht0+<}].
The field $\Phi(y;t)$ thus corresponds to polarization in the $xy$ isospin plane.

Simultaneously, this edge isospin texture carries electric charge. According to \eq{kappa},
the charge density per unit length in the $y$ direction,
integrated over a cross-section of constant $y$, reads
\beqar
    \kappa^\x{1D}[\Phi](y;t)
    &\equiv&\int^0_{-\i}\dx x\, \kappa[\nb_0(x|\Phi(y;t))]\nn\\
    &=&\f1{4\pi}\int_{-\i}^0\dx x\,\sin\tht_0(x) \n_x\tht_0(x) \n_y\Phi(y;t)\nn\\
    &=&\f1{2\pi}\n_y\Phi(y;t).
\lbl{eq:kappa1D}
\eeqar
The associated electric current in the $y$ direction equals
\beq
    j^\x{1D}[\Phi](y;t)
    =-\f1{2\pi}\dot{\Phi}(y;t),
\lbl{eq:j1D}
\eeq
as follows from the continuity equation
\[
    \dot{\kappa}^\x{1D}+\n_y j^\x{1D}=0.
\]

Therefore, the single field $\Phi(y;t)$ carries both isospin and charge degrees of freedom,
``locked'' to each other, and the Luttinger liquid \eqn{L1D} represents an isospin ``helical'' liquid~\cite{helicalLL}.

The theory is fully characterized by two parameters:
the velocity $v$ of the linear gapless excitation spectrum $\om=vk$
and the dimensionless parameter $\Kc$ describing the effective strength of interactions~\cite{Gia}.
We remind the reader that in a generic Luttinger liquid $\Kc=1$ corresponds to a noninteracting system;
$0<\Kc<1$ is the range of repulsive interactions, the stronger the interactions, the smaller $\Kc$;
and $1<\Kc$ is the range of attractive interactions, the stronger the interactions, the larger $\Kc$.

From \eqs{Ft}{Fy}, the parameters are expressed in terms of the functions $F_{t,y}(\hr_z)$ [\eqs{Ftf}{Fyf}] as
\beqar
    v(\hr_z)&=&\q{\rho u}\q{\f{F_y(\hr_z)}{F_t(\hr_z)}}
\lbl{eq:v}
    ,\\
    \Kc(\hr_z)&=&\f{u}{2\pi\eps_\odot} \f1{\q{F_t(\hr_z) F_y(\hr_z)}}.
\lbl{eq:Kc}
\eeqar
Their asymptotic expressions at the phase transition point $h_z=u$ and at large $h_z$,
following \eqs{Ftfa}{Fyfa}, are
\beqar
    v(\hr_z)&=&
        \q{\rho u}\lt\{\ba{ll}
        \q{4\pi} (\hr_z-1)^{\f14},& h_z\rarr u+0,\\
        \q{\f{16}3}\q{\hr_z},& h_z\gg u.
     \ea\rt.
\lbl{eq:va}
\\
    \Kc(\hr_z)&=&
        \f{u}{2\pi\eps_\odot}
            \lt\{\ba{ll}
        \q{\f{16}{\pi}}(\hr_z-1)^{\f14},& h_z\rarr u+0,\\
        \q{12}\,\hr_z,& h_z\gg u.
     \ea\rt.
\lbl{eq:Kca}
\eeqar
The dependence of the parameters $v(\hr_z)$ and $\Kc(\hr_z)$ on the normalized field $\hr_z$ is plotted in \figr{vK}
and follows from the behavior of the functions $F_y(\hr_z)$ and $F_t(\hr_z)$, plotted in \figr{F}.

Both the velocity $v(\hr_z)$ and interaction parameter $\Kc(\hr_z)$ are growing functions of the field $\hr_z$
i.e., increase as the magnetic field $B$ decreases.
Not too close to the phase transition $h_z=u$, when $h_z-u\gtrsim u$,
they scale as
\beqar
    v&\sim&\q{\rho \max{\{u,h_z\}}}\sim
    e_*^2 \q{\f{\max{\{u,h_z\}}}{\eps_\odot}}, \lbl{eq:vest}\\
    \Kc&\sim&\f{\max{\{u,h_z\}}}{\eps_\odot}\ll 1. \lbl{eq:Kcest}
\eeqar
Thus, in the whole range of applicability of the QHFM theory,
when the energy scales $u,h_z\ll\eps_\odot$ of the SU(2)-asymmetric terms
are much smaller than that $\eps_\odot$ of the SU(2)-symmetric interactions,
set by the Coulomb energy [\eq{epsod}],
the interaction parameter $\Kc\ll 1$ is small
and the Luttinger liquid is {\em strongly interacting}.

Moreover, since $F_t(\hr_z\rarr1+0)\sim \f1{\q{\hr_z-1}}$ diverges while $F_y(\hr_z=1)=\f{\pi}2$ becomes constant,
both the velocity $v\sim\q{\rho u}(\hr_z-1)^{\f14}$ and the interaction parameter $\Kc\sim\f{u}{\eps_\odot}(\hr_z-1)^{\f14}$
approach zero at the transition point, $h_z\rarr u+0$.
The Luttinger liquid therefore becomes {\em infinitely} strongly interacting at the transition point.
We note though that the energy window of applicability of this low-energy theory narrows accordingly,
see \eq{L1Dapplic}; for larger fluctuations, the Luttinger liquid model becomes invalid.

The Luttinger liquid models for the edge excitations in the form of \eq{L1D} were
obtained in Ref.~\ocite{Pik} for the double-layer system with inverted band structure
and in Ref.~\ocite{F4} for the F phase of the $\nu=0$ state in graphene.
However, in Ref.~\ocite{Pik}, the expression for the coefficient $F_t(\hr_z)$ at the time-derivative term
does not diverge at the phase transition $h_z\rarr u+0$.
This divergence is physical and to be expected, since the 1D model should fail at the phase transition, where the bulk modes become gapless.
In Ref.~\ocite{F4}, only the scaling of the parameters $F_{t,y}(\hr_z)$, $v(\hr_z)$, $\Kc(\hr_z)$
at the phase transition $h_z\rarr u+0$ was determined, which does agree with our asymptotic results,
whereas we obtain explicit analytical expressions \eqn{Ft}, \eqn{Fy}, \eqn{v}, and \eqn{Kc} for them at all $u<h_z$.

\begin{figure}
\includegraphics[width=.48\textwidth]{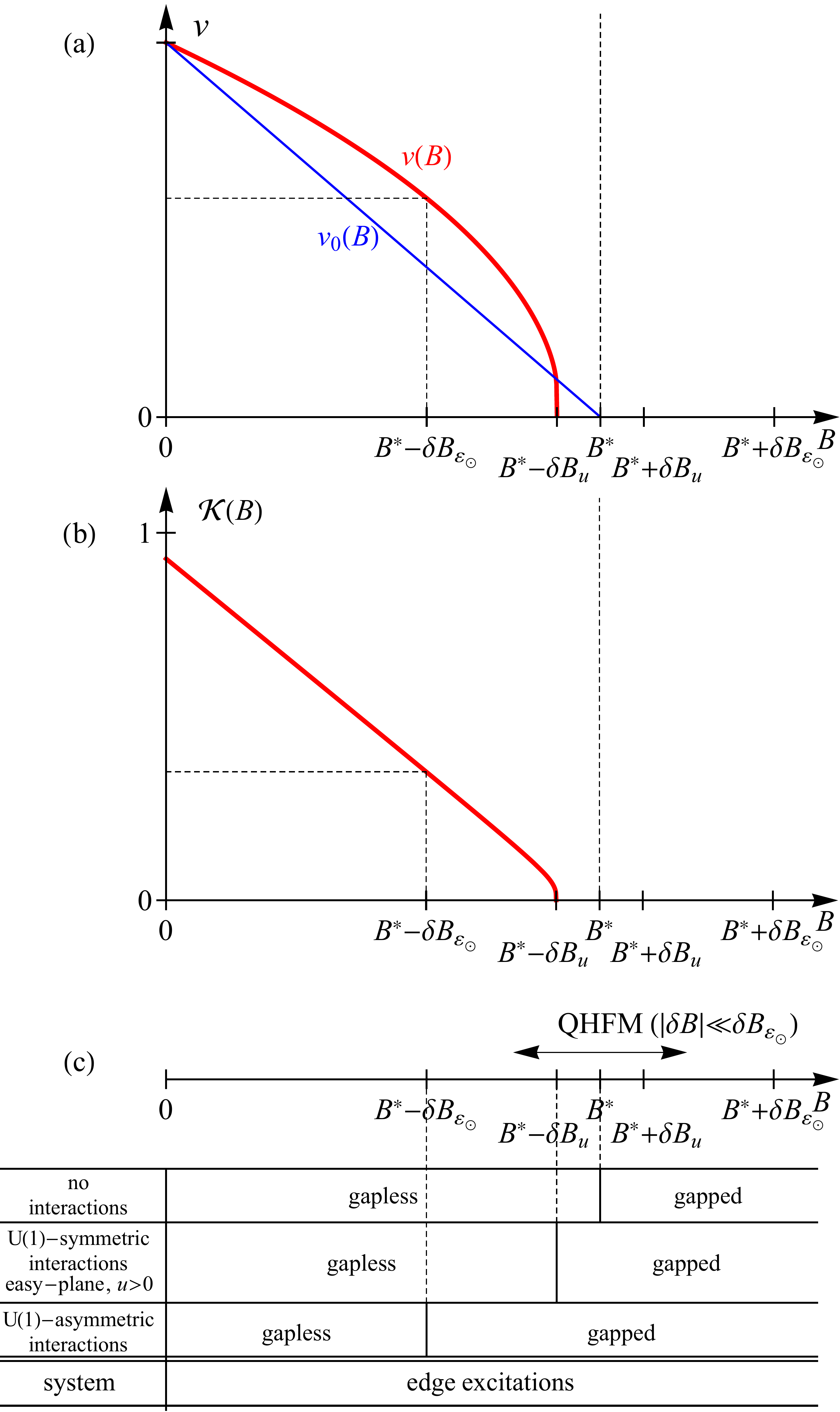}
\caption{
(Color online) Schematic plots of the magnetic-field dependencies of the velocity $v(B)$ (a) and interaction parameters $\Kc(B)$ (b)
of the helical Luttinger liquid describing the edge excitations of the TnT phase of a QHTI
for U(1)-symmetric interactions with easy-plane anisotropy $u>0$.
At small fields $B\ll B^*$, the effective interactions are weak, $\Kc(B)\approx 1$, and
the velocity $v(B)\approx v_0(B)$ is close to its bare value.
In the QHFM regime close to the single-particle topological phase transition,
the effective interactions are strong, $\Kc(B)\ll 1$.
The effective interactions are thus highly tunable.
(c) Summarized properties of the edge excitations.
The TT phase with gapped edge excitations ensues
at lower magnetic fields with switching on the interactions
and lowering their symmetry.
}
\lbl{fig:vK(B)}
\end{figure}

The helical Luttinger liquid  model for the edge excitations above was
derived in a controlled way in the QHFM regime, realized in the vicinity of the topological phase transition at $B=B^*$.
However, the single-particle TnT phase is realized in the whole range $0\leq B<B^*$ of magnetic fields.
It is safe to argue that if, in the presence of interactions,
the Luttinger liquid persists in the region of strong effective interactions near the transition point,
it also persists at all magnetic fields down to zero, i.e., in the range $0\leq B< B^*-\de B_u$.

Consider the ``bare'' velocity $v_0(B)$ of the counter-propagating single-particle edge states
at the crossing point of their energy curves, $\e(p)=h_z$.
It is largest at zero field $B=0$ and monotonically decreases down to zero $v_0(B\rarr B^*-0)=0$
at the single-particle phase transition point, as shown schematically in \figr{vK(B)}.
The effective strength of the interactions {\em at the edge} can be characterized by the dimensionless parameter $e_*^2/v_0(B)$,
which is also roughly equivalent (for a ``sharp enough'' edge) to the ratio $\eps_\odot/h_z$
characterizing the strength of interactions {\em in the bulk}.
The effective strengths of interactions in the bulk and at the edge are thus in accord with each other.

We remind that we work under the assumption of weak bare Coulomb interactions, \secr{hspace},
which can also be formulated as $v_0(B=0)\gg e_*^2$.
Thus, at zero or small fields $B\ll B^*$,
the edge states are weakly interacting and the corresponding low-energy theory
can be derived using the standard procedure~\cite{Gia} based on the linearization
of the spectrum and will have the form of a Luttinger liquid.
The velocity $v(B=0)\approx v_0(B=0)$ of the collective excitations of this Luttinger liquid
will be close to the bare velocity and the interaction parameter $\Kc\approx 1$ will be close to unity,
corresponding to weak effective interactions.

On the other hand, in the QHFM regime, as we have shown above,
the effective interactions in the Luttinger liquid are strong, with $v(B)\ll e_*^2$ and $\Kc(B)\ll 1$ [\eqs{vest}{Kcest}].
The regime of intermediate effective interaction strength
with $v\sim e_*^2$ and $\Kc\sim 1$ occurs at the verge of the QHFM regime,
at such fields $B=B^*-\de B_{\eps_\odot}$,
\beq
    \de B_{\eps_\odot}\equiv\f{\eps_\odot}{|\pd_B h_z(B^*)|},
\lbl{eq:dBeC}
\eeq
where $h_z(B)\sim\eps_\odot(B)$.
In this intermediate-strength regime, the weakly and strongly interacting Luttinger liquids must continuously connect.

Thus, the helical Luttinger liquid describing the edge excitations of a QHTI
persists in almost the whole range of the noninteracting TnT phase, $0\leq B< B^*-\de B_u$,
and is highly tunable there:
the magnetic field allows one to tune the effective interactions
between weak at lower $B$ and infinitely strong at higher $B$, close to the topological transition
to the intermediate phase $\nb^\i=\nb^*(\vphi_0)$ at $B=B^*-\de B_u$.
The corresponding behavior of the velocities $v(B)$ and interaction parameter $\Kc(B)$
in the whole range is plotted schematically in \figr{vK(B)}.

However, we remind that the derived Luttinger liquid theory for the edge
was obtained under the specific assumption of $\Ux(1)$ isospin symmetry [\eqs{D}{Dchi}] of the many-body Hamiltonian $\Hh$ [\eq{H}].
This $\Ux(1)$ symmetry is inherited by the Lagrangian (\ref{eq:L1D}) of the Luttinger liquid,
which is invariant under rotations of the angle field:
\[
    \Phi(y;t)\rarr\Phi(y;t)+\phi.
\]
We now turn to the analysis of the effects that break $\Ux(1)$ symmetry, \secsr{U1nH}{U1nEc}.

\subsection{Broken U(1) symmetry \lbl{sec:U1nlliq}}

\subsubsection{Broken physical symmetry, single-particle effect}

As explained in \secr{U1nH}, there are two categories of $\Ux(1)$-asymmetric terms:
ones that break the physical symmetry and exist already at the single-particle level, $\Ec_{1\os}(\nb)$ [\eq{Ecp}],
and the ones that preserve the physical symmetry and can arise only from interactions, $\Ec_{2\os}(\nb)$ [\eq{Ec2}].
If small, these terms can be incorporated into the Luttinger liquid model (\ref{eq:L1D}).

Proceeding along the same lines as in \secr{lliq},
we expand the term
\beqarn
    \Ec_{1\os}(\nb)
        &=&-h_\p[\sin\tht_0(x)+\Oc(\de\tht)]\cos(\vphi-\vphi_{1\os})\\
        &\rarr& -h_\p \sin\tht_0(x) \cos(\vphi-\vphi_{1\os})
\eeqarn
of the first category
about the ground state [\eq{dtht}] and keep only the zero-order term in $\de\tht$.
Due to the constraining function $\sin\tht_0(x)$, the approximation (\ref{eq:vphiapprox}) for the variable $\vphi(\rb;t)$ may then be used.
This yields the contribution
\beq
    -\Lf^\x{1D}_{1\os}[\Phi]=-h_\p^\x{1D} \int\dx y\, \cos(\Phi(y;t)-\vphi_{1\os}),
\lbl{eq:L1Dp}
\eeq
to the 1D Lagrangian of the edge excitations, where
\beqar
    h_\p^\x{1D}&=&\f{h_\p l_u}{s} F_\p(\hr_z),\nn\\
    F_\p(\hr_z)&=&\int_{-\i}^0\dx\xr\,\sin\tht_0(\xr;\hr_z)=\x{ln}\f{\q{\hr_z}+1}{\q{\hr_z}-1}.
\lbl{eq:hp1D}
\eeqar

The Lagrangian $\Lf^\x{1D}[\Phi]+\Lf^\x{1D}_{1\os}[\Phi]$
describes the sine-Gordon model~\cite{Gia}, the properties of which are well-studied.
Its excitation spectrum is gapped at $\Kc<2$, i.e.,
including all repulsive interactions, $0<\Kc<1$, the noninteracting case $\Kc=1$,
and an (irrelevant for the considered system) range $1<\Kc<2$ of attractive interactions.

Thus, as expected, if the physical symmetry is broken,
there is no topological protection already at the single-particle level ($\Kc=1$).
As a result, in the presence of repulsive interactions ($\Kc<1$),
the system is in a TT phase with broken U(1) symmetry and gapped edge excitations for all magnetic fields $B\geq0$.

\subsubsection{Preserved physical, but broken $\Ux(1)$ symmetry, interaction effect}

Similarly, expanding the term of the second category about the ground state,
and keeping only the zeroth-order term in $\de\tht$,
\beqarn
    \Ec_{2\os}(\nb)
        &=&\f12u_{2\os}[\sin^2\tht_0(x)+\Oc(\de\tht)]\cos2(\vphi-\vphi_{2\os})\\
        &\rarr&\f12u_{2\os} \sin^2\tht_0(x) \cos2(\vphi-\vphi_{2\os}),
\eeqarn
and performing the substitution (\ref{eq:vphiapprox}),
we obtain a contribution
\beq
    -\Lf^\x{1D}_{2\os}[\Phi]=u^\x{1D}_{2\os}\int\dx y\, \cos2(\Phi(y;t)-\vphi_{2\os})
\lbl{eq:L1D2}
\eeq
to the 1D Lagrangian of the edge excitations, where
\[
    u^\x{1D}_{2\os}=\f{u_{2\os} l_u}{s} F_y(\hr_z).
\]
Note that the coefficient $u^\x{1D}_{2\os}$ appears to be given by the same function $F_y(\hr_z)$ [\eq{Fy}]
as for the gradient term.

Therefore, for preserved physical symmetry (inversion),
the edge is described by the Lagrangian $\Lf^\x{1D}[\Phi]+\Lf^{1D}_{2\os}[\Phi]$.
This is also a sine-Gordon model,
but \eq{L1D2} differs from \eq{L1Dp} by the numerical factor (2 instead of 1) in the cosine argument.
As a result, the edge ground state breaks U(1) symmetry (the field acquires a finite expectation value $\ln\Phi\rn\neq0$)
and the edge excitations become gapped in a different range of interaction strength, namely at $\Kc<\f12$.
According to \secr{lliqa},
the quantum phase transition at the field $B_{2\os}\sim B^*-\de B_{\eps_\odot}$ [\eq{dBeC}] such that $\Kc(B_{2\os})=\f12$
occurs at the verge of the QHFM regime, where $h_z(B_{2\os})\sim\eps_\odot(B_{2\os})$.
The system has gapless edge excitations for lower fields $B<B_{2\os}$
and gapped excitations for all higher fields $B>B_{2\os}$.

\section{Role of U(1) symmetry for topological protection \lbl{sec:U(1)}}

One of the key questions raised in the studies of topological systems
is how interactions affect the topological properties.
A common understanding is that a noninteracting topological insulator
is generally not guaranteed to remain such in the presence of interactions.
Our present findings allow us to specify the symmetry requirements
for the topological protection in the presence of interactions.

We remind that the QHTIs we consider are protected by some physical symmetry, see \secr{QHTIs}.
Due to this symmetry, two relevant LLs (\figr{LLs}) have different transformation properties
and are thus not coupled at the single-particle level.
As a result, the projected {\em single-particle} Hamiltonian within these two LLs
possesses $\Ux(1)$ symmetry with respect to uniaxial isospin rotations [\eqs{n}{Dchi}].

Aggregating the results of \secsr{edgeexc}{lliq},
we conclude that this {\em effective continuous} $\Ux(1)$ symmetry is
the one responsible for the topological protection {\em in the presence of interactions} in (at least) this class of systems, QHTIs.
If U(1) symmetry is preserved, the TnT phase with gapless edge charge excitations persists for any effective strength of interactions,
even in the QHFM regime of strong interactions, realized in the vicinity of the single-particle topological phase transition.

However, we find that the physical symmetry alone is generally not sufficient
to protect the TnT phase in the presence of interactions.
The interactions preserving the physical symmetry can still break $\Ux(1)$ symmetry
and thereby destroy the TnT phase for strong enough effective interactions.

More precisely, ``preserved U(1) symmetry'' means that {\em both} (i) the bulk ground state
and (ii) the interacting projected Hamiltonian are U(1)-symmetric.
Accordingly, the two mechanisms by which U(1) symmetry can be broken
correspond to violation of one of these conditions:

(1) First, as demonstrated in \secr{bulk},
even the $\Ux(1)$-symmetric interactions (condition (ii) satisfied)
with the right properties (``easy-plane'' anisotropy)
can result in a bulk ground state
with spontaneously broken $\Ux(1)$ symmetry (condition (i) violated):
the $\nb^\i=\nb_0^*(\vphi_0)$ phase.
As demonstrated in \secr{edgeexc}, this phase has gapped edge charge excitations
and is thus TT.

(2) Second, as demonstrated in \secsr{U1nH}{U1nEc},
interactions can (depending on the physical symmetry)
contain terms that preserve the physical symmetry, but break the effective $\Ux(1)$ symmetry (condition (ii) violated).
As then shown in \secr{U1nlliq}, these terms result in a phase transition to the TT phase
with broken $\Ux(1)$ symmetry at the edge and gapped excitation spectrum.
At the same time, the U(1)-symmetric bulk ground state $\nb^\i=\nb_z$ may still persists beyond this transition
(condition (i) satisfied).
We mention that similar types of interactions have earlier been considered~\cite{helicalLL}
in the studies of helical Luttinger liquids for the edge of 2D topological insulators protected by time-reversal symmetry.

The transitions from the TnT phase to either of these TT phases occur
upon increasing the magnetic field $B$,
as the single-particle phase transition point $B^*$ is approached and
the effective interactions get stronger,
but at fields {\em lower} than $B^*$: at $B^*-\de B_u$ [\eq{dBu}]
and $B_{2\os}\sim B^*-\de B_{\eps_\odot}$ [\eq{dBeC}], respectively,
as summarized in \figr{vK(B)}(c).
Thus, these are {\em interaction-induced topological quantum phase transitions},
enabled due to the tunability of the effective interactions by the magnetic field, \secr{lliqa}.

\section{Conclusion\lbl{sec:concl}}

In this work, we studied the effect of electron interactions
on the topological properties of {\em quantum Hall topological insulators}.
Due to the crossing of Landau levels at the single-particle topological phase transition,
its vicinity is automatically the regime of strong effective interactions.
An appealing theoretical aspect of such a system is that it can be studied
in a controlled way within the framework of quantum Hall ferromagnetism.

A particular attention was paid to establishing the requirements for the topological protection in this interacting system.
We find that this question is ultimately related to the effective symmetry of the system:
the continuous U(1) symmetry is a necessary condition for the topologically nontrivial phase
to persist in the regime of strong effective interactions.

If U(1) symmetry is preserved, the edge of the topologically nontrivial phase is described by the helical Luttinger liquid.
The effective interactions of this Luttinger liquid are highly tunable by the magnetic field $B$:
they are weak (for weak bare Coulomb interactions) at small $B$
and grow as $B$ is increased, becoming strong in the quantum Hall ferromagnet regime
in the vicinity of the single-particle topological phase transition.

The U(1) symmetry may be broken, however, either spontaneously or by the interactions that are explicitly U(1)-asymmetric.
In either scenario, this eventually results in a phase transition to a topologically trivial phase with gapped edge excitations,
which can be achieved by tuning the interaction strength by the magnetic field.

The tunability of interactions, the accessibility of the regime of strong effective interactions even
in a system with weak bare interactions, and the possibility to realize interaction-induced topological phase transitions
are among the properties that make quantum Hall topological insulators an attractive class of systems
for investigating the interplay of interactions and topology,
both theoretically and experimentally.

\section{Acknowledgements}

We  acknowledge financial support by the DFG (SPP 1666 and SFB 1170 ``ToCoTronics''), the Helmholz Foundation (VITI),
and the ENB Graduate school on ``Topological Insulators''.

\end{document}